\documentclass[a4paper,fleqn,usenatbib]{mnras}
\pdfoutput=1

\usepackage[T1]{fontenc}
\usepackage{ae,aecompl}

\usepackage{graphicx}	
\usepackage{amsmath}	
\usepackage{amssymb}	
\usepackage[utf8]{inputenc}
\usepackage{epsfig}
\usepackage{capt-of}
\usepackage{url}
\usepackage{multicol}
\usepackage{epstopdf}
\usepackage{caption}
\usepackage{textcomp}
\usepackage[usenames,dvipsnames]{color}
\usepackage{float,lscape}
\usepackage{pdfpages}

\title[KROSS: The Tully-Fisher Relation at $z \sim 1$]{The KMOS Redshift One Spectroscopic Survey (KROSS): The Tully-Fisher Relation at $z \sim 1$}

\author[Tiley et al.]{Alfred L. Tiley,$^{1}$
John P. Stott,$^{1,2}$
A. M. Swinbank,$^{3,2}$
Martin Bureau,$^{1}$
\newauthor Chris M. Harrison,$^{3}$
Richard Bower,$^{2,3}$
Helen L. Johnson,$^{3}$
Andrew J. Bunker,$^{1,4}$
\newauthor Matt J. Jarvis,$^{1,5}$
Georgios Magdis,$^{1,6}$
Ray Sharples,$^{3,7}$
Ian Smail,$^{3,2}$
\newauthor David Sobral$,^{8,9,10}$
Philip Best$^{11}$
\\
$^{1}$Sub-department of Astrophysics, Department of Physics, University of Oxford, Denys Wilkinson Building, Keble Road, Oxford OX1 3RH, UK\\
$^{2}$Institute for Computational Cosmology, Durham University, South Road, Durham, DH1 3LE, UK\\
$^{3}$Centre for Extragalactic Astronomy, Department of Physics, Durham University, South Road, Durham DH1 3LE, UK\\
$^{4}$Affiliate Member, Kavli Institute for the Physics and Mathematics of the Universe, 5-1-5 Kashiwanoha, Kashiwa, 277-8583, Japan\\
$^{5}$Department of Physics, University of the Western Cape, Bellville 7535, South Africa\\
$^{6}$Institute for Astronomy, Astrophysics, Space Applications and Remote Sensing, National Observatory of Athens, GR-15236 Athens, Greece\\
$^{7}$Centre for Advanced Instrumentation, Department of Physics, Durham University, South Road, Durham, DH1 3LE, UK\\
$^{8}$Instituto de Astrof\'{\i}sica e Ci\^{e}ncias do Espa\c{c}o, Universidade de Lisboa, OAL, Tapada da Ajuda, 1349-018 Lisboa, Portugal\\
$^{9}$Department of Physics, Lancaster University, Lancaster, LA1 4YB, UK\\
$^{10}$Leiden Observatory, Leiden University, P.O. Box 9513, NL-2300 RA Leiden, The Netherlands\\
$^{11}$SUPA, Institute for Astronomy, Royal Observatory of Edinburgh, Blackford Hill, Edinburgh, EH9 3HJ, UK
}

\date{Accepted XXX. Received YYY; in original form ZZZ}

\pubyear{2015}

\begin{document}
\label{firstpage}
\pagerange{\pageref{firstpage}--\pageref{lastpage}}
\maketitle

\begin{abstract}
We present the stellar mass ($M_{*}$), and K-corrected $K$-band absolute magnitude ($M_{K}$) Tully-Fisher relations (TFRs) for sub-samples of the 584 galaxies spatially resolved in H$\alpha$ emission by the KMOS Redshift One Spectroscopic Survey (KROSS). We model the velocity field of each of the KROSS galaxies and extract a rotation velocity, $V_{80}$ at a radius equal to the major axis of an ellipse containing 80\% of the total integrated H$\alpha$ flux. The large sample size of KROSS allowed us to select 210 galaxies with well measured rotation speeds. We extract from this sample a further 56 galaxies that are rotationally supported, using the stringent criterion $V_{80}/\sigma > 3$, where $\sigma$ is the flux weighted average velocity dispersion. We find the $M_{K}$ and $M_{*}$ TFRs for this sub-sample to be $M_{K} / \rm{mag}= (-7.3 \pm 0.9) \times [(\log(V_{80}/\rm{km\ s^{-1}})-2.25]- 23.4 \pm 0.2$ , and $\log(M_{*} / M_{\odot})= (4.7 \pm 0.4) \times [(\log(V_{80}/\rm{km\ s^{-1}}) - 2.25] + 10.0 \pm 0.3$, respectively. We find an evolution of the $M_{*}$ TFR zero-point of $-0.41 \pm 0.08$ dex over the last $\sim $8 billion years. However, we measure no evolution in the $M_{K}$ TFR zero-point over the same period. We conclude that rotationally supported galaxies of a given dynamical mass had less stellar mass at $z \sim 1$ than the present day, yet emitted the same amounts of $K$-band light. The ability of KROSS to differentiate, using integral field spectroscopy with KMOS, between those galaxies that are rotationally supported and those that are not explains why our findings are at odds with previous studies without the same capabilities. 
\end{abstract}




\section{Introduction}
\label{sec:introduction}

With the dawn of integral field spectroscopy and integral field units (IFUs) it is now possible to gain both the imaging and spatially resolved spectral information of galaxies to study their morphological, chemical, and dynamical properties and evolution in just a single observation. IFUs reveal spatially resolved information on galaxies' internal dynamics, metallicities, star formation, and stellar mass among other things. Work by surveys such as the ATLAS$^{3\rm{D}}$ Project \citep{Cappellari:2011}, a multi-wavelength galaxy survey using the Spectrographic Areal Unit for Research on Optical Nebulae \citep[SAURON, ][]{Bacon:2001} to kinematically classify galaxy morphology, have arguably transformed the paradigm with which we describe the formation and evolution of (early-type) galaxies \citep[see e.g.][]{Cappellari:2011III}. More beneficial still was the advent of IFUs that operated in the near-infrared (near-IR) (see e.g. Cambridge IR PAnoramic
Survey Spectrograph [CIRPASS], \citealt{Parry:2000,Smith:2004}, Fibre Large Array Multi Element Spectrograph [FLAMES]-GIRAFFE \citealt{Pasquini:2002}, Spectrograph for INtegral Field Observations in the Near Infrared [SINFONI] \citealt{Bonnet:2004}), allowing a comparison of the well understood restframe optical properties of galaxies at $z \sim 0$ with those at $z \sim 1-2$.  

The benefits to our understanding of galaxy evolution that single IFUs such as SAURON and SINFONI have provided are significant. In the present day however, a new era of surveys using Multi-Object Spectrographs (MOSs) - instruments consisting of multiple optical fibre bundles, IFUs or multi-slit spectrographs that can be simultaneously deployed - is allowing spatially-resolved observations of ever increasing numbers of galaxies within much shorter timescales than previously possible with single IFU instruments. The KMOS Redshift One Spectroscopic Survey \citep[KROSS;][]{Stott:2016}, a joint undertaking between the University of Oxford and Durham University, is one such survey. Using the K-band Multi-Object Spectrograph \citep[KMOS; see][]{Sharples:2013} on UT1 of the Very Large Telescope (VLT), KROSS has observed $795$ star-forming galaxies at $z \sim 1$ in the $YJ$-band. H$\alpha$ was detected in 719 of those galaxies observed. 584 of these detections were resolved. KROSS has studied the spatially-resolved dynamics, star formation properties, and metallicities of these 584 galaxies.

At $z \sim 1$, we begin to probe the epoch of peak star formation in the Universe \citep[see e.g.][]{Lilly:1996,Madau:1996,Hopkins:2006}, a key era for galaxy mass assembly. The primary causes of this increased star formation are hotly debated, as are the dominant mechanisms for mass growth. It has been thought that the increased star formation rates in galaxies at intermediate redshifts is due to a larger galaxy merger rate than in the present day Universe \citep[see e.g.][]{Bridge:2007}. However there are recent theoretical claims that suggest an alternative explanation. At the epoch in question, numerical simulations predict that galaxies are being fed by constant streams of cold gas from their surroundings \citep{Dekel:2009aa}. These streams then induce gas instabilities, triggering star formation. Thus gas accretion has a direct and significant effect on the star formation of galaxies at this redshift \citep[for a recent review see][]{Almeida:2014}. With this debate as a backdrop it is essential to constrain how the relationship between the stellar, gaseous and dark mass in galaxies has varied over cosmic time, and determine whether this is related to the global fall of star formation activity with decreasing look-back time since $z \sim 1-3$ \citep[e.g. ][]{Sobral:2014}. This work employs the Tully-Fisher relation (TFR) as a useful tool with which to explore this issue.

The TFR is a fundamental scaling relation describing the interdependence of luminous and dark matter in galaxies. It provides a simple means of tracing the evolution of the mass-to-luminosity ratio of populations of galaxies at different epochs. First devised by \citet{Tully:1977aa}, it was initially only considered for disk galaxies, which are predominantly rotationally supported, allowing their rotation velocity to be utilised as a proxy for their total dynamical mass. At first it was used solely as a cosmological distance indicator, however it was soon employed as a tool to probe the nature of various populations of galaxies. 

The TFR of local (i.e. low-$z$) late-type galaxies is well studied \citep[see e.g.][]{TullyPierce2000,Bell:2001aa,Masters:2008aa,Lagattuta:2013aa}. However the Tully-Fisher relation at intermediate redshifts was until relatively recently unknown. There have been several studies of the Tully-Fisher relation of small numbers of intermediate redshift galaxies (i.e. $z \sim 1-2$) \citep[see e.g.][]{Conselice:2005aa,Flores:2006,Kassin:2007b,Kassin:2007a,Puech:2008,ForsterSchreiber:2009,Cresci:2009,Miller:2011,Gnerucci:2011,Miller:2012,Swinbank:2012b,Sobral:2013} but until now it has not been possible to observe large enough numbers of galaxies at this epoch in order to compose statistically large samples with which to compare to local samples. With the use of KMOS, the KROSS survey is in a position to be able to address this problem. In this work we present the K-corrected absolute $K$-band ($M_{K}$) and stellar mass ($M_{*}$) TFRs for the KROSS galaxies - the largest sample of galaxies of its kind at $z \sim 1$.  

Circular motion is one of the main assumptions of the TFR. Thus, to maintain the validity of this assumption, we must only consider those galaxies that are predominantly rotationally supported in any comparison of TFRs. Whilst this is true of late-type galaxies in the local Universe, it is not neccessarily true of all the hotter ``disky" galaxies we see at $z \sim 1$ \citep[see e.g.][]{Genzel:2006}.  In this respect KROSS has the following advantages; firstly, such a large sample as KROSS allows us to make a meaningful comparison between the TFR at $z \sim 1$ and the local Universe by selecting for rotationally supported galaxies whilst maintaining a reasonable sample size. Secondly, there is a particular benefit to using IFU observations, as opposed to slit or fibre observations, in that it allows us to trace the entire two-dimensional kinematics whilst avoiding susceptibility to centering inaccuracies. This spatially resolved information is unique to IFU observations and allows us to separate those galaxies that are rotationally supported from those that are pressure supported with much greater ease and certainty than previously possible with slit studies, for example. These matters are discussed further in \S\ref{subsec:subsample}, \S\ref{subsec:results}, and \S\ref{sec:discussion}. However, it can be seen from the outset that, in comparison to previous studies, KROSS, with its large sample of IFU galaxy observations, is in a strong position from which to cast a more definitive light on the evolution of the Tully-Fisher relation over the last $\sim 8$ Gyr.

This paper is structured as follows: In \S\ref{sec:KROSSsurvey} we describe the KROSS survey; we detail our methods of data acquisition and reduction using KMOS, and describe the velocity field modelling. We also discuss the extraction of absolute magnitudes and stellar mass values via fitting of the spectral energy distributions (SEDs) of the KROSS galaxies. In \S\ref{sec:KROSSTF} we outline the selection criteria for three samples drawn from KROSS, namely the {\it parent} sample and sub-samples {\it all} and {\it disky}. We present the $M_{K}$ and $M_{*}$ Tully-Fisher relations for the sub-samples, and outline the methods used to construct and fit the relations. \S\ref{sec:discussion} comprises a discussion and interpretation of the resultant relations, including a comparison to existing lower and higher redshift Tully-Fisher studies. Finally, in \S\ref{sec:conclusions} we give our main conclusions and outline our intentions for future work that will build and expand on the results presented here.

A cosmology of $\Omega_{\Lambda}=0.73$, $\Omega_{m}=0.27$, and $H_{0}=72$ kms$^{-1}$ Mpc$^{-1}$ is used throughout this work. All magnitudes are quoted in the Vega system. All stellar masses are calculated assuming a Chabrier initial mass function (IMF), as detailed in \citet{Chabrier:2003}. Masses extracted from the literature were converted to a Chabrier IMF \citep[based on offsets taken from][]{Madau:2014} as

\begin{multline}
\log M_{*,\rm{C}}=\log M_{*,\rm{K}}-0.034=\log M_{*,\rm{S}}-0.215 \\ 
 =\log M_{*,\rm{dS}}-0.065
\end{multline}

\noindent where $M_{*,\rm{C}}$, $M_{*,\rm{K}}$, $M_{*,\rm{S}}$, and $M_{*,\rm{dS}}$ are the galaxy stellar masses calculated assuming a Chabrier, Kroupa \citep{Kroupa:2001}, Salpeter \citep{Salpeter:1955}, and ``diet-Salpeter" \citep{Bell:2001aa,Bell:2003} IMF respectively. There are a number of differing stellar mass conversion factors used in the literature \citep[see e.g.][]{Karim:2011,Papovich:2011,Zahid:2012,Speagle:2014}. To account for this, in this work we incorporate an uncertainty of $\pm 0.06$ dex in stellar mass in any measured offset between $M_{*}$ TFRs where masses were originally derived assuming an IMF other than Chabrier.  

\section{The KMOS Redshift One Spectroscopic Survey: Description of the Survey}
\label{sec:KROSSsurvey}

\subsection{KMOS}

KMOS \citep{Sharples:2013} is mounted on UT1 of the VLT, Cerro Paranal, Chile. It consists of 24 arms that are deployable in the focal plane of the telescope over a circular area of diameter 7.2$'$. Each arm has a pick-off mirror that directs the incident light to an image slicer, forming an integral field unit (IFU). Each IFU has a field of view 2.8$''$ x 2.8$''$ and divides this area into 14 slices, each acting like a classical slit. Each slice is further divided into 14 spatial pixels (spaxels). The result is 14 x 14 spaxels within the field of view, each of size 0.2$''$ x 0.2$''$ and each with a full spectrum associated with it. 

KROSS galaxies were observed with KMOS in the $YJ$ band, which covers a wavelength range of approximately $1.02$--$1.36 \mu$m. Resolving power across the $YJ$ band ranges from $R \sim 3000$ at shorter wavelengths to $R \sim 4000$ at longer wavelengths. KMOS allows for simultaneous integral field spectroscopy of $24$ targets in a given field of view, and thus proves a valuable tool in conducting large surveys of the kinematic properties of galaxies over a large range of redshifts. Importantly, its wavelength range is such that galaxies at intermediate redshifts may be observed in the well-understood optical restframe. For a more detailed description of KMOS, see the KMOS User's Manual \footnote{\url{https://www.eso.org/sci/facilities/paranal/instruments/kmos/doc.html}}.

\subsection{Survey Aims}

Using KMOS, KROSS aimed to detect the H$\alpha$ emission line (redshifted in to the $YJ$ band at $z\sim1$) from the warm ionised gas within galaxies at $z\sim1$. Combined with other emission lines, such as the [NII] doublet, the internal dynamics of the galaxies can be studied along with other properties such as their chemical abundances, star formation rates, and ionisation mechanisms. 

Now complete, KROSS has observed $795$ star forming galaxies at $0.8<z<1$. For a detailed description of the sample selection and statistics see \citet{Stott:2016}. Here we give a brief summary. The sample of galaxies was selected using a magnitude cut $K<22.5$, and a colour cut $r$-$z<1.5$. These selections were made with the intention of selecting so-called ``blue cloud" galaxies \citep[see ][]{Bell:2004}, characterised by their blue colour and ongoing star-formation.  Some fainter, redder, more passive (or dusty) galaxies were also included in the sample selection, but these were given a lower priority for observations. Target galaxies for KROSS are spread between several fields, namely the Extended \textit{Chandra} Deep Field-South Survey (ECDFS) field, the Special Selected Area field (SA22), the COSMOlogical evolution Survey (COSMOS) field, and the UKIRT Infrared Deep Sky Survey (UKIDSS) Ultra Deep Survey (UDS) field. KROSS detected H$\alpha$ in $719$ galaxies. $584$ of these detections were resolved.

\subsection{Data Reduction}
For a detailed description of the reduction process see \citet{Stott:2016}. Here we present a summary.

The ESO Recipe EXecution tool ({\sc esorex}) and the Software Package for Astronomical Reduction with KMOS ({\sc spark}) pipeline \citep{Davies:2013aa} were utilised in order to reconstruct the datacube from each IFU observation. The pipeline performs initial corrections to the data using dark, flat, and arc frames, as well as an additional illumination correction. A telluric correction to each observation was made using an observed standard star.
Observations were taken in an ABAABA nod-to-sky pattern, where A represents time on source, and B time on sky. Upon reconstruction of the datacubes, each AB pair was then further reduced separately. 

An initial A-B sky subtraction was made of the cubes, using the temporally closest sky. Following this an attempt was made to remove any remaining residual sky using a designated sky cube per spectrograph. This ``sky'' cube was further median-collapsed to a single spectrum, reducing the noise between sky emission lines, and subtracted from the spectrum of each spaxel in the object cube. See \citet{Stott:2016} for a description of the residual sky subtraction. 

The reduced cubes of the same object from several observations were then combined via a 3-sigma-clipped average, using the header information for alignment.

\subsection{Modelling Velocity Fields}
\label{subsec:modelvels}

For a detailed description of how the velocity fields of the KROSS galaxies were constructed, and modelled see \citet{Stott:2016}. Essentially, we simultaneously fit the H$\alpha$, [NII]6548, and [NII]6583 emission lines with three Gaussians. The fit uses the Levenberg-Marquardt technique to perform an uncertainty weighted, least-squares minimisation between the data and model. The intensity, central velocity and width of each of the Gaussians are left as free parameters, however the central velocity and width of the H$\alpha$ and [NII] lines are coupled. We construct the mean velocity fields by plotting the central velocity (in km s$^{-1}$), in each spaxel, of the Gaussian fit to the H$\alpha$ emission. We plot all velocities relative to the systemic recession velocity calculated from the known spectroscopic redshift \citep[taken from various surveys, see Table 1 of][]{Stott:2016}. For both the flux map and velocity field, if the signal-to-noise ratio, $S/N<5$ in a given $0.1'' \times 0.1''$ spaxel, we consider a larger area of $0.3'' \times 0.3''$. If the $S/N$ is still insufficient the area is enlarged again to $0.5'' \times 0.5''$ and finally $0.7'' \times 0.7''$.

\begin{figure*}
\centering
\begin{minipage}[]{1\textwidth}
\label{fig:Amap}
\centering
\includegraphics[width=1\textwidth]{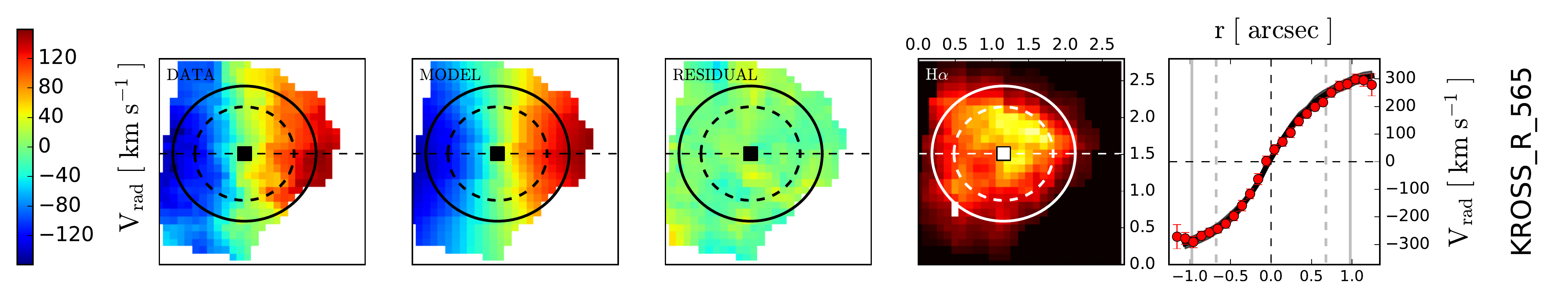}
\end{minipage}
\begin{minipage}[]{1\textwidth}
\label{fig:Bmap}
\centering
\includegraphics[width=1\textwidth]{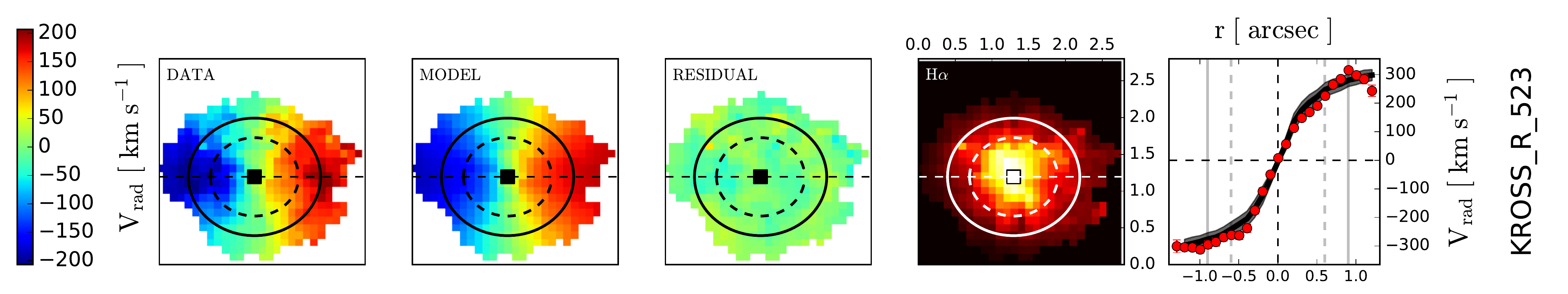}
\end{minipage}
\begin{minipage}[]{1\textwidth}
\label{fig:Dmap}
\centering
\includegraphics[width=1\textwidth]{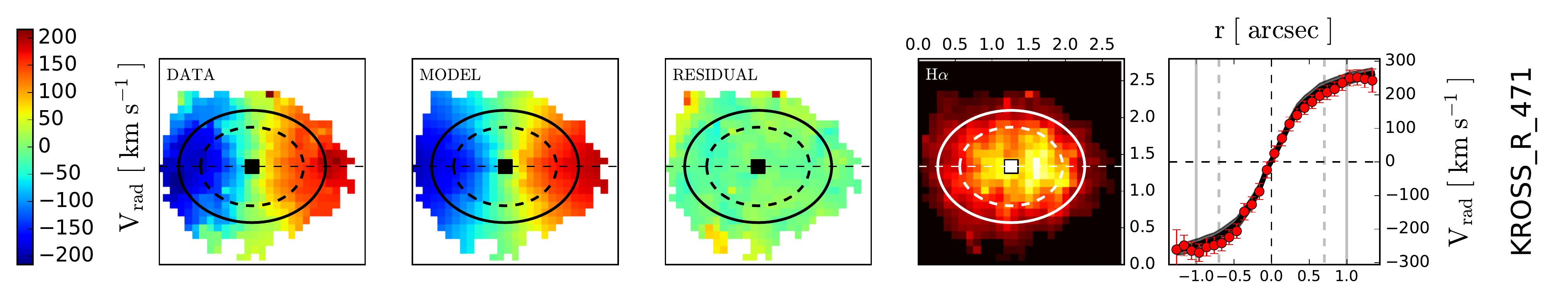}
\end{minipage}
\caption{Three examples of KROSS galaxy velocity fields that are well modelled by the \citet{Courteau:1997aa} arctangent function. For each galaxy the observed (``DATA") and best fitting model (``MODEL") velocity fields are displayed (with $0.1'' \times 0.1''$ spaxels) along with the residual (``RESIDUAL") between the two. Also included is the associated integrated H$\alpha$ flux map (``H$\alpha$"), constructed by integrating under the best fit to the H$\alpha$ emission line in each spaxel. The dashed and solid ellipses, centred on the dynamical centre of the best fitting model velocity field, contain 50\% and 80\% of the total H$\alpha$ flux respectively. The axial ratio of the ellipse is determined by initially fitting a two dimensional Gaussian to the entire flux map. On the far right, we display the extracted rotation curve (red points) from the inclination corrected observed velocity field. This is constructed by extracting velocities from each pixel along a 0.7$''$ ($\sim$the typical seeing) wide strip, in 0.1$''$ spaxel steps, along the position angle axis (horizontal in each map). Each spaxel's velocity is corrected for inclination. We also plot the rotation curve extracted from the corresponding model velocity field as a solid black line. We include the $\pm1\sigma$ bootstrap uncertainties from the model as a shaded grey region. These reflect the 1$\sigma$ uncertainty in $V_{0}$, $V_{\rm{max}}$, and $r_{\rm{dyn}}$  of the model. Further, we include $r_{50}$ and $r_{80}$ as the vertical dashed and solid grey line respectively. It should be stressed that the black curve is not a fit to the extracted rotation curve but rather it is the curve extracted from the best fitting model field. Both the observed and model rotation curve are plotted only for reference. It should also be noted that the extracted rotation curve is more susceptible to noise within the data. In this respect, the residual map is a much clearer indicator of the quality of the arctangent model fit.}%
\label{fig:goodvels}
\end{figure*}

The velocity fields were modelled by fitting a two-dimensional modification of the well known \citet{Courteau:1997aa} arctangent disk model for galaxy rotation curves, given as

\begin{equation}
V(r)=\frac{2}{\pi} V_{\text{max}} \arctan{\left(\frac{r}{r_{\text{dyn}}}\right)}\,\,\,,
\end{equation}

\noindent where $V(r)$ is the rotation velocity at radius $r$, $V_{\text{max}}$ is the rotation velocity at infinite radius, and $r_{\text{dyn}}$ is the characteristic radius associated with the arctangent turn over. In practice we fit a model with seven parameters to the velocity field. The parameters include ($x_{0}$,$y_{0}$) (the dynamical centre in $x$ and $y$ spaxel space), the position angle $\phi$ (in spaxel space), inclination $i$, $r_{\text{dyn}}$, and $V_{\text{max}}$.  We also include a systemic velocity parameter $V_{0}$, which allows for a best-fit systemic recession velocity that differs from that calculated from the known spectroscopic redshift. We constrain the dynamical centre to lie within $0.7''$ (equivalent to the typical seeing of KROSS observations) of the peak of the H$\alpha$ integrated flux. We define the line-of-sight velocity at each spaxel as 

\begin{equation}
V=V_{0}+(\sin{i} \cos{\theta} V(r))\,\,\,,
\end{equation}

\noindent where 

\begin{equation}
\cos{\theta}=\frac{(\sin{\phi} (x_{0}-x))+(\cos{\phi} (y-y_{0}))}{r}\,\,\,,
\end{equation}

\noindent and the radial distance from the dynamical centre for each spaxel is given as

\begin{equation}
r=\sqrt{(x-x_{0})^{2} + \left( \frac{y-y_{0}}{\cos{i}}\right)^{2}}\,\,\,.
\end{equation}

\noindent We use a genetic algorithm \citep{Charb:1995} to find the best fitting model.

As an example, Figure \ref{fig:goodvels} shows three observed velocity fields that are well modelled by the arctangent function, the corresponding model fields, as well as the associated H$\alpha$ flux maps. For illustration purposes we also display rotation curves extracted from the corresponding observed and model fields along the position angle axis of the galaxy. 

\subsection{Extracting Rotation Velocities}
\label{subsec:extractV}

When extracting a rotation velocity from the velocity field of each of the KROSS galaxies we consider two important factors. Firstly, for the TFR it is neccessary to consider the rotational velocity at radii that probe the flat part of a galaxy's rotation curve. Ideally this would mean considering the rotation velocity of the galaxy at the maximum radius of the (galaxy) disk so that the total dynamical mass is being probed. In practice however a compromise must be struck with regards to the second factor - that H$\alpha$ emission in galaxies is only detected out to finite radii. Therefore we must measure the rotation velocity at a radius that samples a galaxy's rotation curve at or beyond the turnover in order to sample the majority of the dynamical mass (and avoid the rapidly changing inner parts of the galaxy's rotation curve); but we must also choose a radius for which a large enough number of the KROSS galaxies have sufficiently extended H$\alpha$ emission.

Whilst the simple \citet{Courteau:1997aa} arctangent disk model used in this work is a satisfactory description of a disk-like galaxy's rotation curve, it is clear that the asymptotic velocity, $V_{\text{max}}$ has little physical meaning; considering the finite spatial extent of the H$\alpha$ emission it is obvious that, in all cases, a small change to the extrapolated velocity curve can lead to large changes in $V_{\text{max}}$. Thus it is sensible to extract a velocity at a more physically motivated choice of radius.

The characteristic radius, $r_{\text{dyn}}$ in the arctangent model defines where the arctangent curve begins to turn over and has no direct relation to the mass distribution of the galaxy or the corresponding radius at which the rotation curve becomes flat. It is therefore not a suitable choice. Previous studies have measured velocities at a circular radius containing 80\% of the red (e.g. $i$-band) stellar light \citep[see e.g.][]{Pizagno:2007}. For an exponential disk, this radius corresponds to $3.03$ times the disk scale radius and is thus a well motivated choice, physically.  We would therefore proceed in extracting velocities from the KROSS velocity fields in a similar manner. However, the KROSS sample lacks a homogeneous set of high resolution near-infrared imaging for the stellar light of the galaxies. The signal-to-noise of continuum detections is also insufficient in the majority of KROSS cubes to reliably measure the galaxy size. These measurements are ongoing and will be the subject of future publications by KROSS (Harrison et al., in prep). We instead return to the H$\alpha$ emission for which we have emission maps for all the KROSS galaxies.     
    
We extract a rotation velocity, $V_{80}$ for each galaxy from the best fitting arctangent model at a radius equal to the major axis of an ellipse containing 80\% of the total integrated H$\alpha$ flux, $r_{80}$. Naively, there is little reason to expect the spatial distribution of the H$\alpha$ emission to correspond to the underyling mass distribution, as is true with the red stellar continuum light. However, previous studies have shown that the radial extent of H$\alpha$ emission in galaxies at $z \sim 1$ is in agreement, or slightly more extended than the stellar light; \citet{Nelson:2012} find $<r_{e}(\rm{H}\alpha)/r_{e}(R)> = 1.3 \pm 0.1$ where $r_{e}(\rm{H}\alpha)$ and $r_{e}(R)$ are the H$\alpha$ and $R$-band effective radius respectively. We determine $r_{80}$ by growing ellipses (on the model integrated flux maps) outwardly from the best fitting dynamical centre. The axial ratio of the ellipse is determined by initially fitting a two dimensional Gaussian to the entire flux map, such that the ellipse reflects the overall spatial shape of the H$\alpha$ emission. 

\begin{figure*}
\centering
\begin{minipage}[]{1\textwidth}
\centering
\includegraphics[width=0.34\textwidth]{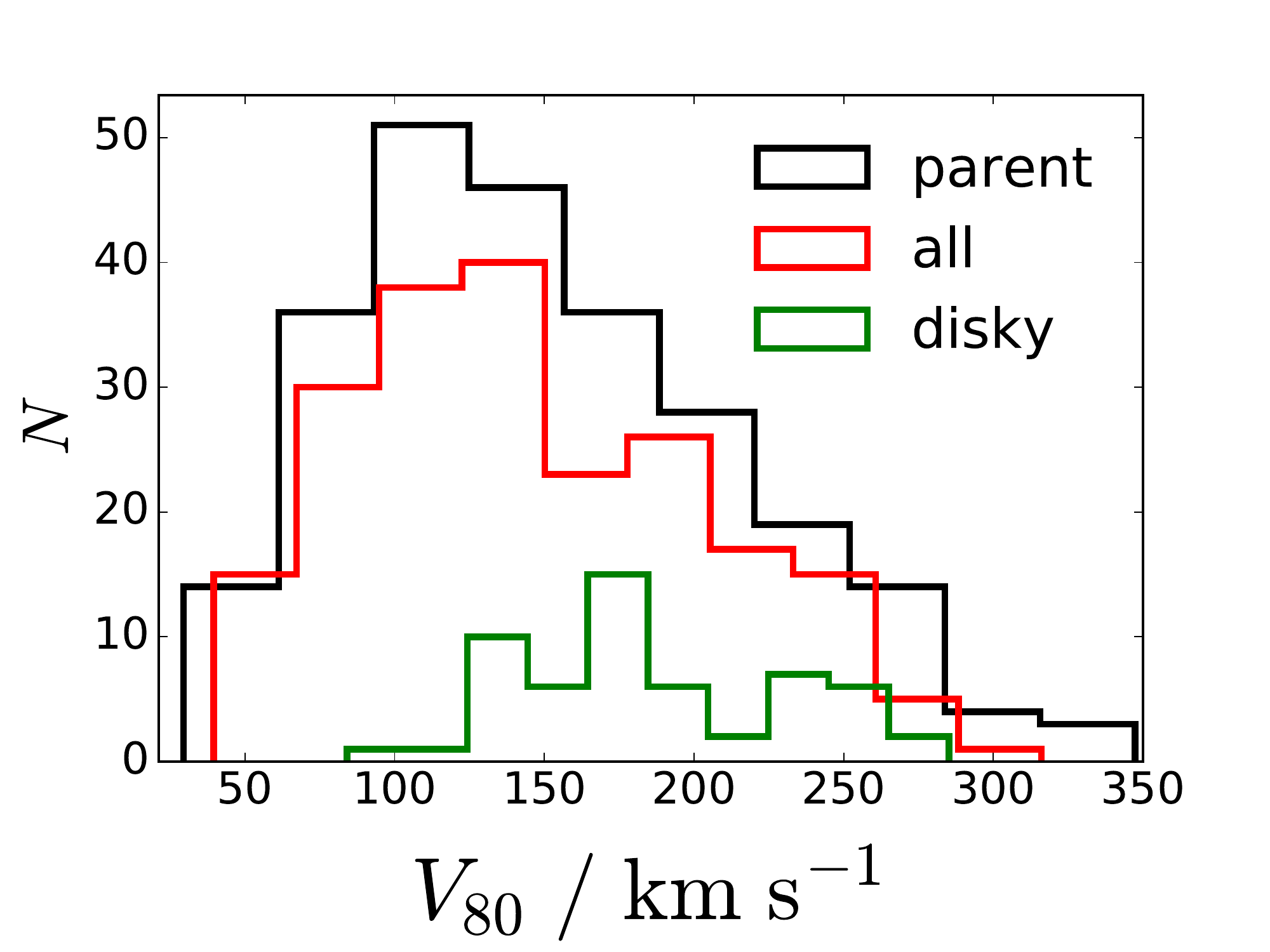}\includegraphics[width=0.34\textwidth]{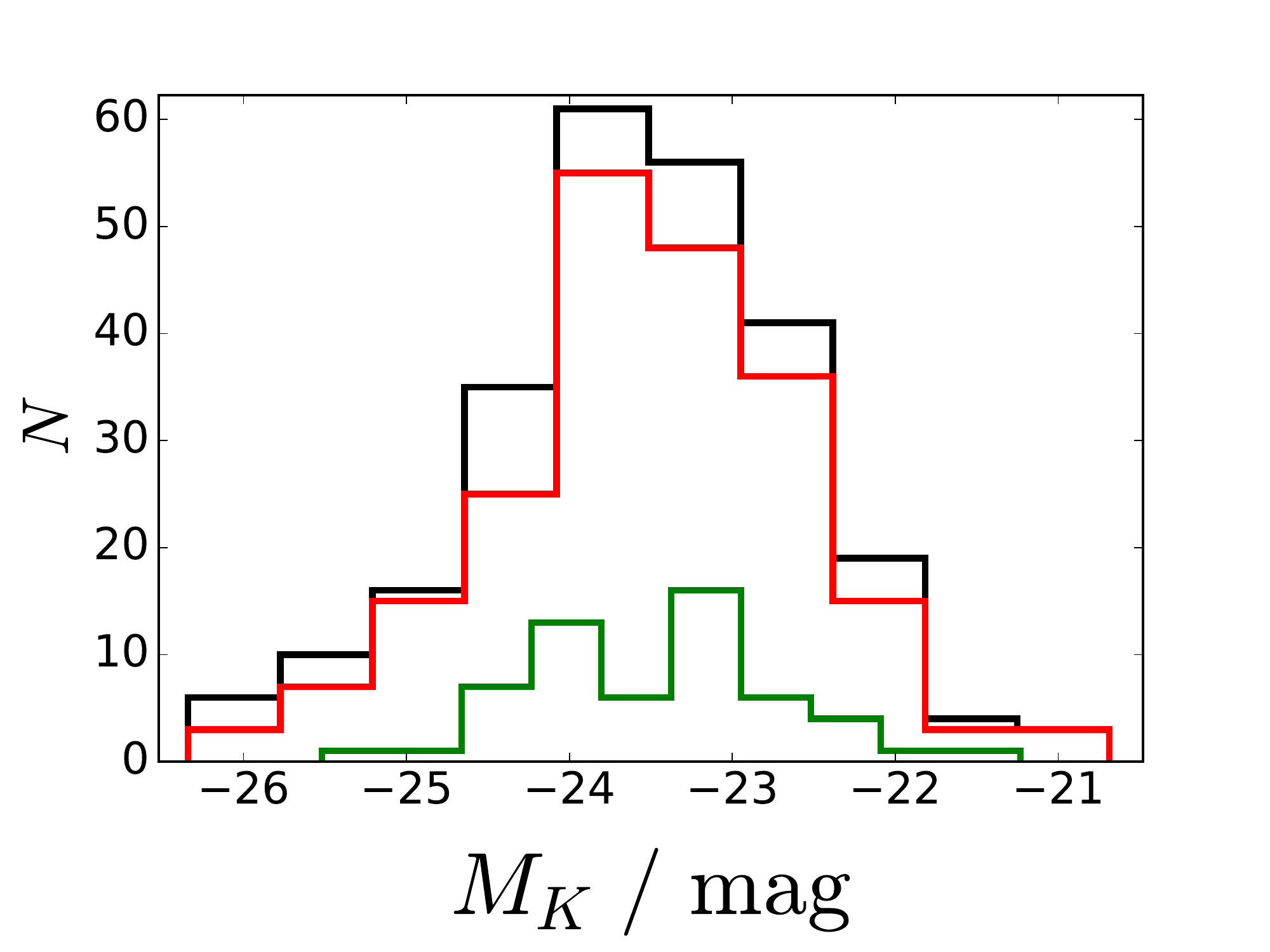}\includegraphics[width=0.34\textwidth]{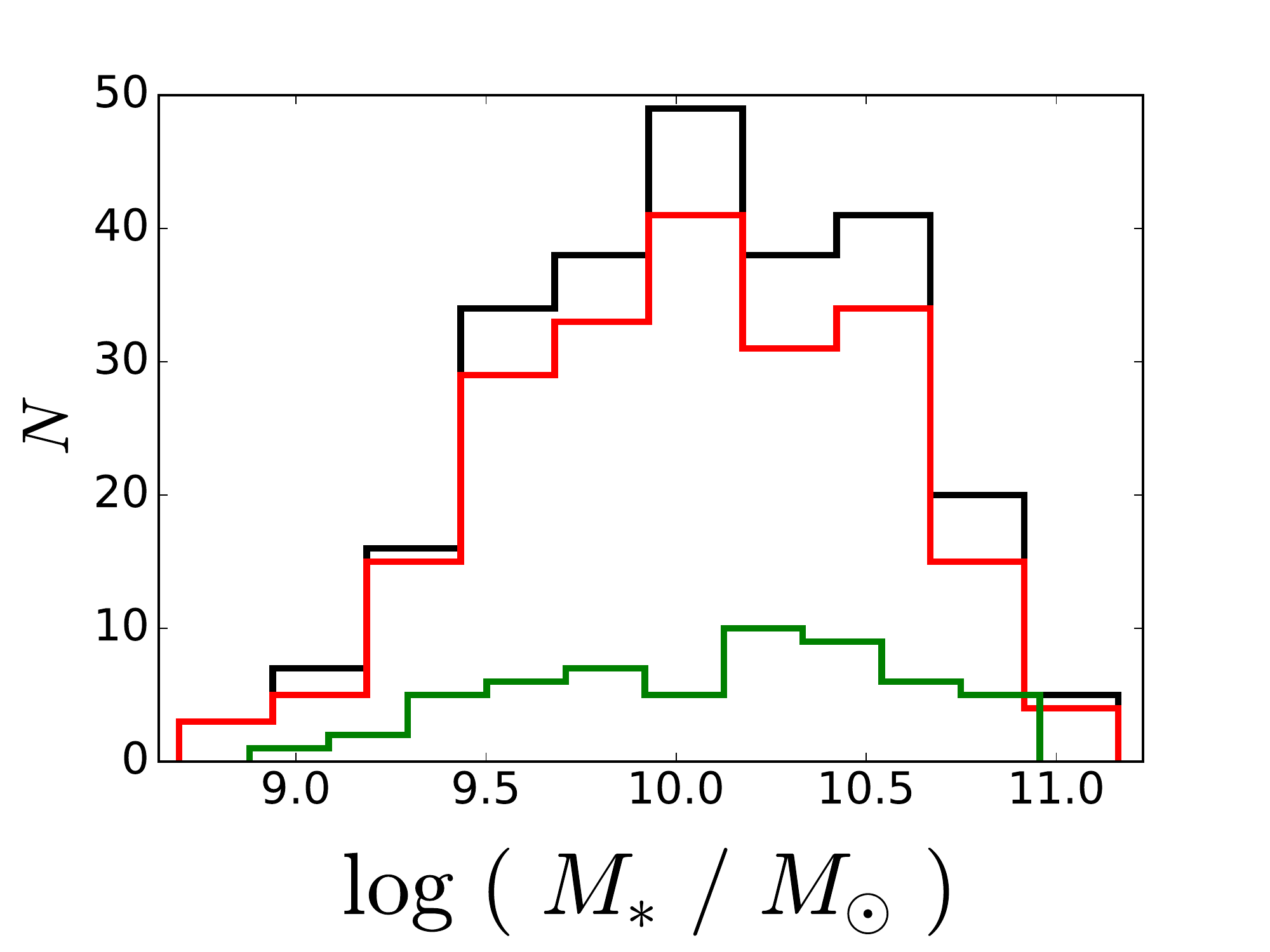}
\end{minipage}
\caption{%
Distributions of $V_{80}$, $M_{K}$, and $\log{(M_{*}/M_{\odot})}$ for the KROSS \textit{parent} sample, and sub-samples \textit{all} and \textit{disky}. The spread, and the position of the peak in all three distributions remain approximately constant between the \textit{parent} sample and sub-sample \textit{all}. Considering the \textit{disky} sub-sample it can be seen that the peak position remains relatively constant in comparison to the \textit{parent} sample and sub-sample \textit{all} in both the $M_{K}$ and $\log{(M_{*}/M_{\odot})}$ distributions. However, the \textit{disky} sub-sample is biased towards higher values of $V_{80}$ than the \textit{parent} sample and sub-sample \textit{all}.%
     }%
\label{fig:hists}
\end{figure*}

It should be noted at this point that, despite the overall light distribution of H$\alpha$ generally agreeing with the red stellar continuum, H$\alpha$ emission traces the ongoing star formation in galaxies and thus tends to be more clumpy than the continuum light, particularly at intermediate redshifts as demonstrated from the model KROSS H$\alpha$ maps. In some cases, the best fitting dynamical centre is displaced with respect to the peak of the integrated H$\alpha$ flux. Here there is the potential for the value of $r_{80}$ to be at odds with the radius containing 80\% of the stellar continuum light. This will introduce an unknown degree of scatter in to the KROSS TFR. To investigate this further, for those galaxies for which we are able to measure at least a continuum centre (i.e. the spatial position of the peak of the continuum, extracted from the KROSS cubes), we compare the difference between the dynamical centre as derived from the modelling of the velocity fields with the continuum centre. We find that in most cases the dynamical and continuum centres agree within $0.3''$. The typical seeing radius of any given KROSS observation is $\sim 0.35''$ (only $1.5$ spaxels in the original KMOS spaxel scale of $0.2''$). We also measure no significant change in the scatter of the TFR of our initial sub-sample (sub-sample {\it all}, see \S\ref{subsec:subsample}) after excluding those galaxies with dynamical centres that differ from the continuum centre by more than $0.3''$.  We are therefore satisfied that our measurements of the KROSS TFRs would not be significantly improved by forcing the dynamical centres of the model velocity fields to be coincident with the continuum centres. In light of this, and since continuum centre values are not available for all 584 KROSS galaxies with resolved H$\alpha$, we thus proceed with the analysis using the best fitting model velocity fields, where the dynamical centre is constrained to lie within $0.7''$ of the peak of the H$\alpha$ flux.  

In the same vein, in Appendix \ref{sec:centres} we also consider the effect those systems with asymmetric rotation curves (i.e. systems for which the H$\alpha$ emission extends up to or beyond $r_{80}$ on only one side of the rotation curve) have on the TFR scatter. One might expect the asymmetry in these systems to reduce the accuracy to which the dynamical centre may be determined. However, we show that exclusion of these systems from our initial sub-sample does not significantly change the scatter in the TFR. We therefore proceed with the analysis including all galaxies for which the H$\alpha$ emission extends up to or beyond $r_{80}$ on at least one side.

Examples of the velocity field modelling and extracted rotation curves are given in Figure \ref{fig:goodvels}. Also displayed are the corresponding integrated model H$\alpha$ flux maps along with the ellipses used to determine $r_{80}$ (and, for reference, $r_{50}$ - the radius equal to the major axis of an ellipse containing half of the integrated H$\alpha$ flux). The same plot for each of the 56 KROSS galaxies in the \textit{disky} sub-sample (see \S\ref{subsec:subsample} for sub-sample selection) are shown in Appendix \ref{sec:ssA}.

\subsection{SED fitting: Stellar Masses and Absolute Magnitudes}

Stellar masses and K-corrected absolute $K$-band magnitudes were derived by fitting the spectral energy distributions (SEDs) of each of the KROSS galaxies using {\sc HyperZ} \citep[see][]{Bolzonella:2000} and {\sc Le Phare} \citep[see][]{Arnouts:1999,Ilbert:2006} respectively (the former is used to maintain homogeneity with \citealt{Stott:2016} but does not compute absolute magnitudes). We used photometry spanning (where available) the optical to mid-infared ($u$, $B$, $V$, $R$, $I$, $J$, $H$, $K$, and IRAC $ch1$--$ch4$). For a full description of the catalogs utilised for each field observed by KROSS see Swinbank et al. (2016, in prep.). Each fitting routine generates model SEDs from the \citet{Bruzual:2003aa} stellar population synthesis models. We fit for extinction, metallicity, age, star formation, and stellar mass. Both routines allows for three main types of star formation history, namely a `single burst' of star formation, an exponential decline in star formation over the age of a galaxy, or constant/``boxy" star formation. 

We note that we do not directly observe the rest-frame $K$-band for all of the KROSS galaxies; in such cases the $K$-band K-corrections may suffer from an added degree of uncertainty. Throughout this work we adopt a uniform stellar mass uncertainty of $\pm 0.2$ dex. 

\section{The KROSS Tully-Fisher Relation}
\label{sec:KROSSTF}

\subsection{Defining Sub-Samples}
\label{subsec:subsample}

Of the 584 KROSS galaxies with resolved H$\alpha$ emission, we define a \textit{parent} sample, for the purposes of this work, of 251 galaxies that were detected in H$\alpha$, have a non-zero rotation velocity (derived in the manner described in \S\ref{subsec:extractV}), with a fractional uncertainty $\Delta_{V_{80}}\ /\ V_{80} < 0.3$, and associated K-corrected absolute $K$-band magnitudes and stellar mass values from SED fitting. To ensure that those galaxies included in the \textit{parent} sample are at least moderately well described by our simple arctangent model (i.e. have some disk-like rotation) we use an $R^{2}$ goodness of fit test (applied to the two-dimensional fit the velocity field), requiring for each galaxy that 

\begin{equation}
R^{2} \geq 85\%\,\,\,.
\end{equation}

\noindent The choice of $R^{2}$ is subjective but was chosen to remove those galaxies for which the arctangent model is a ``bad" fit whilst maintaining a reasonable sample size.

Of these 251 galaxies, 26 galaxies have H$\alpha$ emission which does not extend out to, or beyond, $r_{80}$. Inclusion of these galaxies will introduce scatter in to the Tully-Fisher relation as a result of extracting a velocity extrapolated from the model rotation curve, beyond the data. For this reasons we exclude them, leaving 225 galaxies. Similarly, we also exclude a further 6 galaxies for which the best fitting dynamical centre is completely spatially offset from any H$\alpha$ emission, as the best fitting models in these cases are obviously poorly constrained by the data. This leaves a total of 219 galaxies.

To avoid making large inclination corrections to the extracted galaxy rotation velocities, and in keeping with previous studies \citep[see e.g.][]{Meyer:2008,Stark:2009}, we make a further cut such that the inclination  $i > 25^{o}$, excluding an extra 4 galaxies -- leaving 215. 

Lastly, we employ kinemetry\footnote{{\sc kinemetry} for {\sc IDL} \url{http://davor.krajnovic.org/idl/}} \citep{Krajnovic:2006} in order to exclude any major-merger candidate systems following the prescription of \citet{Shapiro:2008}. See \citet{Stott:2014} for a previous application of Kinemetry to KMOS data. Briefly, kinemetry describes the moment maps (e.g. surface brightness, velocity map, sigma map) of a given galaxy as a series of concentric ellipses of increasing radii, each with a common centre but with individual position angles and inclinations. The series of ellipses describing a moment, $K$ can be expressed as 

\begin{equation}
K(a,\psi) = A_{0}(a) + \sum_{n=1}^{N} k_{n}(a) \cos[n(\psi - \phi_{n}(r))]\,\,\,,
\end{equation}

\noindent where $a$ is the semi-major axis of each ellipse, $\psi$ is the azimuthal angle in the plane of the galaxy, $A_{0}$ is the systemic velocity of each ellipse. The amplitude and phase coefficients are given as

\begin{equation}
k_{n}=\sqrt[]{A_{n}^{2} + B_{n}^{2}}\ {\rm\ \  and\ \ } \phi_{n} = \arctan(A_{n}/B_{n})\,\,\,,
\end{equation}

\noindent where $A_{n},B_{n}$ are (``kinemetry") constants. 

Each moment map can therefore be described by the values of $A_{n}$, $B_{n}$, and the orientations and semi-major axes of the ellipses. Further, different orders of each of the coefficients describe characteristics of the map. Specifically, for velocity maps, $B_{1}$ describes the bulk rotational motion of the galaxy i.e. for a perfect thin disk, $B_{1}$ describes the entire motion of the map; any non-zero higher order coefficients (i.e. $A_{1}, A_{2}, B_{2}, A_{3}, B_{3}, ...$) represent non-circular motion. Similarly, the only non-zero kinemetry cofficient in the dispersion map of a perfect thin disk is $A_{0}$. Therefore, higher order (i.e. $n = 1, 2, 3, 4, 5, ...$) coefficients here also represent deviation from circular motion. 

\citet{Shapiro:2008} use these higher orders to quantify the assymetry of the H$\alpha$ dynamics of high-redshift star-forming galaxies, in order to distinguish between major mergers and rotating disks. They exclude $A_{1}$ from their analysis since it can represent inflows/outflows into/from a galaxy, usually a result of stellar winds or AGN - not major mergers. They define the asymmetry $v_{\rm{asym}}$, and $\sigma_{\rm{asym}}$ of a galaxy's velocity and sigma map respectively as

\begin{equation}
v_{\rm{asym}} = \left\langle\frac{k_{2}+k_{3}+k_{4}+k_{5}}{B_{1,v}}\right\rangle_{r}\,\,\,,
\end{equation}

\hfill \\
and 

\begin{equation}
\sigma_{\rm{asym}} = \left\langle\frac{k_{1}+k_{2}+k_{3}+k_{4}+k_{5}}{B_{1,v}}\right\rangle_{r}\,\,\,,
\end{equation}
 
\noindent where $B_{1,v}$ is the $B_{1}$ kinemetry coefficent of the velocity map and the average is over all radii, $r$ of the kinemetry ellipses. Using templates of high-redshift disks and mergers they define an empirical delineation such that major mergers obey

\begin{equation}
K_{\rm{asym}}=\sqrt[]{v^{2}_{\rm{asym}} + \sigma^{2}_{\rm{asym}}} > 0.5\,\,\,.
\end{equation}   

From our sample we exclude 5 systems with a kinemetry asymmetry parameter, $K_{\rm{asym}} > 0.5$. We hereon refer to the remaining 210 galaxies as sub-sample \textit{all}. 

Next, we define a further sub-sample, \textit{disky}, that contains only those galaxies from sub-sample \textit{all} that are primarily rotationally supported. The KROSS sample will contain a number of galaxies that can be deemed disk-like but that are much more turbulent and chaotic than the spiral galaxies we see in the local Universe \citep[see e.g.][]{Schreiber:2006,Swinbank:2012,Stott:2016}. So whilst these galaxies may have disk-like structures and significant rotational support, they are also likely to have significant dispersion support as well. Since the Tully-Fisher relation assumes circular motion and relies on the assumption that the galaxies in question are rotationally supported, we define a ratio of rotation to dispersion support, $V_{80}/\sigma$ in order to exclude those galaxies that violate the assumption of circular motion. $\sigma$ is the flux weighted average value of the velocity dispersion map of each galaxy, after correcting for the instrumental resolution and local velocity gradient from beam smearing (see \citealt{Stott:2016} for more details). 

A cut was made to sub-sample \textit{all} such that $V_{80}/\sigma > 3$ in order to select galaxies that were predominantly rotationally supported (the choice of which value of $V_{80}/\sigma$ to cut by is discussed in \S\ref{subsubsec:TFRoffset_vs_z}). The resultant \textit{disky} sub-sample contains 56 galaxies. This may seem a large reduction in the number of galaxies from \textit{parent} to \textit{disky} but the cut is vital in order to ensure validity in any measured evolution of the Tully-Fisher relation; whilst it is valid to assume rotation support dominates in late-type galaxies in the local Universe, this assumption is not valid for all of the galaxies in the KROSS sample. Thus, in order to compare to $z \sim 0$ TFRs, we must select for the minority of galaxies within KROSS that are significantly rotation dominated. This is an issue that has not been sufficiently addressed in previous studies and is discussed further in \S\ref{subsec:results} and \S\ref{sec:conclusions}. 

A summary of the selection criteria for the samples and sub-samples defined in this work is given in Table \ref{tab:sampselec}, along with the number of galaxies in each. 

\begin{table}
\centering
\begin{tabular}{ | l | l | l |}
\hline
Sample & $N_{\rm{gal}}$ & Selection \\
\hline
\textit{parent} & 251  & Detected in H$\alpha$, $V_{80} > 0$, $\Delta_{V_{80}} / V_{80} < 0.3$,\\%
 & & $R^{2} \ge 85$\%, $M_{K}$ and $M_{*}$ from SED fitting\\
\hline
\textit{all} & 210 & Member of {\it parent},\\%
 & &  sufficient H$\alpha$ radial extent, dynamical centre \\ 
 & &  coincident with H$\alpha$ emission, $i > 25^{o}$,\\
 & &  $K_{\rm{asym}} \leq 0.5$\\
\hline
\textit{disky} & 56 & Member of {\it all}, $V_{80}/\sigma > 3$\\%
\hline
\end{tabular}
\caption{A summary of the selection criteria for samples and sub-samples defined in this work}
\label{tab:sampselec}
\end{table}

The distributions of $V_{80}$, $M_{K}$, and $\log{(M_{*}/M_{\odot})}$ for the \textit{parent} sample, sub-sample \textit{all}, and the \textit{disky} sub-sample can be seen in Figure \ref{fig:hists}. To quantify any biases between the distributions we conducted a Kolmogorov-Smirnov (K-S) two-sample test between the \textit{parent} sample and sub-samples \textit{disky} and \textit{all}. We define a null hypothesis that the two samples in question are drawn from the same distribution. We reject the null hypothesis if the $p$-value, $p < 0.05$. The resultant $p$-values can be seen in Table \ref{tab:pvalues}. It can be seen from Figure \ref{fig:hists} that the spread, and the position of the peak, of the $M_{K}$ and $\log{(M_{*}/M_{\odot})}$ distributions remain approximately constant between the \textit{parent} sample and sub-samples \textit{all} and \textit{disky} - the only difference being the number of galaxies in each sample, and a moderate truncation in the range of stellar masses and absolute magnitudes. Correspondingly, Table \ref{tab:pvalues} shows that for the $M_{K}$ and $\log{\ (M_{*}/M_{\odot})}$ distributions, we do not reject the null hypothesis that both \textit{parent} and \textit{all}, and \textit{parent} and \textit{disky} are drawn from the same distribution. Considering the distribution of $V_{80}$, the average value of $V_{80}$ remains roughly constant between the \textit{parent} sample and sub-sample \textit{all}. However there is an apparent increase in the average between sub-sample \textit{all} and sub-sample \textit{disky}. This is confirmed in Table \ref{tab:pvalues}, where it can be seen that we reject the null hypothesis that the $V_{80}$ values of \textit{parent} and \textit{disky} are drawn from the same distribution. This is in line with expectation as the \textit{disky} sub-sample comprises only those galaxies that are predominantly rotationally supported.

\begin{table}
\centering
\begin{tabular}{ | l | l | l |}
\hline
Distribution & \textit{parent} vs. \textit{all} & \textit{parent} vs. \textit{disky} \\
\hline
$M_{K}$ & 1.00  & 0.708  \\%
\hline
$\log M_{*}$ & 1.00 & 0.618 \\%
\hline
$V_{80}$ & 0.999 & $1.99 \times 10^{-6}$  \\%
\hline
\end{tabular}
\caption{$p$--values for Kolmogorov-Smirnov (K-S) two-sample tests between the \textit{parent} sample and sub-samples \textit{disky} and \textit{all}. The null hypothesis is that the two samples in question are drawn from the same distribution. The null is rejected for $p < 0.05$.}
\label{tab:pvalues}
\end{table}

\subsection{Fitting the Tully-Fisher Relation}
\label{subsec:fitting}

We find the best forward and reverse straight line fits to each of the Tully-Fisher relations presented in this work using a Markov chain Monte Carlo (MCMC) minimisation technique, with {\sc emcee}\footnote{\url{http://dan.iel.fm/emcee/current/}} \citep{Foreman:2013} in {\sc Python}. The familiar forward fit minimises 

\begin{equation}
\chi_{\rm for}^{2}\equiv\underset{i}{\Sigma}\left(\frac{1}{\sigma_{i}^{2}}\right)\{y_{i}-[m(x_{i}-x_{0})+b]\}~^{2}\,\,\,,
\end{equation}

\noindent where $x_{i}$ and $y_{i}$ are the spectral velocity and flux data respectively, $x_{0}$ is a ``pivot" point, chosen in order to minimise uncertainty in the intercept $b$ of the straight line (in practice we set $x_{0}$ to the median value of $x_{i}$), and $m$ is the gradient of the line. $\sigma_{i}$ is defined as

\begin{equation}
\sigma_{i}^{2}\equiv\sigma_{y,i}^{2}+m^{2}\sigma_{x,i}^{2}+\sigma_{\text{int}}^{2}\,\,\,,
\end{equation}

\noindent where $\sigma_{y,i}$ and $\sigma_{x,i}$ are the uncertainty of an
individual data point in $y$ and $x$ respectively, and
$\sigma_{\text{int}}$ is a measure of the intrinsic scatter in the
Tully-Fisher relation. $\sigma_{\text{int}}$ is determined by
adjusting its value such that the reduced chi-squared value is equal
to 1. It should be stressed that this measure is highly dependent on the accuracy of $\sigma_{x,i}$ and $\sigma_{y,i}$. As such, it is better thought of as the scatter unaccounted for by uncertainties. 

The total scatter in the relation is defined as

\begin{equation}
\sigma_{\rm tot}^{2}=\frac{\chi_{\rm for}^{2}}{\underset{i}{\Sigma}(1/\sigma_{i}^{2})}\,\,\,.
\end{equation}

\noindent For the reverse fit, the figure of merit to minimise is

\begin{equation}
\chi_{\rm rev}^{2}=\underset{i}{\Sigma}\left(\frac{1}{\zeta_{i}^{2}}\right)\left[x_{i}-(My_{i}+B+x_{0})\right]^{2}\,\,\,,
\end{equation}

\noindent where similarly $M$ and $B$ are respectively the gradient and intercept of the straight line and

\begin{equation}
\zeta_{i}^{2}\equiv\sigma_{x,i}^{2}+M^{2}\sigma_{y,i}^{2}+\zeta_{\rm int}^{2}\,\,\,.
\end{equation}

\noindent Again $\zeta_{\rm int}$ is the intrinsic scatter and the total scatter is defined as

\begin{equation}
\zeta_{\rm tot}^{2}=\frac{\chi_{\rm rev}^{2}}{\underset{i}{\Sigma}(1/\zeta_{i}^{2})}\,\,\,.
\end{equation}

\noindent The reverse fit parameters can be directly compared to the forward fit parameters by defining the equivalent slope, and intercept as $m^{'} \equiv 1/M$, and $b^{'} \equiv -B/M$, $\sigma_{\rm int}^{'} \equiv M\zeta_{\rm int}$, and $\sigma_{\rm tot}^{'} \equiv M\zeta_{\rm tot}$, respectively \citep{Williams:2010aa}. Since the values of $\zeta_{\rm tot}$ or $\zeta_{\rm int}$ represent respectively the total and intrinsic scatter in $\log{V_{80}}$, we do not tabulate the equivalent forward fit values but instead report them directly.

In all cases measured the best reverse fit slope was significantly (i.e. greater than three times the $1\sigma$ uncertainty) steeper than that of the forward fit. Given the comparable magnitude of uncertainty in both the abscissa ($\log{V_{80}}$) and ordinate ($M_{K}$ and $\log(M_{*})$) values of the TFRs present in this work, we choose not to favour either the best forward or reverse fit and instead take the bisector of the two as our best measurement of the TFR  in each case. 

\subsection{Results}
\label{subsec:firstTFR}

The $M_{K}$, and $\log M_{*}$ Tully-Fisher relations for sub-sample \textit{all} and the \textit{disky} sub-sample can be seen in Figures \ref{fig:TF1}  and \ref{fig:TF2} respectively. The corresponding free fit parameters are presented in Table \ref{tab:freefitpars}, while Table \ref{tab:fixedfitpars} presents the fitted parameters for the case where the slope of the fit to each KROSS sub-sample (and several samples taken from the literature) is fixed to the local Universe value (see \S\ref{subsec:compz0} and \S\ref{subsec:results}). Uncertainties are quoted at a 1$\sigma$ level. Uncertainties in the TFR offsets between KROSS and the $z \sim 0$ comparison sample include the uncertainty in the $z \sim 0$ TFR zero-point and in converting between stellar masses calculated assuming different IMFs. 

\subsubsection{Comparing to the Local Universe}
\label{subsec:compz0}

In all plots a $z \sim 0$ TFR is displayed for comparison, however it should be stressed that caution must be taken when directly comparing these with the KROSS relations. When comparing the TFR between any two samples of galaxies it is very important to consider the systematic bias that may be introduced as a result of the methods of measurement used. This is less of a problem when comparing absolute magnitudes and stellar masses as these are, by definition, corrected for redshift, extinction etc. However when considering the measure of a galaxy's rotation it can pose a problem. For example, the difference between a galaxy's rotation inferred from an IFU observation and a long slit observation may be significant \citep[e.g. ][]{Schreiber:2006}; factors such as slit orientation, resolution, and sensitivity could all introduce bias in to the measured rotation. Similarly, the choice of emission line used to trace the gas dynamics is also significant as different lines trace different phases of the gas, which may each extend out to different radii within a galaxy. Some lines suffer more absorption by dust in the line of sight (e.g. [O II] is more affected by dust than H$\alpha$). As an attempt to account for their effect on the measured intercept of the $z \sim 0$ comparison relations we combine the data from several studies from the literature, which use different measures of galaxy rotation and samples of galaxies from a range of different environment. We then compare the KROSS TFR to the best (free) fitting relation of the combined $z \sim 0$ data. 

The $M_{K}$ comparison relation comprises data from \citet[][TP00]{TullyPierce2000}, who use the linewidth of the integrated H{\small I} emission profile of galaxies from a range of cluster environments to derive their rotation velocity; and \citet[][V01]{Verheijen:2001aa}, who use the integrated H{\small I} emission profiles and H{\small I} rotation curves of a volume limited sample of late-type galaxies in the Ursa Major Cluster. 

The $M_{*}$ comparison relation comprises the data of \citet[][P05]{Pizagno:2005}, who use rotation velocities derived from long slit spectroscopy of H$\alpha$ emission; \citet[][R11]{Reyes:2011}, who use long slit spectroscopy of H$\alpha$ emission from a sub-sample of a large sample of disc galaxies selected from the Sloan Digital Sky Survey Data Release 7; and \citet[][RH04]{Rhee:2004}, who re-analyse the optical emission line rotation curves of the the \citet{Kent:1986} sample of Sb and Sc spiral galaxies. 

Aside from using a composite $z \sim 0$ sample, the obvious solution is to use the same measurement of rotation for all galaxy samples considered for comparison. In this way, if there is systematic bias introduced as a result of the measurement method, the same bias will be present in both relations. In this case, any measured relative offset between the relations of the two samples will be intrinsic. In the absence of a directly comparable $z \sim 0$ TFR, we make do with the average of several different methods. With the advent of several large IFU surveys at $z \sim 0$ it will soon be possible to compare the observations of KROSS at $z \sim 1$ to a similarly observed sample in the local Universe. This issue and future work in relation to it are discussed further in \S\ref{sec:conclusions}.

\subsubsection{The Tully-Fisher Relations}
\label{subsec:resultsTFR}

\begin{figure*}
\centering
\begin{minipage}[]{1\textwidth}
\label{fig:AH}
\centering
\includegraphics[width=0.75\textwidth, trim= 0 0 10 50,clip=True]{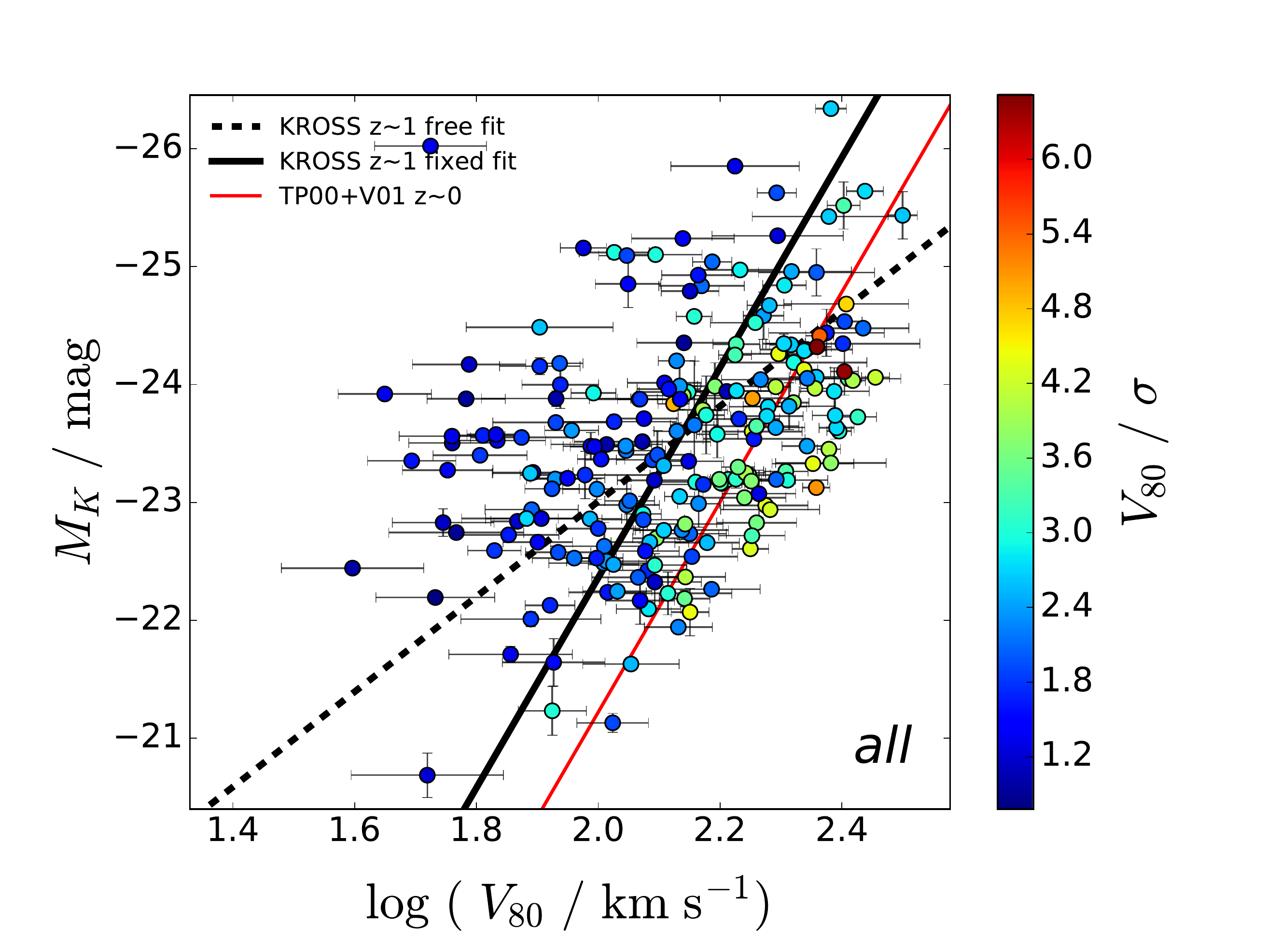}\vspace{0.35cm}
\end{minipage}
\begin{minipage}[]{1\textwidth}
\label{fig:BH}
\centering
\includegraphics[width=0.75\textwidth, trim= 0 0 10 50,clip=True]{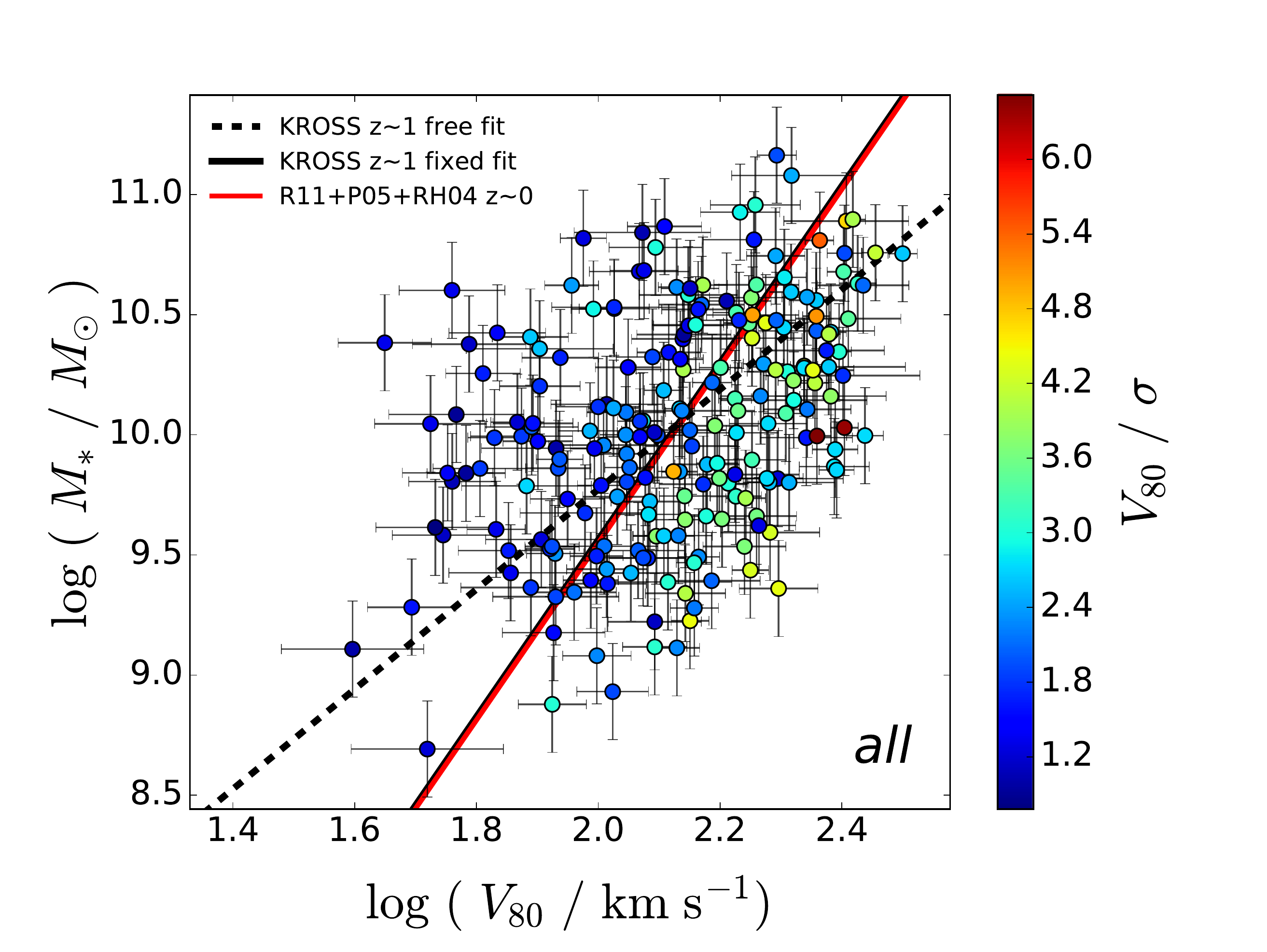}\vspace{0.37cm}
\end{minipage}
\caption{%
The $M_{K}$ and $M_{*}$ Tully-Fisher relations for sub-sample \textit{all} as described in \S\ref{subsec:subsample}. Displayed is the bisector of the best forward and reverse fits to the data, as well as the best fit when the slope is constrained to that of a $z \sim 0$ comparison sample (as described in \S\ref{subsec:compz0}). The data points are colour-coded by their corresponding values of $V_{80}/\sigma$. Those data points with correspondingly high values of $V_{80}/\sigma$ follow a tighter relation than those with lower values. The higher $V_{80}/\sigma$ points are coincident with the $z \sim 0$ relation in the case of the $M_{K}$ TFR. For the stellar mass TFR the same points are offset to lower values of $\log(M_{*}/M_{\odot})$, for a given value of $V_{80}$, in comparison to the $z \sim 0$ relation. Those with lower $V_{80}/\sigma$ tend to be scattered to lower values along the abscissa of the plot. These may correspond to systems with greater pressure support and lower rotation velocities.%
     }%
\label{fig:TF1}
\end{figure*}

Figure \ref{fig:TF1} shows the $M_{K}$ and $\log M_{*}$ TFRs of sub-sample \textit{all}. The fit parameters for the bisector and fixed-slope linear fits are shown in Table \ref{tab:freefitpars} and Table \ref{tab:fixedfitpars} respectively. From the free fitting we infer large intrinsic scatter in stellar mass and absolute $K$-band magnitude, in comparison to the local Universe, for both the $M_{K}$ and $\log M_{*}$ TFRs respectively ($\sigma_{\rm{int}}=0.84 \pm 0.04$ mag and $\sigma_{\rm{int}}=0.38 \pm 0.02$ dex). For the $z \sim 0$ comparison sample we find an intrinsic scatter of $\sigma_{\rm{int}}=0.36 \pm 0.04$ mag and $\sigma_{\rm{int}}=0.16 \pm 0.01$ dex for the $M_{K}$ and $\log M_{*}$ relation respectively. 

The free fitted slope of each of the sub-sample \textit{all} TFRs is much shallower than that of the respective comparison relation, within uncertainties ($-4.0 \pm 0.5$ and $2.1 \pm 0.2$ for the $M_{K}$ and $\log M_{*}$ relation respectively). Fixing the slopes to that of the $z \sim 0$ comparison relations we find an increase in the inferred intrinsic scatter (1.43 $\pm$ 0.08 mag and 0.58 $\pm$ 0.03 dex for the $M_{K}$ and $M_{*}$ relation respectively) for both the {\it all} TFRs. Considering the $\log M_{*}$ {\it all} fixed slope relation we find no evidence for a significant offset of the TFR between $z \sim 1$ and $z \sim 0$. Considering the $M_{K}$ TFR for the same sample however, for a given rotation velocity we measure an offset of $-$1.1 $\pm$ 0.1 mag between the KROSS and the $z \sim 0$ comparison relations. 

\begin{figure*}
\centering
\begin{minipage}[]{1\textwidth}
\label{fig:AK}
\centering
\includegraphics[width=0.75\textwidth,trim= 0 0 0 30,clip=True]{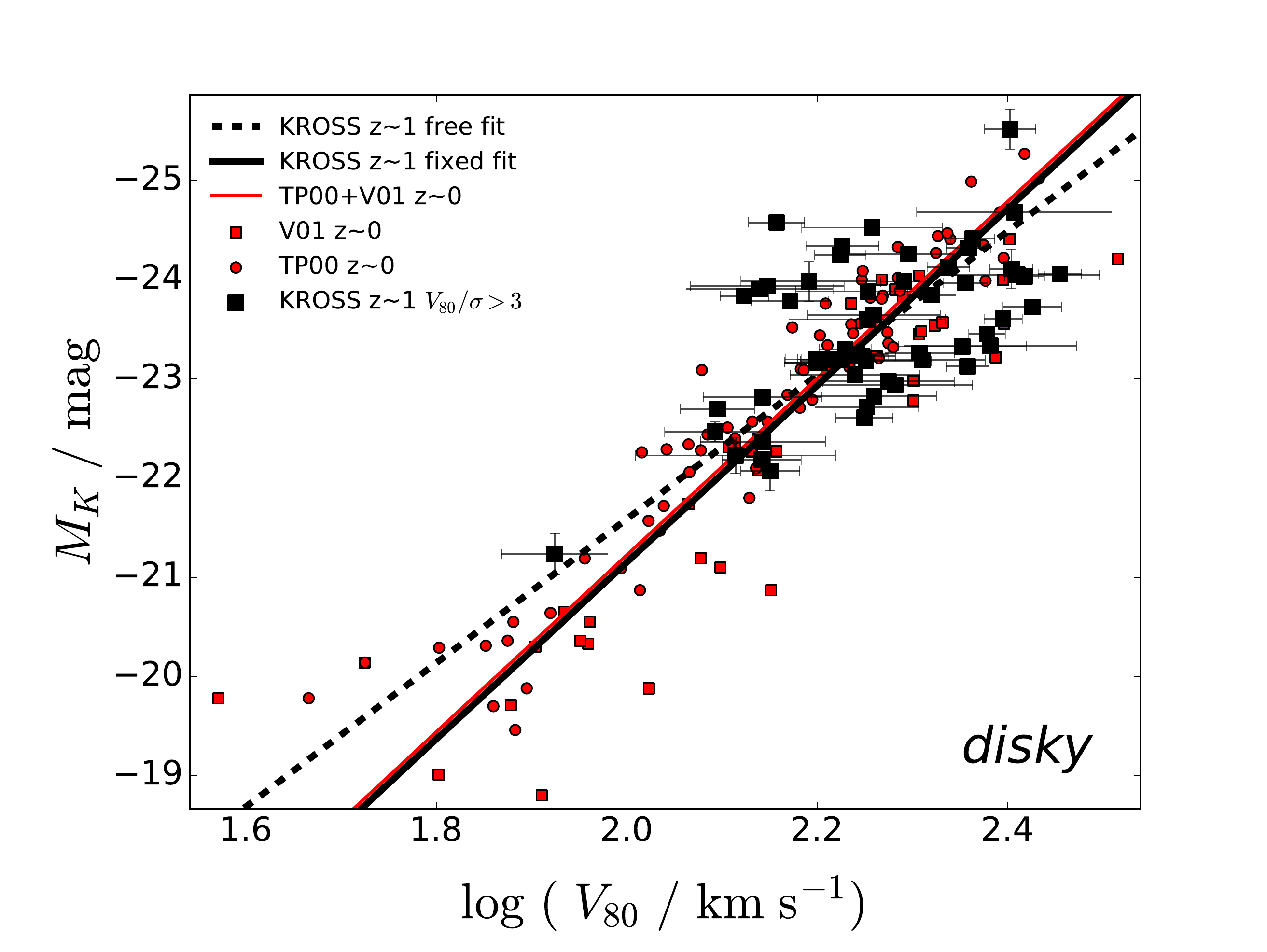}\vspace{0.35cm}
\end{minipage}
\begin{minipage}[]{1\textwidth}
\label{fig:BK}
\centering
\includegraphics[width=0.75\textwidth,trim= 0 0 0 30,clip=True]{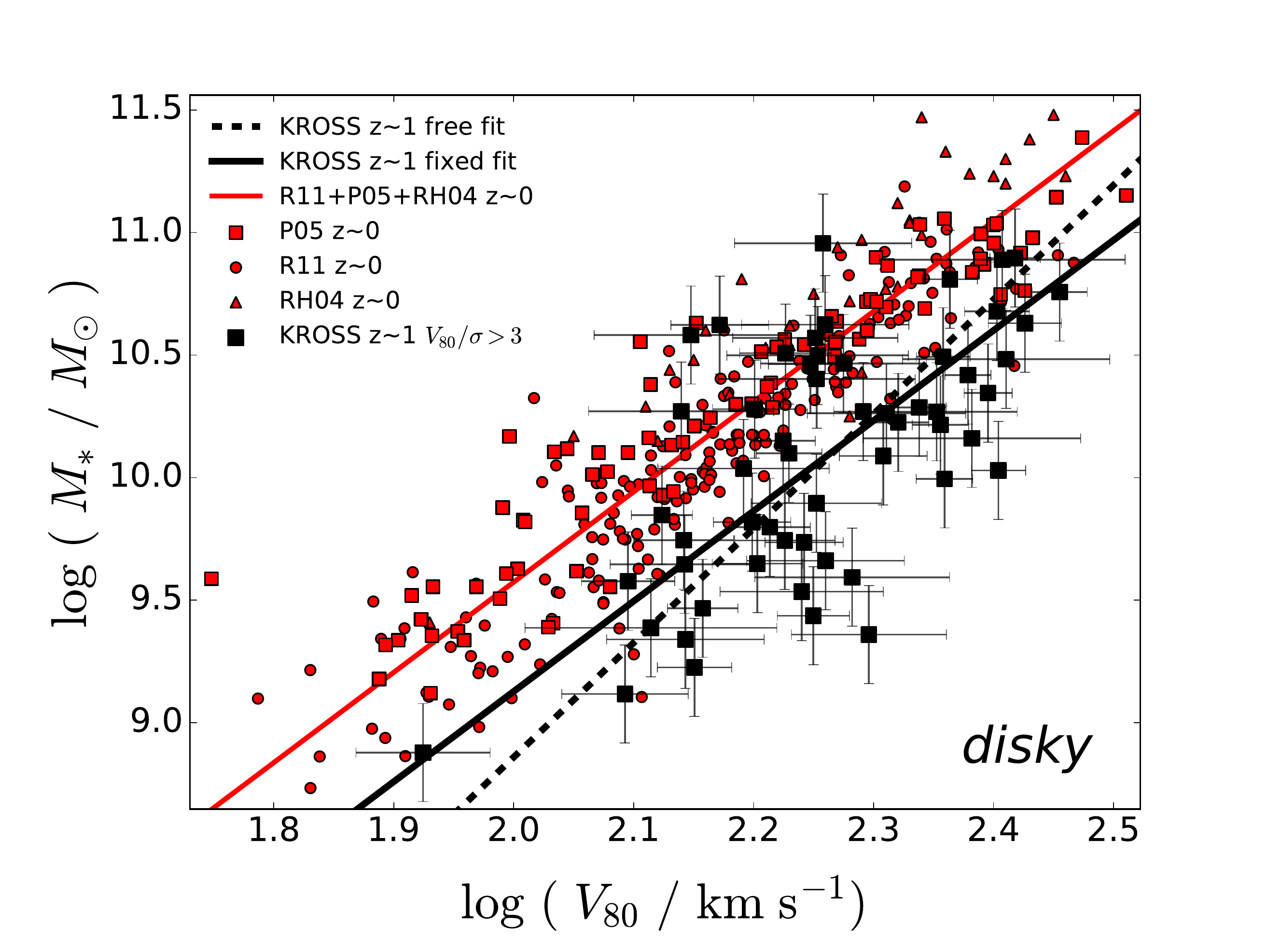}\vspace{0.37cm}
\end{minipage}
\caption{%
The $M_{K}$ and $M_{*}$ Tully-Fisher relations for the \textit{disky} sub-sample as described in \S\ref{subsec:subsample}. Displayed is the bisector of the best forward and reverse fits to the data, as well as the best fit when the slope is constrained to that of a $z \sim 0$ comparison sample (as described in \S\ref{subsec:compz0}). There is a clear offset between the KROSS and the $z \sim 0$ $M_{*}$ TFRs. Conversely, the zero-point of the KROSS $M_{K}$ TFR is in agreement with that of the $z \sim 0$ comparison relation.%
     }%
\label{fig:TF2}
\end{figure*}

Importantly, the TFRs of sub-sample {\it all} exhibit large scatter in $\log{V_{80}}$ ($\zeta_{\rm{int}}=0.153 \pm 0.009$, or equivalently $\sigma_{\rm{int}}^{'}=1.9 \pm 0.1$ mag or $\sigma_{\rm{int}}^{'}=0.81 \pm 0.5$ dex). As is apparent from the colour-coding in Figure \ref{fig:TF1}, this scatter dramatically reduces with increasing $V_{80}/\sigma$ i.e. the scatter is reduced for the more disk-like galaxies in the sample. Since the TFR assumes purely rotational motion, it is more informative to examine the TFR of galaxies with high $V_{80}/\sigma$. We thus look to the TFRs, displayed in Figure \ref{fig:TF2}, of the \textit{disky} sub-sample, which contain galaxies selected to be disk-like with significant rotation support (as described in \S\ref{subsec:subsample}). The fit parameters for the free and fixed-slope linear fits are shown in Table \ref{tab:freefitpars} and Table \ref{tab:fixedfitpars} respectively. We see a reduction in the intrinsic scatter compared to that of sub-sample \textit{all} ($\sigma_{\rm{int}}=0.57 \pm 0.06$ mag and $\sigma_{\rm{int}}=0.25 \pm 0.05$ dex for the $M_{K}$ and $M_{*}$ relation respectively). However, this is still large compared to the $z \sim 0$ comparison samples. The free fit bisector slopes of both TFRs for the \textit{disky} galaxies ($-7.3 \pm 0.9$ and $4.7 \pm 0.4$ for the $M_{K}$ and $M_{*}$ relation respectively) are much steeper than the free fit slopes of sub-sample \textit{all}, with the slope of the \textit{disky} sub-sample stellar mass TFR steeper even than the $z \sim 0$ comparison sample, within uncertainties. The $M_{K}$ TFR for \textit{disky} galaxies remains slightly shallower than the slopes at $z \sim 0$, within uncertainties. 

Fixing the slopes to that of the $z \sim 0$ comparison relations, we see a change in the offset, towards dimmer magnitudes or lower stellar masses for a given rotation velocity, across both TFRs in comparison to sub-sample \textit{all}. For the $M_{K}$ TFR, this brings the KROSS relation in line with the $z \sim 0$ comparison sample (an offset of $0.1 \pm 0.1$ mag). However, we measure a large offset ($-0.41 \pm 0.08$ dex), towards lower stellar masses for a given rotation velocity, between the $z \sim 0$ and KROSS $M_{*}$ TFRs. This offset is larger than the measured intrinsic scatter and greater than three times the 1$\sigma$ uncertainty on the offset. We note also that the measured offsets of both the $M_{*}$ and $M_{K}$ {\it disky} TFR, with respect to $z \sim 0$, are robust to a change in the radial position at which the velocity is extracted for each galaxy; extracting a velocity $V_{70}$ and $V_{90}$ (i.e. at a radius containing respectively 70\% and 90\% of the H$\alpha$ flux) changes the measured offset by only $\pm 0.08$ dex and $\pm 0.2$ mag for the $M_{*}$ and $M_{K}$ {\it disky} relations, respectively. Thus we find evidence of a significant evolution in the zero-point of the stellar mass TFR for disk-like galaxies between $z \sim 1$ and $z \sim 0$; for the same galaxies we find no evolution in the zero-point of the $M_{K}$ TFR over the same period. Again, the measured intrinsic scatters increase in comparison to the free fit, but are reduced in comparison to the fixed fit of sub-sample \textit{all}. 

If the zero-point of the stellar mass Tully-Fisher relation has evolved as measured since $z \sim 1$ to the present day, this implies a decrease, by a factor of $\sim 0.36$, in the {\it dynamical} mass-to-stellar mass ratio of disk-like galaxies over the last $\sim$8 billion years i.e. $(M_{\rm{dyn}}/M_{*})_{z=0} \sim 0.36 \times (M_{\rm{dyn}}/M_{*})_{z=1}$; for a given dynamical mass, galaxies at $z \sim 1$ have less stellar mass than they do in the local Universe. Further, given that there is no apparent evolution in the zero-point of the $M_{K}$ TFR, this also implies an increase in the absolute $K$-band {\it stellar} mass-to-light ratio, by a factor of $\sim 2.75$ since $z \sim 1$. There is more $K$-band light emitted per stellar mass at $z \sim 1$ than at $z \sim 0$. This is not outside of reasonable expectation given the comparatively greater specific star formation rates of galaxies in the $z \sim 1$ Universe (see \S~\ref{sec:discussion} for further discussion on this point).

\subsection{Evolution}
\label{subsec:results}

The offset between the KROSS $z \sim 1$ stellar mass TFR and that of the $z \sim 0$ comparison sample is at odds with the findings of \citet{Miller:2011aa} who studied the stellar mass Tully-Fisher relation for 129 disk-like galaxies out to $z \sim 1.3$ based on slit spectroscopy. They find intrinsic scatter comparable to that of low redshift studies and, most importantly, they find almost no evolution in the stellar mass Tully-Fisher relation from $z \sim 1$ to $z \sim 0.3$. Similarly \citet{Miller:2012}, using resolved spectra - taken with Keck I Low Resolution Imaging Spectrograph \citep{Oke:1995} - of 42 morphologically selected star-forming galaxies at $1 < z < 1.7$, find no evidence for an evolution of the offset of the TFR compared with the local universe.

To investigate this further and put the results of this work in context, we use the data of \citet{Miller:2011aa}, \citet{Miller:2012} and other intermediate redshift studies, by \citet{Flores:2006}, \citet{Puech:2008}, \citet{Cresci:2009}, and \citet{Gnerucci:2011} in order to directly measure and compare the offsets of each study with our composite $z \sim 0$ comparison sample and with the predicted TFR zero-point evolution according to theory and simulations. We find the best fit to each of the studies' data, constraining the slope to that of our local Universe comparison relation. We thus obtain a homogeneous measure of any evolution of the stellar mass TFR zero-point between each data sample and $z \sim 0$. The resulting measurements are presented in Table \ref{tab:fixedfitpars} and displayed in the left panel of Figure \ref{fig:TFevolution}. In \S\ref{subsubsec:comparison_studies} we describe each of the previous studies we use (in addition to the work of \citealt{Miller:2011,Miller:2012}) in our comparison and the predictions from theory and simulation. In \S\ref{subsubsec:TFRoffset_vs_z} we discuss the emergent picture of the evolution of the stellar mass TFR zero-point since $z \sim 3$. 

\subsubsection{Comparison Studies}
\label{subsubsec:comparison_studies}

Here we summarise the work of \citet{Flores:2006}, \citet{Puech:2008}, \citet{Cresci:2009}, and \citet{Gnerucci:2011}. We also discuss the work of \citet{Dutton:2011}, who use semi-analytical modelling (SAM) to predict the TFR zero-point evolution as a function of redshift. Lastly we describe the process by which we derive similar predictions from the Evolution and Assembly of GaLaxies and their Environments (EAGLE) simulation \citep{Schaye:2015,Crain:2015,McAlpine:2015}.

\citet{Flores:2006} and \citet{Puech:2008} used the multi-integral field spectrograph FLAMES-GIRAFFE \citep[see e.g.][]{Pasquini:2002} at the VLT to study the [OII]3727 emission line kinematics of galaxies at $z \sim 0.6$. Both studies apply the kinematic classification scheme devised by \citet{Flores:2006} to their galaxy sample. The scheme uses optical images combined with the velocity and dispersion fields of each galaxy to place it in one of three kinematic categories; \\

\noindent - {\it rotating disks} (RD): the axis of rotation of the velocity field is aligned with the optical major axis, and the peak of the dispersion field is close to the galaxy's dynamical centre;

\noindent - {\it perturbed rotation} (PR): the axis of rotation is aligned with the optical major axis but the peak of the dispersion is misaligned with the dynamical centre;

\noindent - {\it complex kinematics} (CK): both the velocity and sigma fields differ greatly from that expected of a rotating disk. \\

\noindent In order to draw parallels with our own sub-samples {\it all} and {\it disky} ($V_{80}/\sigma > 3$), we examine the offset between our $M_{*}$ $z \sim 0$ comparison TFR and those of both \citet{Flores:2006} and \citet{Puech:2008} - first including galaxies within all three of their kinematic categories, and secondly considering just RDs. We note that the authors of both studies focus mainly on the RD samples in their discussions.

\citet{Cresci:2009} constructed the $M_{*}$ TFR of 18 star-forming galaxies at $z \sim 2.2$, using spatially resolved H$\alpha$ emission line dynamics as observed with the SINFONI \citep{Eisenhauer:2003} integral field spectrograph at the VLT. The galaxies were selected from the high-$z$ galaxy Spectroscopic Imaging survey in the NIR with SINFONI \citep[SINS;][]{ForsterSchreiber:2009} based on the prominence of ordered rotation versus more complex dynamics in each system. Combining the empirical kinemetry methods of \citet{Shapiro:2008} (see Section \ref{subsec:subsample}) with visual inspection of the velocity maps of each galaxy, the authors deemed all 18 galaxies to exhibit ordered rotation. We include 16 of these 18 galaxies, for which there are definitive stellar mass values, in our comparison. 2 galaxies, for which the authors state only upper limits on the stellar mass, are excluded. 

Finally, \citet{Gnerucci:2011} also used SINFONI to study the spatially resolved kinematics of 33 galaxies at $z \sim 3$ from the AMAZE \citep[Assessing the Mass-Abundance redshift Evolution; ][]{Maiolino:2008a,Maiolino:2008b} and LSD \citep[Lyman-break galaxies Stellar populations and Dynamics; ][]{Mannucci:2008,Mannucci:2009} projects. Of the 33 galaxies observed, they studied the TFR of 11 that displayed ordered rotation (as quantified by the goodness of fit of a plane to the velocity map). We include these 11 in our comparison.  

We also compare our measured offsets with predictions from the galaxy formation models of \citet{Dutton:2011}, which consist of N-body simulations of baryonic (stellar and gas) discs growing in Navarro-Frenk-White haloes \citep{Navarro:1997} that evolve with redshift. A further additional comparison is made to the $M_{*}$ TFR zero-point evolution as predicted by the EAGLE simulation. EAGLE comprises a state-of-the-art suite of cosmological hydronamical galaxy formation simulations performed in volumes ranging from 25--100 comoving Mpc$^{3}$. It has been shown to pass a large range of observational tests in both the local and higher redshift Universe. We compute the predicted TFR zero-point evolution by drawing samples of galaxies from the EAGLE public data release\footnote{\url{http://icc.dur.ac.uk/Eagle/database.php}} at redshifts, $z=0, 0.5, 0.87, 1.0, 1.5, 2, 3, \rm{and}\ 3.98$ respectively. To facilitate a reasonable comparison to observations, at each redshift we include only those galaxies with star formation rates $> 1 M_{\odot} \rm{yr}^{-1}$. We first find the best fit to the TFR of the extracted $z=0$ EAGLE sample. We then perform fixed-slope fits to the TFRs of the successively higher redshift samples with the slope constrained to that of the $z = 0$ sample.

\subsubsection{Zero-Point Evolution}
\label{subsubsec:TFRoffset_vs_z}

From the left panel of Figure \ref{fig:TFevolution} it can be seen that our analysis of the \citet{Miller:2011aa} and \citet{Miller:2012} samples agrees with their findings - we find no evidence for any significant (i.e. greater than 3 times the 1$\sigma$ uncertainty) evolution of the stellar mass TFR offset for the $z < 1.7$ redshift range of their data; the maximum offset we find is $-0.09 \pm 0.08$ dex, for galaxies in the redshift range $z = 0.5$--$0.8$. We find a significant negative offset in $\log(M_{*})$ with respect to the $z \sim 0$ TFR for the \citet{Cresci:2009}  sample ($-0.33 \pm 0.09$ dex). Similarly we measure a large offset ($-0.6 \pm 1.0$ dex) between the \citet{Gnerucci:2011} sample and our $z \sim 0$ comparison sample, although the uncertainty is very large. The general trend is marginally consistent with that predicted by \citet{Dutton:2011}, and EAGLE. However, the zero offset, with respect to $z \sim 0$, that we measure using the highest redshift sample of \citet{Miller:2011} and the sample of \citet{Miller:2012} disagree with both predictions. 

Next we consider the samples of \citet{Flores:2006} and \citet{Puech:2008}, where we initially include galaxies within all three of their kinematic categories (samples Flores+06 all, and Puech+08 all). For the \citet{Flores:2006} sample, we find a marginally significant offset (0.27 $\pm$ 0.09 dex) towards larger stellar masses for a given rotation velocity in comparison to the $z \sim 0$ relation. Whilst the \citet{Puech:2008} relation is consistent with no offset, within three times the 1$\sigma$ uncertainties. However, the measured offsets are shifted towards much lower stellar mass values when considering RD galaxies only; we find offsets, with respect to $z \sim 0$, of $-0.2 \pm 0.2$ dex and $-0.2 \pm 0.1$ dex for the RD sample of \citet{Flores:2006} and \citet{Puech:2008}, respectively. These measurements are at odd with the predicted zero-point evolution for this redshift according to the SAM of \citet{Dutton:2011} but agree well with the prediction from EAGLE.    

In a similar vein, considering KROSS sub-sample \textit{all}, we find an offset consistent with that of \citet{Miller:2011,Miller:2012} and with little to no evolution of the zero-point of the stellar mass TFR since $z \sim 1$. This is at odds with the predictions of both \citet{Dutton:2011} and EAGLE. However as discussed above, considering the \textit{disky} KROSS sub-sample, composed of predominantly rotationally supported galaxies, we see a significant offset ($-0.41 \pm 0.08$ dex) to lower stellar masses, with respect to $z \sim 0$, for a given dynamical mass. This offset agrees with the prediction of EAGLE within uncertainties, but is perhaps larger than expected. The intepretation of such an offset is explored further in \S\ref{sec:discussion}. 

These results suggest that previous studies such as \citet{Miller:2011,Miller:2012} detect no evolution in the TFR out to $z \sim 1$ due to the inclusion of galaxies with low ratios of rotation-to-pressure support. With the benefit of IFU observations, a more direct selection of galaxies displaying ordered rotation is possible. Despite the differing methods by which rotating disks or disk-like galaxies are selected in the IFU studies discussed here, in each of the cases where some distinction has been possible we find the $M_{*}$ TFR zero-point of the disk-like galaxies to be significantly more offset to lower stellar mass values for a given rotation velocity in comparison to samples from the corresponding study which do not make a distinction between disk-like and non-disk-like galaxies. Similarly whilst the models of \citet{Dutton:2011} describe the evolution of the baryonic discs of galaxies, they do not make any distinction between those that are pressure supported or rotationally supported. The same may be said of the predictions from EAGLE, in which we only select galaxies based on their star formation rates. This may explain the discrepancy between the predicted evolution of the TFR zero-point and that measured for the \textit{disky} sub-sample in this work.

The offset of the KROSS \textit{disky} sub-sample is representative of a more general dependence of the KROSS stellar mass TFR on $V_{80}/\sigma$, which can be seen in the right panel of Figure \ref{fig:TFevolution}. There is a shift in the TFR offset from positive to negative (compared to $z \sim 0$) with increasing $V_{80}/\sigma$. This can be interpreted as evidence for an offset of the TFR for ``disky" galaxies at $z \sim 1$, with galaxies being scattered in the direction of lower velocities depending on to what extent they are rotationally or pressure supported. We note here that the flattening of the curve for galaxies $V_{80}/\sigma \gtrsim 3$, balanced with the need to maintain a reasonable sample size, lead to the chosen cut of $V_{80}/\sigma > 3$ for {\it disky} KROSS galaxies (the {\it disky} sub-sample essentially comprises all those galaxies deemed to be ``disky", as discussed in \S\ref{subsec:subsample}). Thus the offset of $-0.41 \pm 0.08$ dex from $z \sim 0$ found for this sub-sample is (unsurprisingly) approximately the average of the offsets of the three highest bins in $V_{80}/\sigma$ in Figure \ref{fig:TFevolution}.   

\begin{table*}
\centering
\begin{tabular}{ | l | l | l | l | l | l | l | l | l | l |}
\hline
TFR & Sample & Slope & Intercept & Pivot & $\sigma_{\rm{tot}}$ & $\sigma_{\rm{int}}$ & $\zeta_{\rm{tot}}$ & $\zeta_{\rm{int}}$ \\
\hline
$M_{K}$ & $z \sim 0$ & $-8.9$ $\pm$ 0.3 & $-$23.00 $\pm$ 0.04 & 2.2 & 0.41 mag  & 0.36 $\pm$ 0.04 mag & 0.067 dex & 0.045 $\pm$ 0.008 dex \\%
& KROSS {\it all} & $-4.0$ $\pm$ 0.5 & $-$23.57 $\pm$ 0.09 & 2.14 & 0.86 mag & 0.84 $\pm$ 0.04 mag & 0.169 dex & 0.153 $\pm$ 0.009 dex \\ %
& KROSS {\it disky} & $-7.3$ $\pm$ 0.9 & $-$23.4$\phantom{0}$ $\pm$ 0.2 & 2.25 & 0.62 mag & 0.57 $\pm$ 0.06 mag & 0.086 dex & 0.06$\phantom{0}$ $\pm$ 0.01$\phantom{0}$ dex\\%
\hline
$\log M_{*}$ & $z \sim 0$ & 3.68 $\pm$ 0.08 & 10.20 $\pm$ 0.01 & 2.17 & 0.21 dex & 0.16 $\pm$ 0.01 dex & 0.059 dex & 0.054 $\pm$ 0.003 dex\\%
& KROSS {\it all} & 2.1$\phantom{0}$ $\pm$ 0.2 & 10.06 $\pm$ 0.05 & 2.14 & 0.44 dex & 0.38 $\pm$ 0.02 dex & 0.171 dex & 0.153 $\pm$ 0.009 dex\\%
& KROSS {\it disky} & 4.7$\phantom{0}$ $\pm$ 0.4 & 10.0$\phantom{0}$ $\pm$ 0.3 & 2.25 & 0.37 dex & 0.25 $\pm$ 0.05 dex & 0.08$\phantom{0}$ dex & 0.06$\phantom{0}$ $\pm$ 0.01$\phantom{0}$ dex\\%
\hline
\end{tabular}
\caption{Parameters of the bisector of the forward and reverse straight line fits to the $M_{K}$ and $\log M_{*}$ Tully-Fisher relations of the composite $z \sim 0$ comparison samples and KROSS sub-samples \textit{all} and \textit{disky}. Uncertainties are quoted at a 1$\sigma$ level.}
\label{tab:freefitpars}
\end{table*}

\begin{table*}
\hspace{-10mm}\begin{tabular}{ | l | l | l | l | l | l | l | l | l |l |}
\hline
TFR & Sample & redshift & $N_{\rm{gal}}$ & Slope & Intercept & Pivot & $\sigma_{\rm{tot}}$ & $\sigma_{\rm{int}}$ & Offset \\
\hline
$M_{K}$ & KROSS all & 0.8--1.0 & 210 &$-$8.9  & $-$23.6 $\pm$ 0.1 & 2.14 & 1.53 mag & 1.43 $\pm$ 0.08 mag & $-$1.1 $\pm$ 0.1 mag \\
& KROSS disky & 0.8--1.0 & 56 &$-$8.9  & $-$23.4 $\pm$ 0.1 & 2.25 & 0.79 mag & 0.66 $\pm$ 0.08 mag & $\phantom{-}$0.1 $\pm$ 0.1 mag \\
\hline
$\log M_{*}$ & \citet{Miller:2011aa} & 0.2--0.5 & 129 & 3.68 & $\phantom{0}$9.83 $\pm$ 0.04 & 2.09 & 0.26 dex & 0.20 $\pm$ 0.03 dex & $-$0.04 $\pm$ 0.07 dex\\
& & 0.5--0.8 & &3.68  & $\phantom{0}$9.74 $\pm$ 0.05 & 2.08 & 0.33 dex &  0.26 $\pm$ 0.03 dex & $-$0.09 $\pm$ 0.08 dex\\
& & 0.8--1.3 & &3.68 & 10.34 $\pm$ 0.05 & 2.23 & 0.30 dex & 0.21 $\pm$ 0.05 dex & $-$0.04 $\pm$ 0.08 dex\\
& \citet{Flores:2006} all & 0.41--0.71 & 30 & 3.68  & 10.48 $\pm$ 0.08 & 2.18 & - & - & $\phantom{-}$0.27 $\pm$ 0.09 dex\\
& \citet{Flores:2006} RD & 0.46--0.70 & 9 & 3.68  & 10.4$\phantom{0}$ $\pm$ 0.1 & 2.29 & - & - & $-$0.2$\phantom{0}$ $\pm$ 0.2$\phantom{0}$ dex\\
& \citet{Puech:2008} all & 0.42--0.74 & 54 & 3.68  & 10.34 $\pm$ 0.08 & 2.23 & 0.60 dex & 0.43 $\pm$ 0.09 dex & $\phantom{-}$0.0$\phantom{0}$ $\pm$ 0.1$\phantom{0}$ dex\\
& \citet{Puech:2008} RD & 0.42--0.70 & 12 & 3.68  & 10.52 $\pm$ 0.09 & 2.31 & 0.14 dex & 0 dex & $-$0.2$\phantom{0}$ $\pm$ 0.1$\phantom{0}$ dex\\
& KROSS {\it all} & 0.8--1.0 & 210 & 3.68  & 10.08 $\pm$ 0.04 & 2.14 & 0.66 dex & 0.58 $\pm$ 0.03 dex & $\phantom{-}$0.02 $\pm$ 0.08 dex\\
& KROSS {\it disky} & 0.8--1.0 & 56 & 3.68  & 10.05 $\pm$ 0.05 & 2.25 & 0.37 dex & 0.26 $\pm$ 0.05 dex & $-$0.41 $\pm$ 0.08 dex\\
& \citet{Miller:2012} & 1--1.7 & 42 & 3.68  & $\phantom{0}$9.88 $\pm$ 0.07 & 2.10 & 0.43 dex & 0.26 $\pm$ 0.08 dex & $-$0.03 $\pm$ 0.09 dex\\
& \citet{Cresci:2009} & 1.5--2.5 & 16 & 3.68  & 10.57 $\pm$ 0.06 & 2.37 & 0.23 dex & 0.1$\phantom{0}$ $\pm$ 0.1$\phantom{0}$ dex & $-$0.33 $\pm$ 0.09 dex\\
& \citet{Gnerucci:2011} & 3--3.6 & 11 & 3.68  & $\phantom{0}$9$\phantom{0}$$\phantom{0}$  $\phantom{,}$$\pm$ 1 & 2.11 & 1.11 dex & 0 dex & $-$0.6$\phantom{0}$ $\pm$ 1.0$\phantom{0}$ dex\\
\hline
\end{tabular}
\caption{Parameters of the fixed slope linear fits to the $M_{K}$ and $\log M_{*}$ Tully-Fisher relations of KROSS sub-samples \textit{all} \& \textit{disky}, and samples of varying redshift from the literature. For each fit the slope was constrained to that of the respective $z \sim 0$ comparison relation and the pivot was set as the median value of $V_{80}$. The best fit to the data of \citet{Puech:2008} RD, and \citet{Gnerucci:2011} was consistent with zero intrinsic scatter. Since tabulated uncertainties were unavailable for the \citet{Flores:2006} data, we make no attempt to determine the scatter of these samples. Uncertainties are quoted at a 1$\sigma$ level. Uncertainties in the offsets include the uncertainty in the $z \sim 0$ TFR zero-point and an uncertainty of $0.06$ dex in the conversion between stellar masses derived assuming an IMF other than Chabrier. We do not report $\zeta_{\rm{tot}}$ or $\zeta_{\rm{int}}$ since, in the case of fixed slope, $m$ the forward and reverse total and intrinsic scatters are related as $\zeta_{\rm{tot}}=\sigma_{\rm{tot}}/m$ and $\zeta_{\rm{int}}=\sigma_{\rm{int}}/m$ respectively.}
\label{tab:fixedfitpars}
\end{table*}

\section{Discussion}
\label{sec:discussion}

Using the \textit{disky} sub-sample, we found the best bisector fits to the rest-frame $K$-band and stellar mass TFRs to be as follows

\begin{multline}
M_{K} / \rm{mag}= (-7.3 \pm 0.9) \times [(\log(V_{80}/\rm{km\ s^{-1}})-2.25]\\
- 23.4 \pm 0.2
\end{multline}

\begin{multline}
\log(M_{*} / M_{\odot})= (4.7 \pm 0.4) \times [(\log(V_{80}/\rm{km\ s^{-1}}) - 2.25]\\
+ 10.0 \pm 0.3
\end{multline}

Considering sub-sample \textit{all}, the slope of both the $M_{K}$ and $M_{*}$ TFRs are shallower than those measured in the local Universe. For the \textit{disky} sub-sample, the slope of the $M_{K}$ relation is slightly shallower, within 1$\sigma$ uncertainties, than the $z \sim 0$ comparison relation. Conversely the slope of the {\it disky} stellar mass TFR is slightly steeper, within 1$\sigma$ uncertainties, than the $z \sim 0$ comparison relation. We draw no conclusions from the free-fitted slopes given the still comparatively large scatter, with respect to $z \sim 0$, for the $z \sim 1$ samples.

\begin{figure*}
\begin{minipage}[]{1\textwidth}
\centering
\includegraphics[width=0.5\textwidth, trim= 0 10 10 50,clip=True]{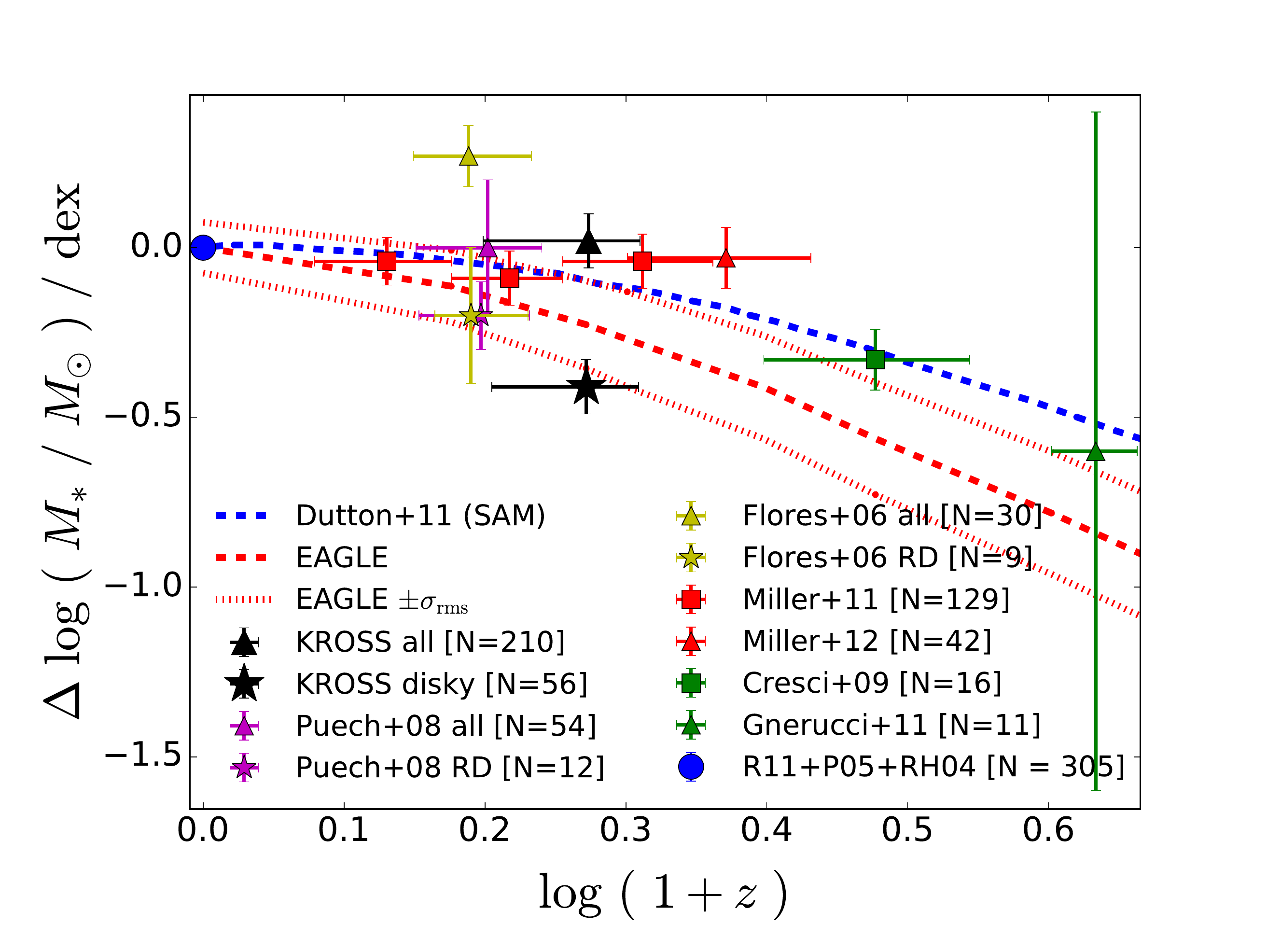}\includegraphics[width=0.5\textwidth, trim= 0 10 10 50,clip=True]{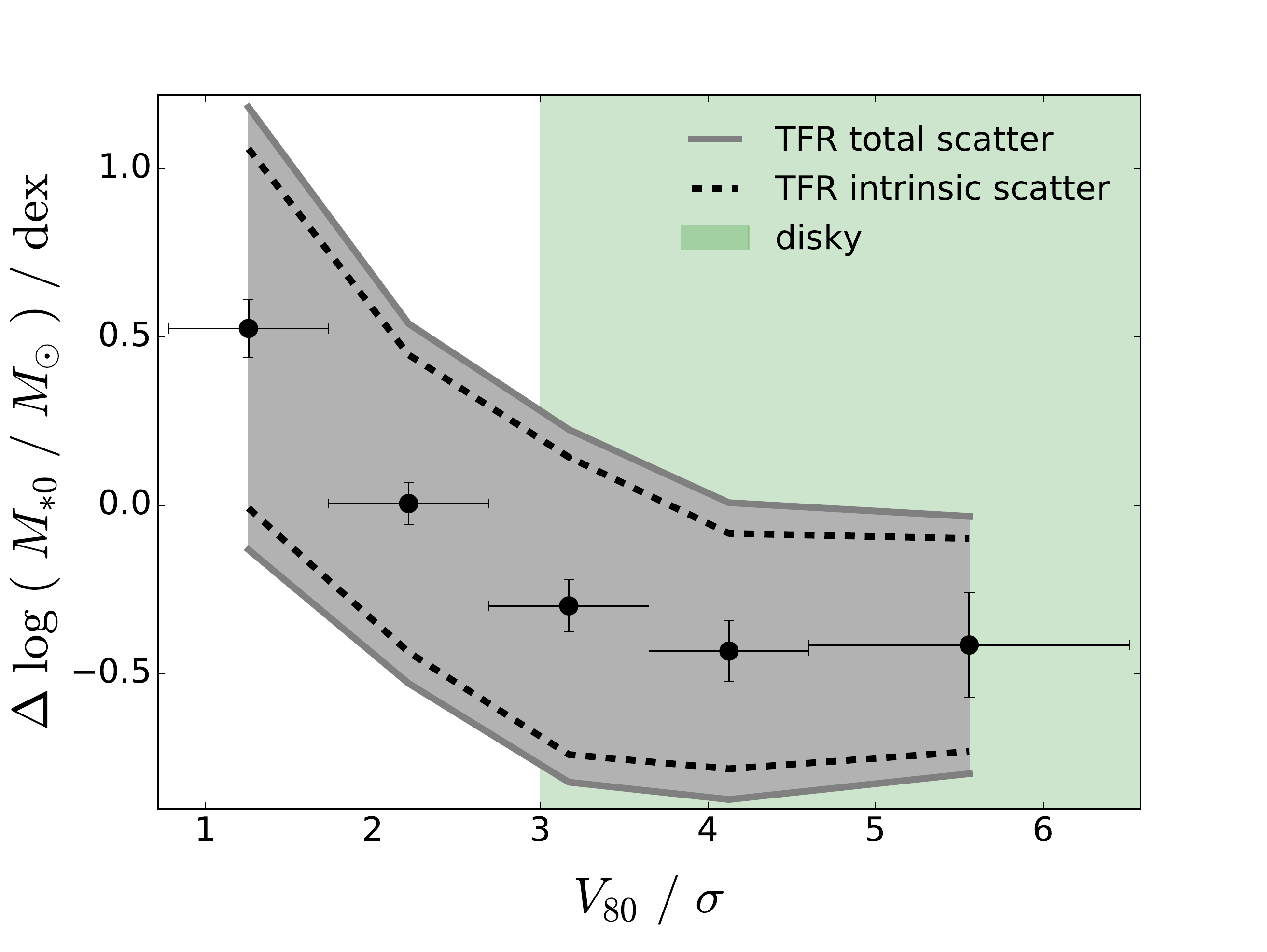}
\end{minipage}
\caption{%
{\bf Left: } The evolution of the stellar mass TFR offset with respect to the $z \sim 0$ comparison sample as described in \S\ref{subsec:compz0}. The corresponding fit parameters are shown in Table \ref{tab:fixedfitpars}. We use the best fit to KROSS sub-samples \textit{all} and \textit{disky}, and several samples, at varying redshifts, from the literature. The slope of the best fit is constrained to be equal to that of the best free fit to the $z \sim 0$ comparison sample. We include the evolution of the TFR zero-point (with respect to $z \sim 0$) with redshift, as predicted by the semi-analytical modelling (SAM) of \citet{Dutton:2011} and the EAGLE simulation. For reference we also plot the rms scatter as measured from the EAGLE samples. We linearly interpolate between each point in order to better highlight the trend in predicted offset as a function of redshift from EAGLE and \citet{Dutton:2011}. The KROSS \textit{disky} sub-sample, and the literature samples which comprise rotating disks, generally agree with the predictions of EAGLE or SAM. We measure little (or in some cases positive) evolution of the TFR zero-point for those samples that do not use IFU observations to differentiate between disk-like and non-disk-like galaxies; these tend to lie above the predictions of EAGLE and \citet{Dutton:2011}. {\bf Right: } The evolution of the KROSS stellar mass TFR zero-point as a function of  $V_{80}\ /\ \sigma$. As with the left panel we find the offset of the TFR with respect to the $z \sim 0$ comparison sample, but in this case for the best fit to KROSS galaxies within bins of varying $V_{80}/\sigma$. We again constrain the slope of the fit to that of the comparison sample. The data points are plotted at the centre of each bin whilst the error bars in $V_{80}/\sigma$ denote the width of the bin. The intrinsic and total scatter are interpolated between points. It can be seen that there is a general trend from positive offsets to negative offsets with increasing $V_{80}/\sigma$. This is consistent with a significantly offset TFR, with respect to $z \sim 0$ for \textit{disky} galaxies (i.e. $V_{80}/\sigma > 3$, indicated by the green shaded area), whilst galaxies are correspondingly scattered to more positive offsets on the TFR depending on the relative importance of pressure support in these systems.%
     }%
\label{fig:TFevolution}
\end{figure*}

Fixing the slope of the KROSS TFRs to that of their respective comparison relations provides evidence to suggest an evolution of the $M_{*}$ TFR zero-point between $z \sim 1$ and $z \sim 0$, for rotationally supported \textit{disky} galaxies. However, we find no evolution in the $M_{K}$ TFR zero-point between $z \sim 1$ and $z \sim 0$ when considering the same {\it disky} sub-sample. We note that both these measurements are robust to the choice of radius at which we extract a rotation velocity (see \S\ref{subsec:resultsTFR}). 

These results imply that disk-like galaxies of a given dynamical mass contain less stellar mass at $z \sim 1$ than at $z \sim 0$, yet emit the same amounts of $K$-band light. The latter implies an increase in the $K$-band stellar mass-to-light ratio by a factor of $\sim 2.75$ since $z \sim 1$. \citet{Anouts:2007} analyse the evolution of the stellar mass-to-light ratio since $z \sim 1.5$ via SED fitting (with visible to mid-IR photometry) for a sample containing tens of thousands of galaxies. They find that the average $K$-band stellar mass-to-light ratio, $M_{*}/L_{K}$ of actively star-forming galaxies increases by 0.27 dex, but with a root-mean-squared (rms) scatter of 0.21 dex. Incorporating the rms scatter, this equates to an increase by a factor of $\sim 1.1$--$3.0$. \citet{Drory:2004} measure the evolution of $M_{*}/L_{K}$ since $z \sim 1.2$ by fitting a grid of stellar populations models of varying star formations histories, ages, and metallicities to the visible and near-IR photometry of a sample of over 500 galaxies selected in the $K$-band. They find an increase, by a factor of $\sim 1.4$--$3.4$, in the $M_{*}/L_{K}$ since $z \sim 1$. Thus, whilst the increase in the $M_{*}/L_{K}$ inferred from this work is towards the higher ends of these ranges, it appears a feasible evolution.  

The measured offset in the $M_{*}$ TFR implies that stellar mass assembly continues despite the drop off in global star formation rate density in the Universe at redshifts below $z \sim 1$ \citep[see e.g.][]{Hopkins:2006,Sobral:2013}. We do not compare the measured offset to that predicted from stellar population models as the the latter will depend strongly on the assumed galaxy age and star formation history at $z \sim 1$. Instead we prefer to compare to the cosmological hydronamical galaxy formation simulations of EAGLE. The measured increase in stellar mass in {\it disky} galaxies since $z \sim 1$ agrees, within uncertainties, with the predicted evolution of the TFR from star-forming (star formation rates > $1 M_{\odot} \rm{yr}^{-1}$) EAGLE galaxies. However, it is on the upper limit of the EAGLE prediction. We therefore consider whether such an evolution in the $M_{*}$ TFR zero-point since $z \sim 1$ is physically feasible. 

The relative increase in stellar mass is most easily reconciled with the conversion of gas mass in to stellar mass in galaxies over the last $\sim$8 billion years, in a secular evolution scenario. However, to determine the likelihood of this scenario, the gas mass fraction of the KROSS galaxies must first be considered. \citet{Stott:2016} inverted the Kennicutt-Schmidt relation \citep{Kennicutt:1998} of KROSS galaxies in order to obtain an estimated gas mass-to-baryonic mass ratio of $35 \pm 7$ percent. Converting all of the gas mass within the KROSS galaxies in to stellar mass would therefore only account for, at most, 0.24 dex of the measured offset in the $M_{*}$ TFR (i.e. an increase by a factor $\sim 1.7$ in the stellar mass since $z \sim 1$). In fact, to reconcile the measured $-0.41$ dex offset in this manner would require  a baryonic gas fraction of $\sim72$ percent in the {\it disky} KROSS galaxies. 

There is of course the possibility of galaxies accreting extra gas over the last $\sim 8$ billion years - indeed the observed specific baryon accretions rate at $z \sim 1$ ranges from $\sim 0.8$--$0.6\ \rm{Gyr}^{-1}$ for galaxies with stellar mass $\log(M_{*}/M_{\odot})=9.3$--$10.7$, decreasing to $\sim 0.2$--$0.1\ \rm{Gyr}^{-1}$ for galaxies of the same stellar mass in the local Universe \citep{Elbaz:2007,Salim:2007,Dutton:2010}. Combining this with a measured depletion timescale for KROSS galaxies of $\sim 1$ Gyr \citep{Stott:2016}, it can be seen there is opportunity for star-forming galaxies to assemble significant amounts of stellar mass since $z \sim 1$ such as those inferred by the offset in the $M_{*}$ TFR measured in this work. 

Our findings are at odds with the studies of \citet{Miller:2011,Miller:2012}, which find no evolution of the TFR at similar redshifts. Excluded from our analysis (due to a lack of tabulated data) is the work of \citet{Conselice:2005aa}, who examined the the evolution of both the $K$-band and stellar mass Tully-Fisher relation out to $z \sim 1.2$ using Keck spectroscopy and near-infrared photometry from the Keck Near-Infrared Camera, the UKIRT Fast-Track Imager, and the Cooled Infrared Spectrograph and Camera for OHS on the {\it Subaru} telescope. They find no significant evolution with respect to the TFR as measured in the local Universe, in agreement with the works of \citeauthor{Miller:2011,Miller:2012}. However the key point is that, unlike these studies, this work is able to select for only those galaxies at $z \sim 1$ that are predominantly rotationally supported. Inclusion of galaxies that are more turbulent sees the evolution of the KROSS TFR zero-point decrease to much smaller offsets with respect to $z \sim 0$ -- in line with the findings of \citeauthor{Miller:2011} (and \citeauthor{Conselice:2005aa}), but at odds with the previous IFU studies of \citet{Flores:2006}, \citet{Puech:2008}, \citet{Cresci:2009}, and \citet{Gnerucci:2011}, all of which predominantly agree with the predictions of the EAGLE simulation and, to a lesser degree, those of \citet{Dutton:2011}. 

In a recent paper by \citet{Teodoro:2016}, the authors construct the $z \sim 1$ stellar mass TFR using a sample of only 18 galaxies, each observed with KMOS, of which 14 are publicly available KROSS observations. They report no evolution of the relation since $z \sim 1$ and present this as evidence that disc galaxies at this epoch closely resemble their kinematically mature counterparts in the local Universe. However, there is an obvious danger to drawing conclusions on the nature of all disc galaxies at $z \sim 1$ from such a small sample. Furthermore, their sample was selected to include only those galaxies that lend themselves best to the modelling of their dynamics and is not statistically representative of the bulk of the population of galaxies at $z \sim 1$. With the much larger sample presented in this work, we show that in fact a minority of star-forming galaxies at $z \sim$ 1 are strongly rotation dominated ($V_{80}/\sigma > 3$) and only with the full KROSS sample do we detect an evolution in the stellar mass TFR.      
\hfill \\

Both Tully-Fisher relations for sub-sample \textit{all} show large scatter. This scatter is reduced in the \textit{disky} sub-sample, but still large in comparison the $z \sim 0$ relations. There are several potential sources contributing to the intrinsic scatter in the KROSS TFRs. The most obvious source of scatter in the TFRs of sub-sample \textit{all} is the inclusion of galaxies that have significant pressure support and thus violate the initial assumption of circular motion of the Tully-Fisher relation. This source of scatter is confirmed as we see reductions in the intrinsic scatter of the $M_{K}$ and $M_{*}$ TFR respectively between sub-sample \textit{all} and the \textit{disky} sub-sample, where we select only those galaxies deemed to be predominantly rotationally supported and disk-like (i.e. moving in a circular motion). The intrinsic scatter of sub-sample \textit{all} is $\sim 2.3$ and $\sim 2.4$ times larger than the $M_{K}$ and $\log M_{*}$ $z \sim 0$ comparison relation respectively. However, the intrinsic scatter reduces to $\sim 1.6$ times ($M_{K}$ and $\log M_{*}$) the value of the $z \sim 0$ relations when only rotationally supported galaxies are considered at $z \sim 1$ (i.e. the \textit{disky} sub-sample). Conceptually, we may consider the rotation velocity as a rough proxy for the total (dynamical) mass. Assuming this is true it is clear then that an erroneously small total mass will be inferred for those galaxies with significant pressure support. As a result those galaxies will be scattered to lower values along the abscissa of the Tully-Fisher relation.

Systems with low $S/N$ H$\alpha$ emission will have noisier observed velocity fields. In these cases, the best fitting arctangent model is less reliable. This may also increase the scatter in the TFR. More importantly, the arctangent model is an unsatisfactory description of the dynamics for those galaxies which, while being predominantly rotationally supported, have a rotation curve that peaks centrally ($r \lesssim 10$ kpc) before flattening out at lower velocity; this is consistent with the (baryonic) disk component of the galaxy dominating dynamically over the (dark) halo component in the inner regions of the galaxy \citep[see e.g.][]{Casertano:1991}. However, in the \textit{disky} sub-sample at least, both types of system are rare with most systems adequately described by the arctangent model (see Appendix \ref{sec:ssA}). 

The \textit{disky} sub-sample contains galaxies determined to be predominantly rotationally supported as decided by a cut to sub-sample \textit{all} as $V_{80}/\sigma>3$ (see \S\ref{subsubsec:TFRoffset_vs_z}). The specific limit was chosen in order to select galaxies that were ``disky" whilst also maintaining a reasonable sample size. However, importantly it is also the value above which the offset of the stellar mass TFR does not undergo further significant evolution. In this respect, the choice of limit is not a subjective choice but is rather driven by the data. It is clear however that a change to the limiting value of $V_{80}/\sigma$ will lead to a change in the measured intrinsic scatter, slope, and offset of the TFR as extra galaxies are included or excluded from the sub-sample. This is apparent in Figure \ref{fig:TFevolution} (Right). It is important also to view the choice of $V_{80}/\sigma>3$ to define ``disky" galaxies in the context of galaxies at $z \sim 0$. Indeed, in the local Universe, we see typical values of $V/\sigma \gtrsim 10$ for the thin disk of spiral galaxies \citep{Genzel:2006}. Thus it is clear that, despite the many different definitions of $V/\sigma$ abound in the literature (see \citealt{Stott:2016} for a discussion), even in the upper limit of the range of $V_{80}/\sigma$ values seen in the KROSS sample, these galaxies are far more turbulent in terms of their gas dynamics than galaxies in the present day. In this respect, we may still attribute some of the intrinsic scatter measured in the \textit{disky} sub-sample to turbulence within the gaseous disk providing pressure support.

In addition to the assumption of circular motion inherent in the TFR, a second important assumption of the relation that must be considered is that the average surface mass density of the galaxies within a sample is constant. In the local Universe this means comparing galaxies of the same morphological type. Previous studies have shown that the Tully-Fisher relations for early-type and late-type galaxies have different slopes \citep[see e.g.][]{Williams:2010aa,Davis:2011aa}. Even between late-type morphologies, studies have shown significantly different slopes \citep[see e.g.][]{Lagattuta:2013aa}. At $z \sim 1$, ``disky" galaxies are just starting to emerge. In this regime it makes less sense to talk about morphological types in the same way as at $z \sim 0$. \citet{Abraham:1999} argues infact that the classic ``Hubble Tuning Fork"  description of galaxy morphology begins to break down from $z \sim 0.5$ onwards. Whilst the KROSS galaxies were selected to be ``normal" star forming galaxies, there are no further morphological selection criteria. Thus it is likely that there will be variation in surface mass density from galaxy to galaxy in the KROSS sample. In practice this will result in an increase in the inferred intrinsic scatter as this is effectively combining several different morphological types, each with different slopes and scatter for the Tully-Fisher relation, and then fitting them with one single slope and scatter. 

One must also consider a further issue related to the this. Although the KROSS sample and $z \sim 0$ comparison sample span similar ranges in rotation velocity and stellar mass (in fact, the comparison sample spans a slightly wider range than the KROSS sample in both respects), we cannot assume that we are ``following" the same population of galaxies between $z \sim 1$ and the local Universe. Indeed, considering a window of stellar mass, galaxies will evolve in to and out of this window as time passes between $z \sim 1$ and $z \sim 0$. The implications of choosing to compare samples within the same (stellar) mass window between redshifts as opposed to, for example, galaxies within a similar range of star formation are complex and will be the subject of future work by the authors. As discussed above, the evidence presented in this work is most easily reconciled with a secular evolution or minor mergers scenario whereby galaxies continually accrete and convert gas mass to stellar mass between $z \sim 1$ and $z \sim 0$.

\section{Conclusions \& Future Work}
\label{sec:conclusions}

We have presented the absolute magnitude $M_{K}$, and stellar mass Tully-Fisher relations for sub-samples drawn from the 584 galaxies observed by KROSS with resolved H$\alpha$ emission. We examine the Tully-Fisher relations for a sub-sample of ``disky" galaxies (\textit{disky} sub-sample) that are predominantly rotationally supported. The selection criteria for both sub-sample \textit{all} and the \textit{disky} sub-sample are described in \S\ref{subsec:subsample}. 

The intrinsic scatter in the TFRs for sub-sample \textit{all} is $\sim 2.3$ times larger than the $M_{K}$ $z \sim 0$ comparison relation and $\sim 2.4$ times larger than $\log M_{*}$ $z \sim 0$ comparison relation. However, the intrinsic scatter reduces to $\sim 1.6$ times ($M_{K}$ and $\log M_{*}$) that of the $z \sim 0$ relation when only rotationally supported ``disky" galaxies are considered (i.e. the \textit{disky} sub-sample). Compared to the local Universe, however, these ``disky" galaxies still have much more turbulent gas dynamics. This turbulence will contribute to the measured intrinsic scatter. The remaining intrinsic scatter could be a result of considering a single Tully-Fisher relation for galaxies with various different morphologies, sizes and/or surface mass densities - effectively an average of several Tully-Fisher relations, all with differing slopes. 

Contrary to some previous studies conducted at similar redshift \citep{Miller:2011,Miller:2012,Teodoro:2016}, but in broad agreement with the predictions of the state-of-the-art hydrodynamical simulations of EAGLE, in comparison to $z \sim 0$ we find an offset of the Tully-Fisher relation for rotationally supported galaxies at $z \sim 1$ to lower stellar mass values ($-0.41$ dex) for a given dynamical mass. Yet we measure no significant offset in the absolute $K$-band TFR over the same period. The ability of KROSS to differentiate between those galaxies with high or low $V_{80}/\sigma$ is why this work detects an evolution of the stellar mass TFR zero-point for \textit{disky} galaxies since $z \sim 1$, whilst some previous studies do not. Assuming no evolution in the surface mass density, the detected zero-point evolution implies a decrease by a factor of $\sim 0.36$ in the dynamical mass-to-stellar mass ratio of disk-like galaxies since $z \sim 1$, and an increase in the $K$-band stellar-mass-to light ratio, by a factor of $\sim 2.75$ over the same period. This may be consistent with a secular evolution scenario whereby gas mass in (and accreted on to) galaxies is converted in to stellar mass over the last 8 billion years. If galaxies have grown mostly via mergers since $z \sim 1$, then we would expect the stellar mass TFR at that epoch to be indistinguishable from those measured using nearby galaxies. In this regime, we would still expect to see an offset, to brighter magnitudes with respect to $z \sim 0$, of the $M_{K}$ TFR, reflecting the comparitively large specific star formation rates of the higher-$z$ galaxies. However, as galaxies grow via mergers (ignoring any non-linear increase in star formation rate as a result of a merger) they would only evolve along the stellar mass TFR, with no evolution of the offset.

As stressed previously, in order to make a direct comparison between the Tully-Fisher relations at $z \sim 1$ and $z \sim 0$ it is essential to compare the different samples of galaxies using the same observational and analytical methods (and thus the same systematic biases). In practice, in order to directly compare the KROSS Tully-Fisher relation to one at $z \sim 0$ we must take IFU observations of galaxies in the local Universe and degrade this data to the same quality as that of the KROSS data e.g. the signal-to-noise ratio, spatial resolution, and spectral resolution must all be equivalent. This degraded data must then be analysed in the same manner as the KROSS data has been analysed, at which point a more direct comparison of the TFRs may be made. 

There are a number of IFU surveys of galaxies at low redshift that are already online, or will be online in the near future, that will provide suitable samples of large numbers of galaxies in the local Universe that we may compare to KROSS. These include Mapping Nearby Galaxies at APO \citep[MaNGA\footnote{\url{https://www.sdss3.org/future/manga.php}};][]{Bundy:2015}, the Calar Alto Legacy Integral Field spectroscopy Area survey \citep[CALIFA\footnote{\url{http://www.caha.es/CALIFA/public_html/?q=node/1}}; see e.g.][]{Sanchez:2012}, and the Sydney-Australian-Astronomical-Observatory Multi-object Integral-Field Spectrograph (SAMI) Galaxy Survey\footnote{\url{http://sami-survey.org/}} \citep[see e.g.][]{Bryant:2015}. Future work will utilise some or all of these surveys in order to gain a clearer understanding of how the Tully-Fisher relation has evolved from the epoch of peak global star formation rate density in the Universe to the present day. 

\section*{Acknowledgements}

We thank the anonymous referee for their constructive review of this work. AT acknowledges support from an STFC Studentship. JPS, AMS, RGB, CMH and IRS acknowledge support from STFC (ST/I001573/1 and ST/L00075X/1). JPS also gratefully acknowledges support from a Hintze Research Fellowship. IRS acknowledges support from an ERC Advanced Investigator programme DUSTYGAL 321334 and a Royal Society/Wolfson Merit Award. AJB gratefully acknowledges the hospitality of the Research School of Astronomy \& Astrophysics at the Australian National University, Mount Stromlo, Canberra where some of this work was done under the Distinguished Visitor scheme. MB acknowledges support from STFC rolling grant ‘Astrophysics at Oxford’ PP/E001114/1. DS acknowledges financial support from the Netherlands Organisation for Scientific research (NWO) through a Veni fellowship, from FCT through a FCT Investigator Starting Grant and Start-up Grant (IF/01154/2012/CP0189/CT0010) and from FCT grant PEst-OE/FIS/UI2751/2014..

We thank Holly Elbert, Timothy Green, and Laura Prichard for their assistance with part of the KROSS observing programme. 

We acknowledge the Virgo Consortium for making their simulation data available. The EAGLE simulations were performed using the DiRAC-2 facility at Durham, managed by the ICC, and the PRACE facility Curie based in France at TGCC, CEA, Bruy\`{e}res-le-Ch\^{a}tel.

Based on observations made with ESO Telescopes at the La Silla Paranal Observatory under the programme IDs 60.A-9460, 092.B-0538, 093.B-0106, 094.B-0061, and  095.B-0035. 




\bibliographystyle{mnras}
\bibliography{TILEY_KROSS.bib} 




\appendix
\section[]{Asymmetric Rotation Curves} \label{sec:centres}

\begin{figure}
\includegraphics[scale=0.35, trim= 10 0 30 0,clip=True]{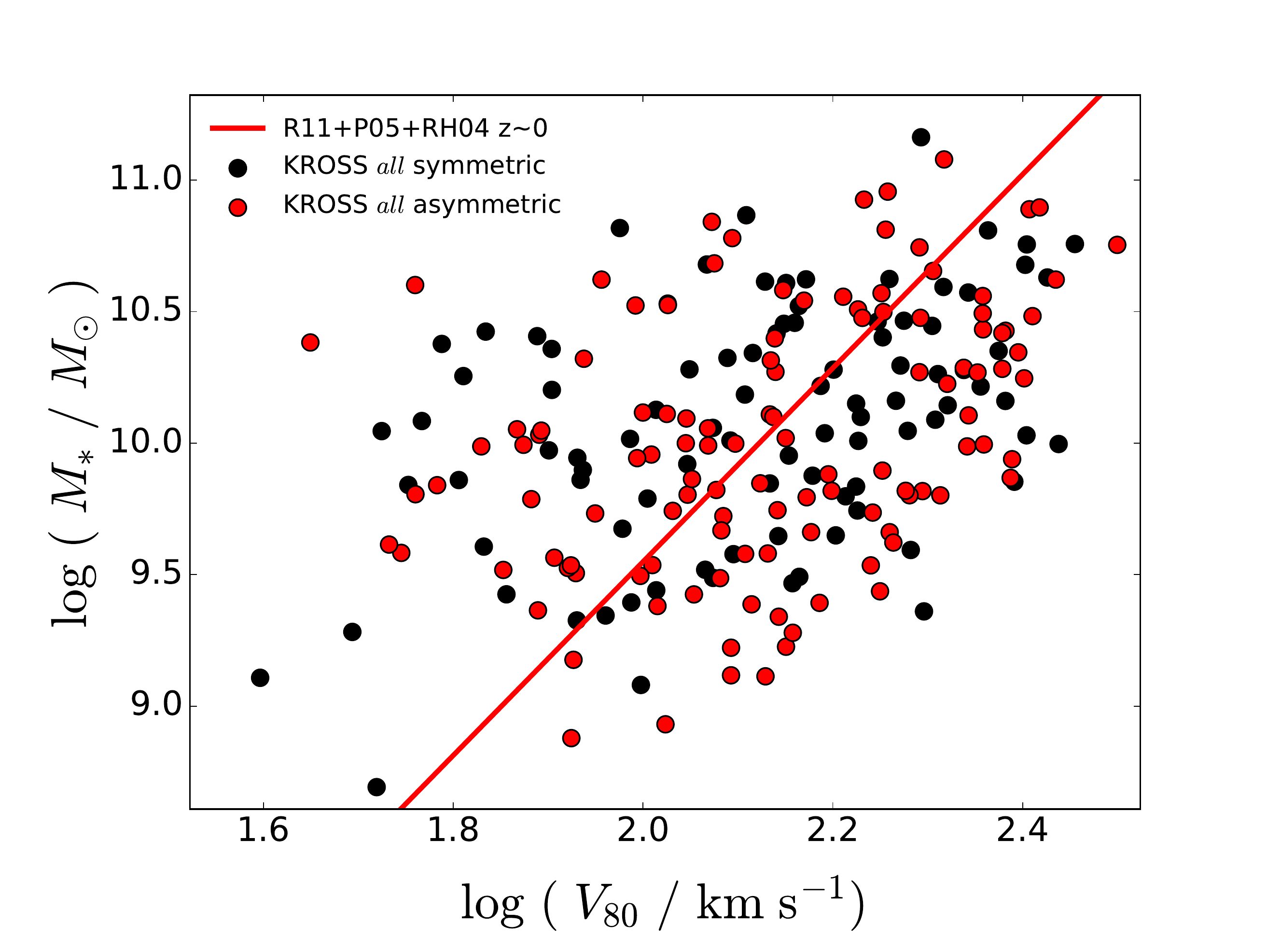}
\caption{%
The TFR for those galaxies in sub-sample {\it all} (as described in Section \S\ref{subsec:subsample}). The points are coloured according to whether they represent galaxies with either a symmetric or asymmetric rotation curve. There is no significant reduction in scatter between those galaxies with symmetric or asymmetric rotation curves. We therefore include both groups of galaxies in our analysis.%
	}%
\label{fig:TFRsymm}
\end{figure} 

The dynamical modelling of KROSS velocity maps is described in Section \S\ref{subsec:modelvels} and in further detail in \citet{Stott:2016}. Here we test the effect on the TFR scatter of those galaxies for which the H$\alpha$ emission extends up to or beyond $r_{80}$ on only one side of the rotation curve i.e. asymmetrical rotation curves. Figure \ref{fig:TFRsymm} shows that the exclusion of such systems in favour of only those with symmetrical (i.e. with H$\alpha$ emission extending up to or beyond $r_{80}$ on both sides) rotation curves does not significantly reduce the scatter in the TFR for sub-sample {\it all} ($\sigma_{\rm{int}}=0.36 \pm 0.04$ for galaxies with symmetric rotation curves versus $\sigma_{\rm{int}}=0.40 \pm 0.04$ for those with asymmetric rotation curves). We therefore only exclude from our analysis those galaxies for which the maximum radial extend of the H$\alpha$ emission is less than $r_{80}$ on {\it both} sides of the rotation curve.

\newpage
\onecolumn
\section[]{Disky Galaxies} \label{sec:ssA}
Figure \ref{fig:diskyfields} shows the best fitting model velocity fields (see \S\ref{subsec:modelvels}) for the KROSS {\it disky} sub-sample (see \S\ref{subsec:subsample}). For each galaxy the observed (``DATA") and best fitting model (``MODEL") velocity fields are displayed alongside the residual (``RESIDUAL") between the two. Also included is the associated integrated H$\alpha$ flux map (``H$\alpha$"), and the extracted rotation curve from the inclination corrected observed and model velocity fields. See Figure \ref{fig:goodvels} for a full description.

\hfill{\\}
\includegraphics[scale=0.45, trim=10 0 0 0, clip=true]{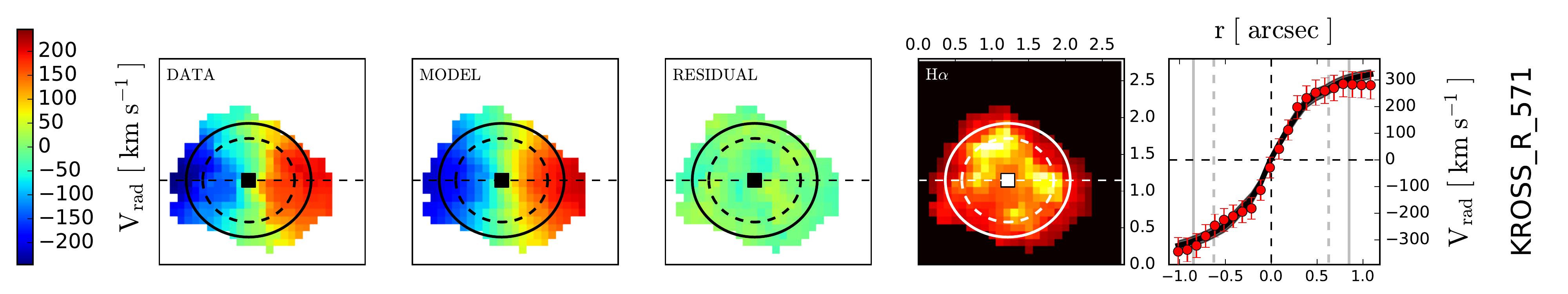}
\includegraphics[scale=0.45, trim=10 0 0 0, clip=true]{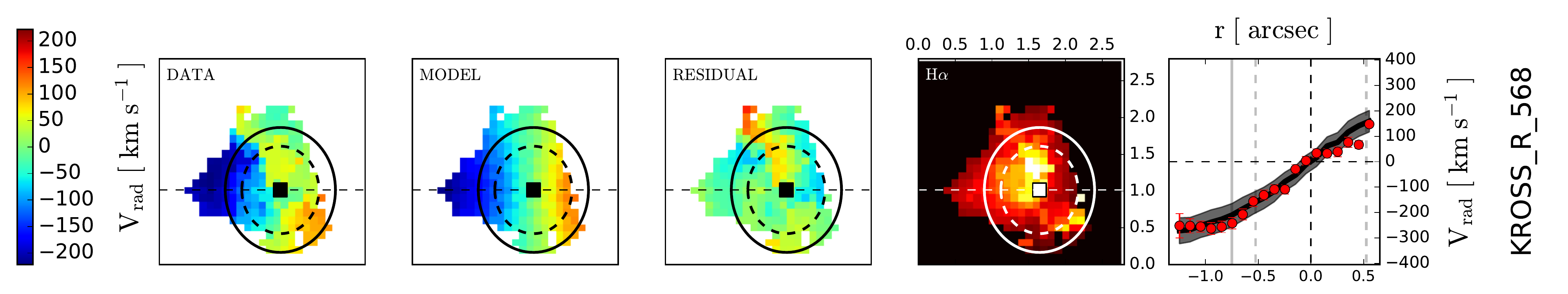}
\includegraphics[scale=0.45, trim=10 0 0 0, clip=true]{PR_ssB_53.pdf}
\includegraphics[scale=0.45, trim=10 0 0 0, clip=true]{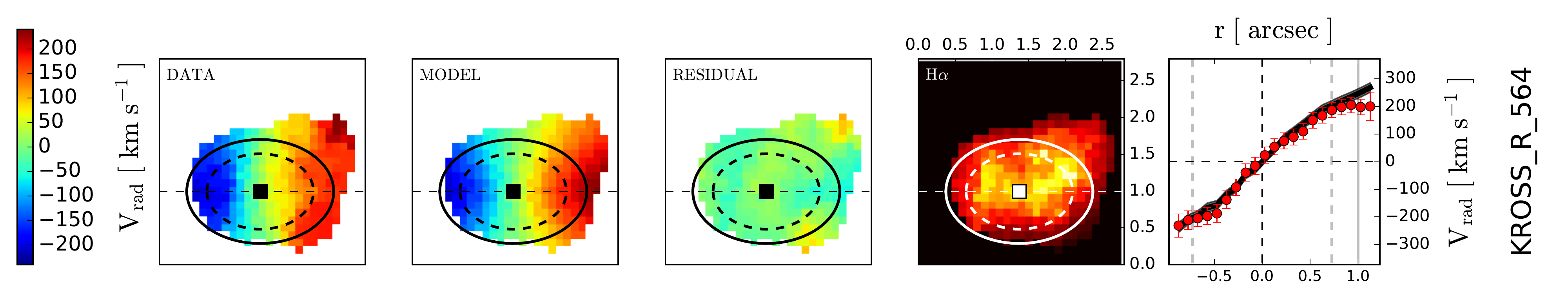}
\includegraphics[scale=0.45, trim=10 0 0 0, clip=true]{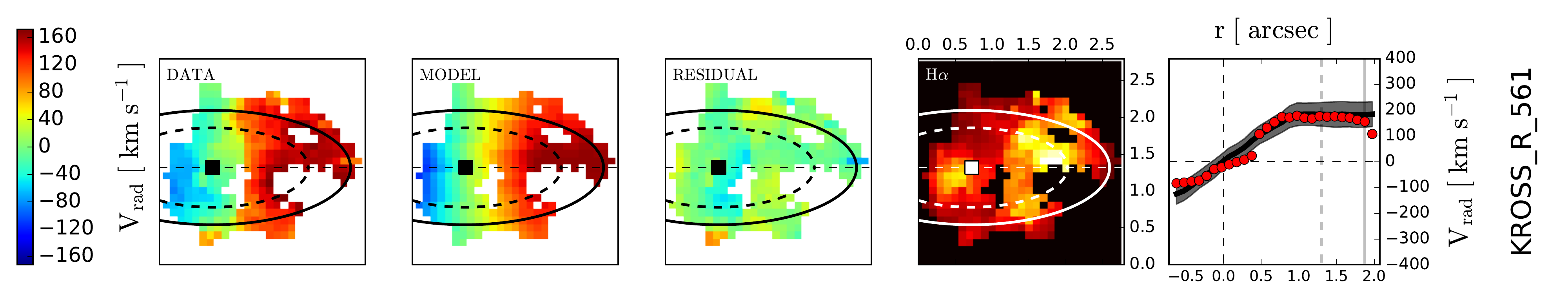}
\includegraphics[scale=0.45, trim=10 0 0 0, clip=true]{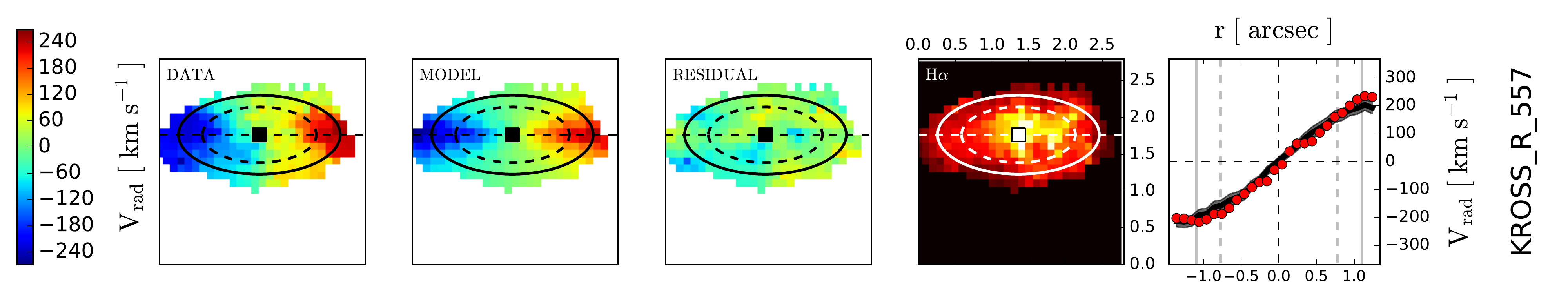}
\includegraphics[scale=0.45, trim=10 0 0 0, clip=true]{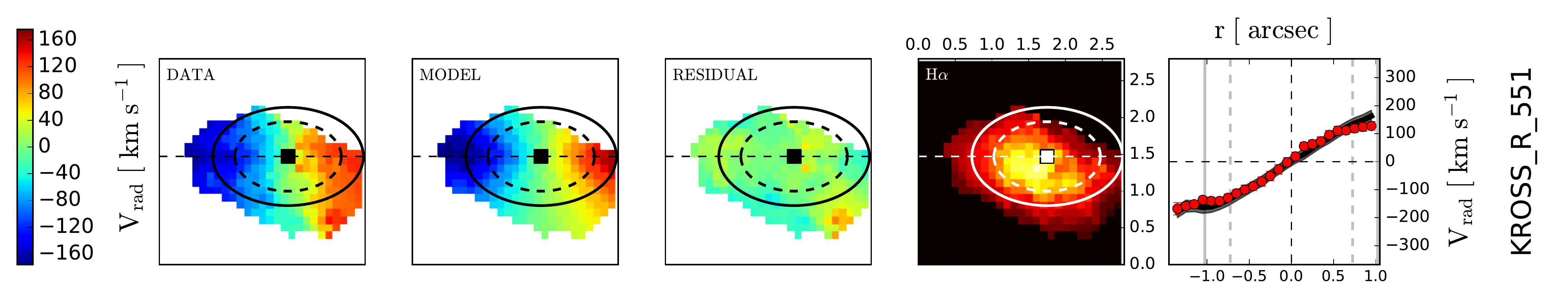}
\includegraphics[scale=0.45, trim=10 0 0 0, clip=true]{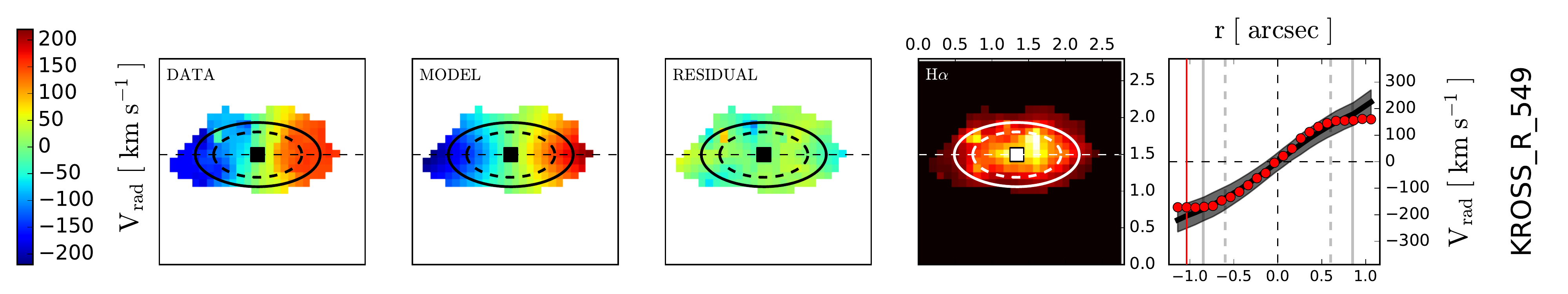}
\includegraphics[scale=0.45, trim=10 0 0 0, clip=true]{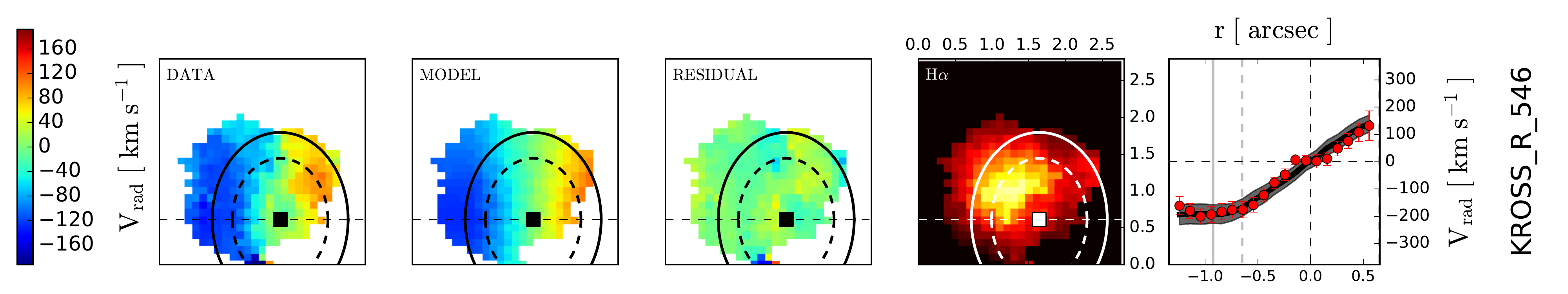}
\includegraphics[scale=0.45, trim=10 0 0 0, clip=true]{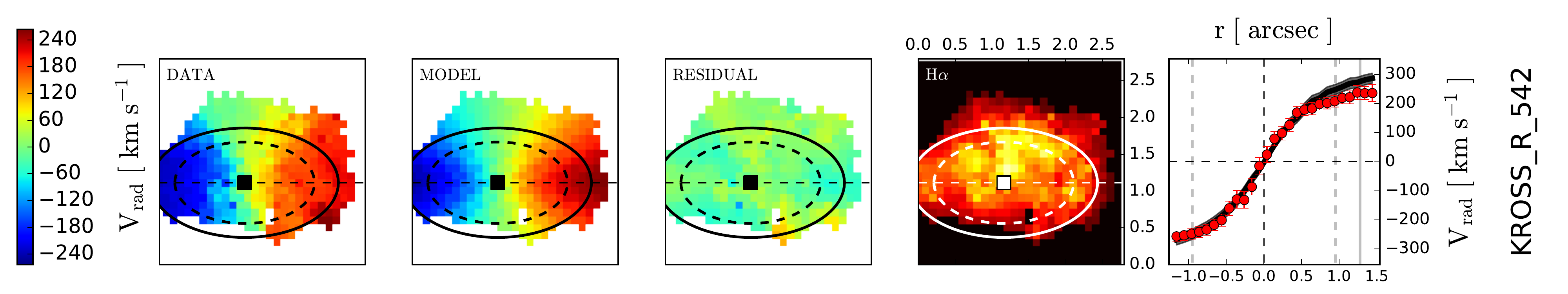}
\includegraphics[scale=0.45, trim=10 0 0 0, clip=true]{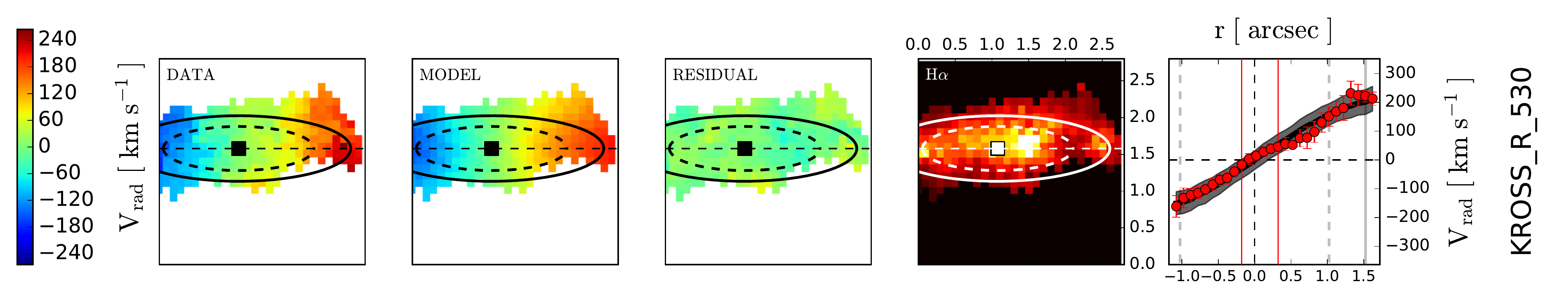}
\includegraphics[scale=0.45, trim=10 0 0 0, clip=true]{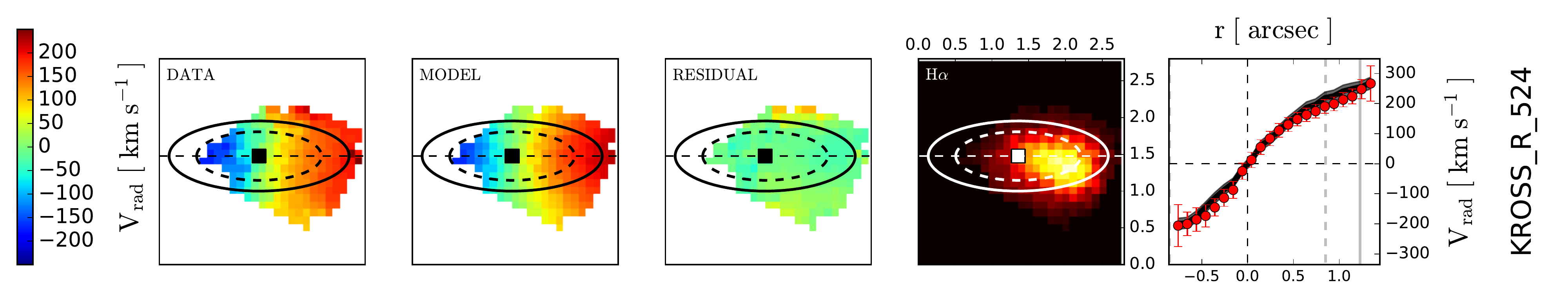}
\includegraphics[scale=0.45, trim=10 0 0 0, clip=true]{PR_ssB_43.pdf}
\includegraphics[scale=0.45, trim=10 0 0 0, clip=true]{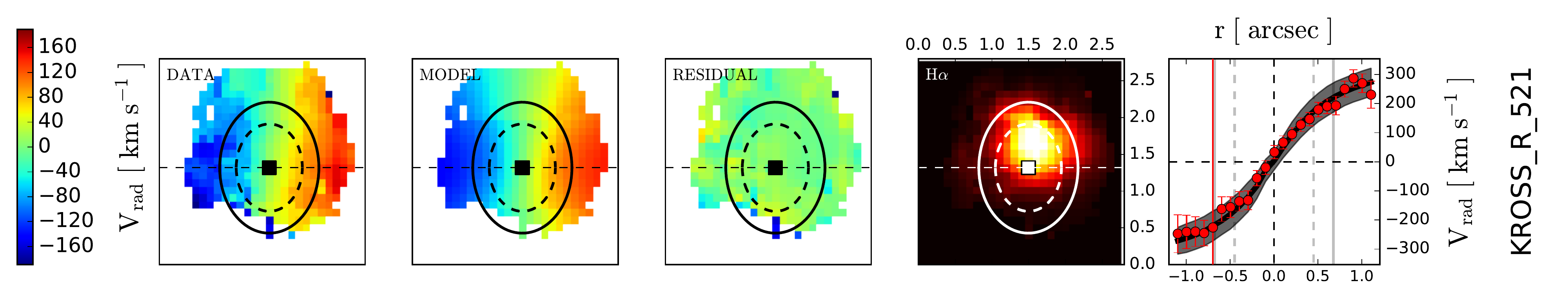}
\includegraphics[scale=0.45, trim=10 0 0 0, clip=true]{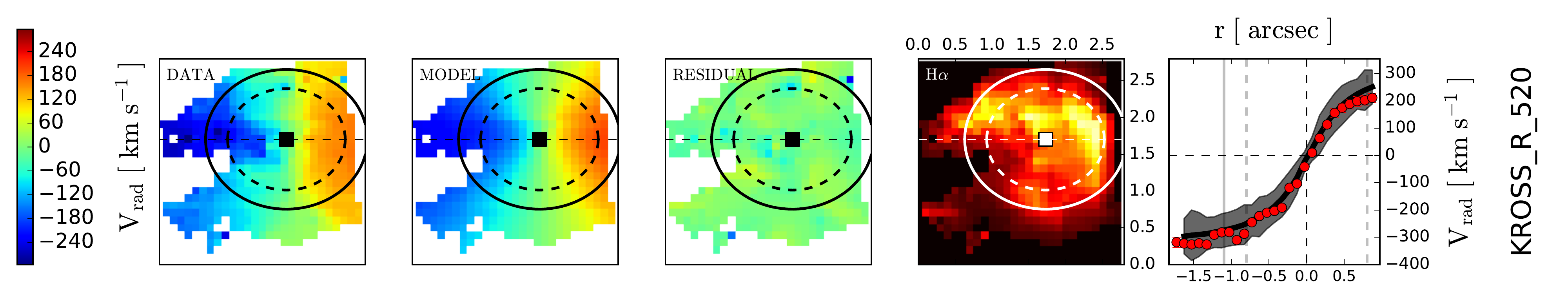}
\includegraphics[scale=0.45, trim=10 0 0 0, clip=true]{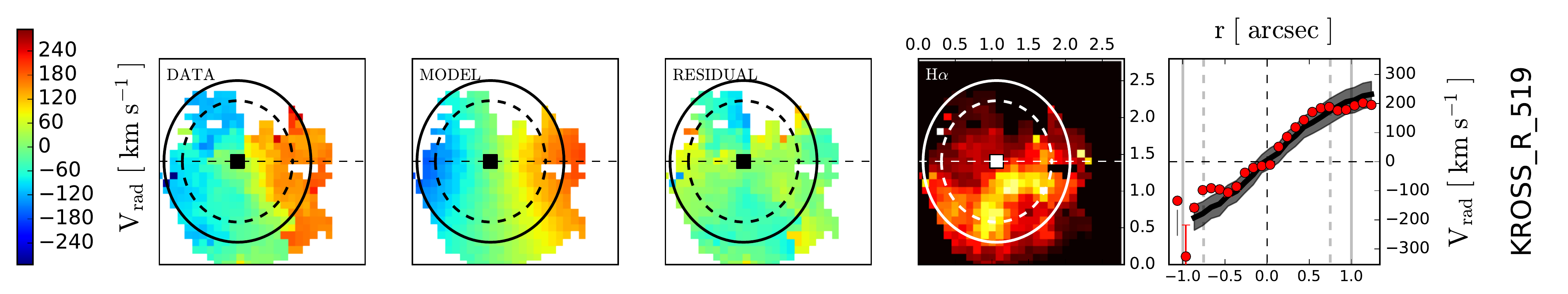}
\includegraphics[scale=0.45, trim=10 0 0 0, clip=true]{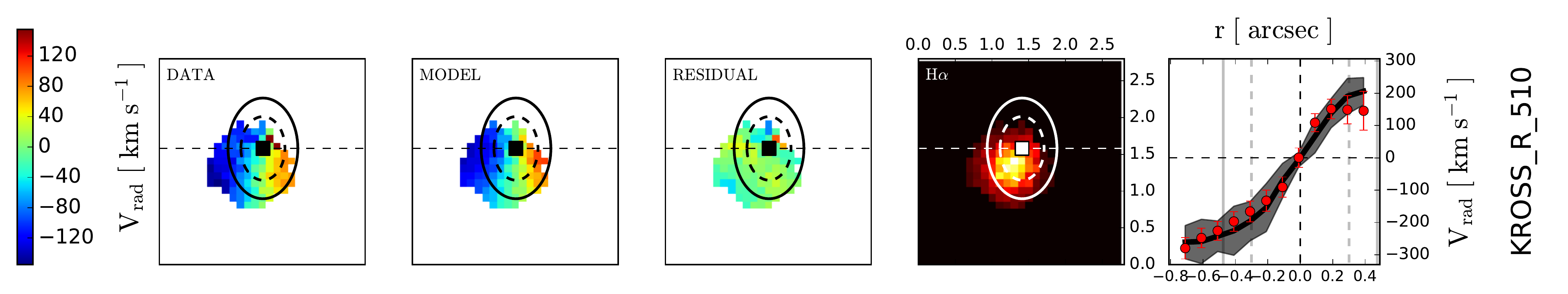}
\includegraphics[scale=0.45, trim=10 0 0 0, clip=true]{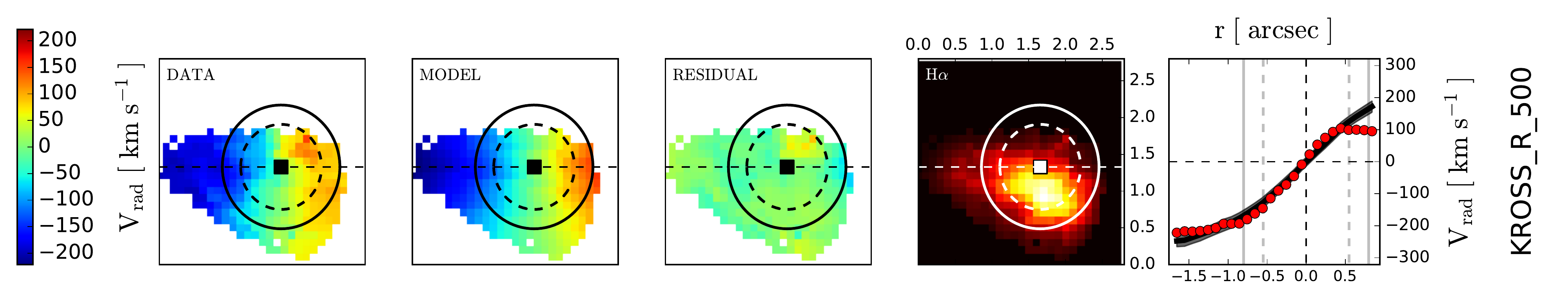}
\includegraphics[scale=0.45, trim=10 0 0 0, clip=true]{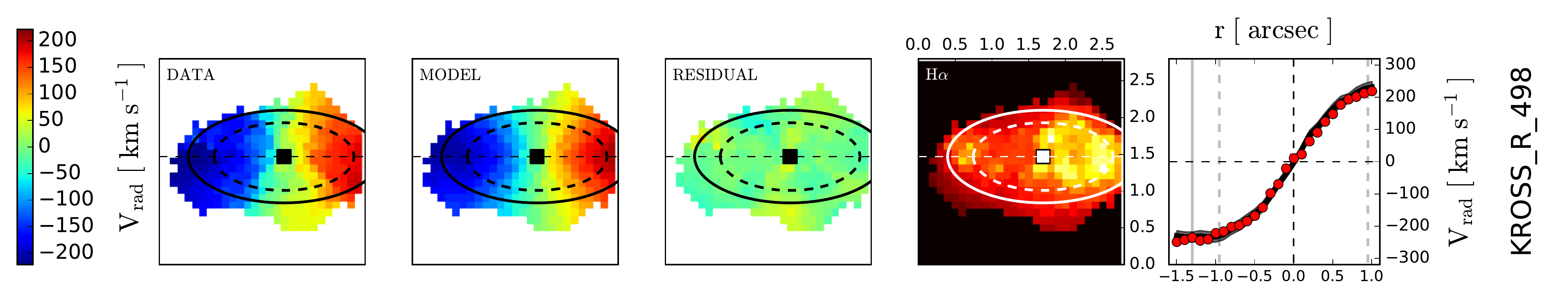}
\includegraphics[scale=0.45, trim=10 0 0 0, clip=true]{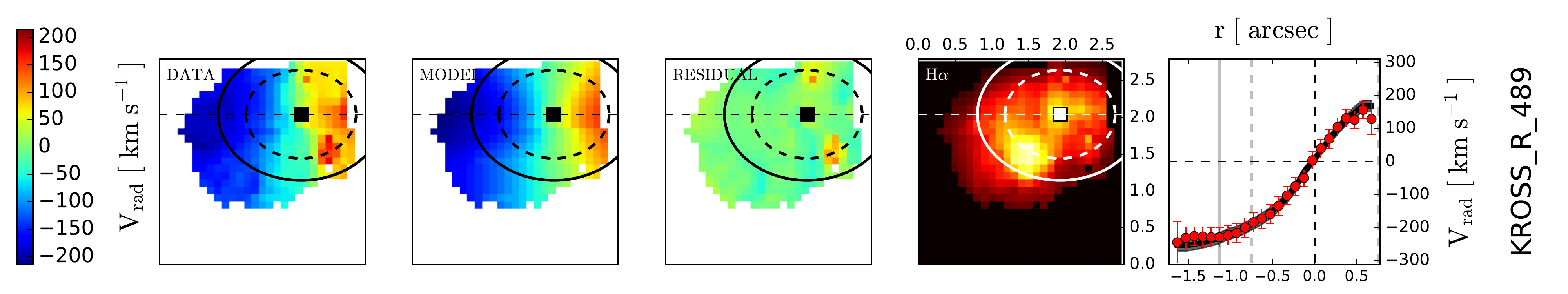}
\includegraphics[scale=0.45, trim=10 0 0 0, clip=true]{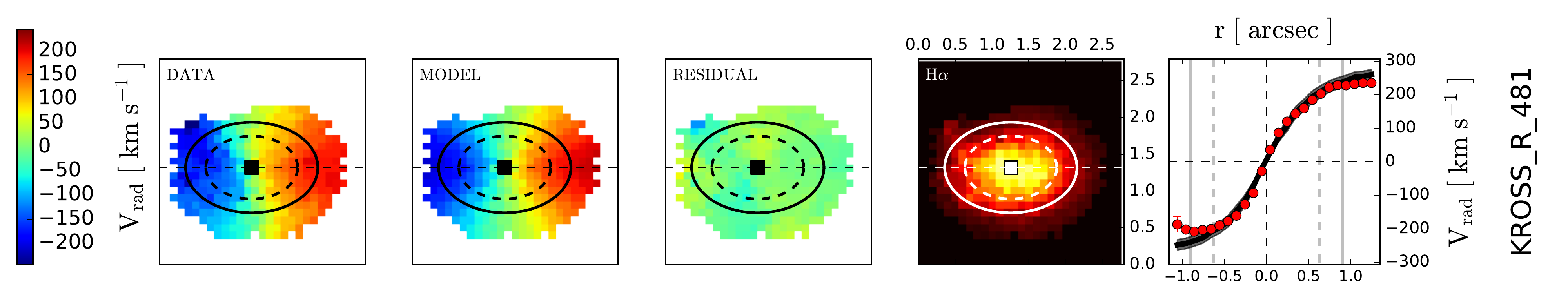}
\includegraphics[scale=0.45, trim=10 0 0 0, clip=true]{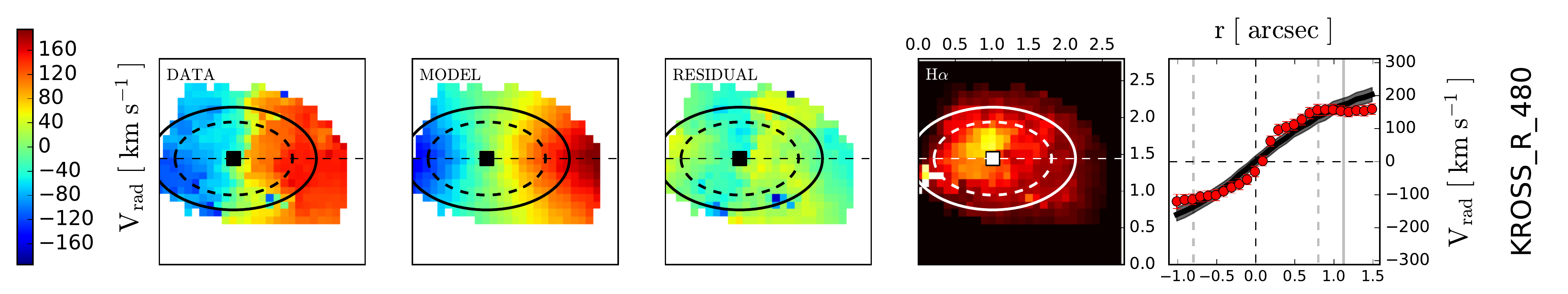}
\includegraphics[scale=0.45, trim=10 0 0 0, clip=true]{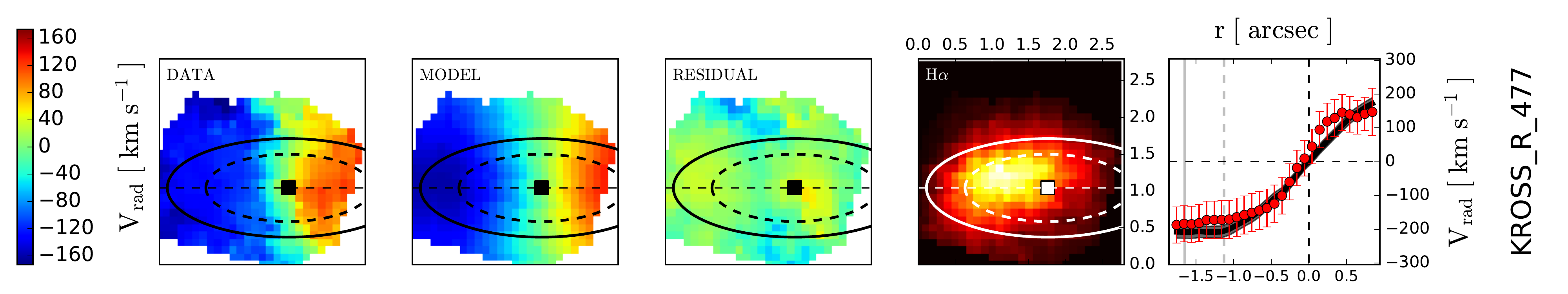}
\includegraphics[scale=0.45, trim=10 0 0 0, clip=true]{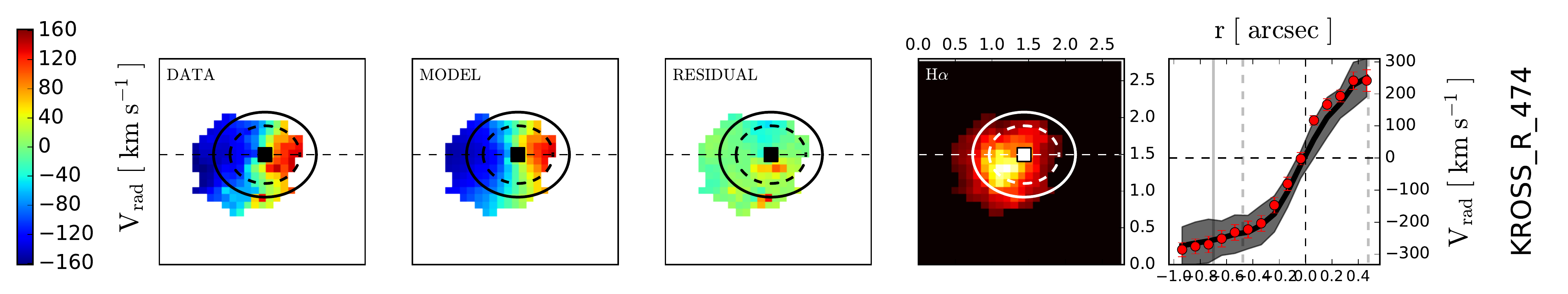}
\includegraphics[scale=0.45, trim=10 0 0 0, clip=true]{PR_ssB_31.pdf}
\includegraphics[scale=0.45, trim=10 0 0 0, clip=true]{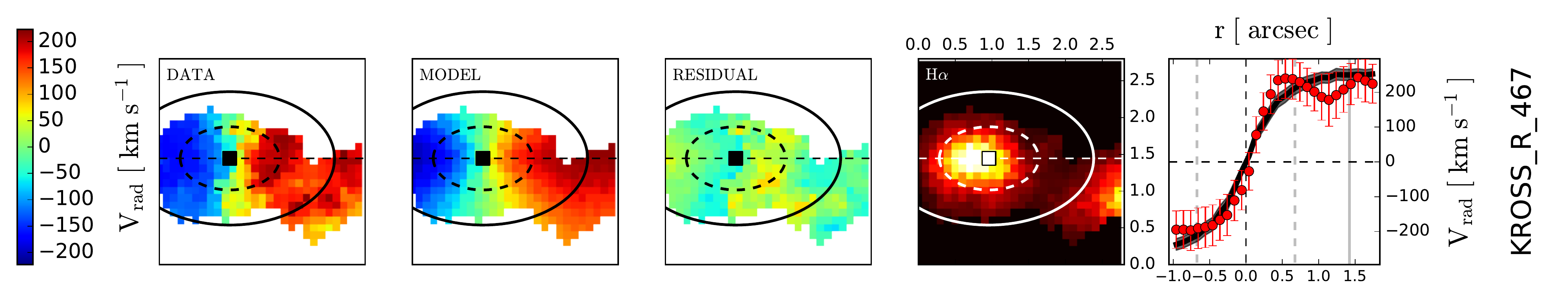}
\includegraphics[scale=0.45, trim=10 0 0 0, clip=true]{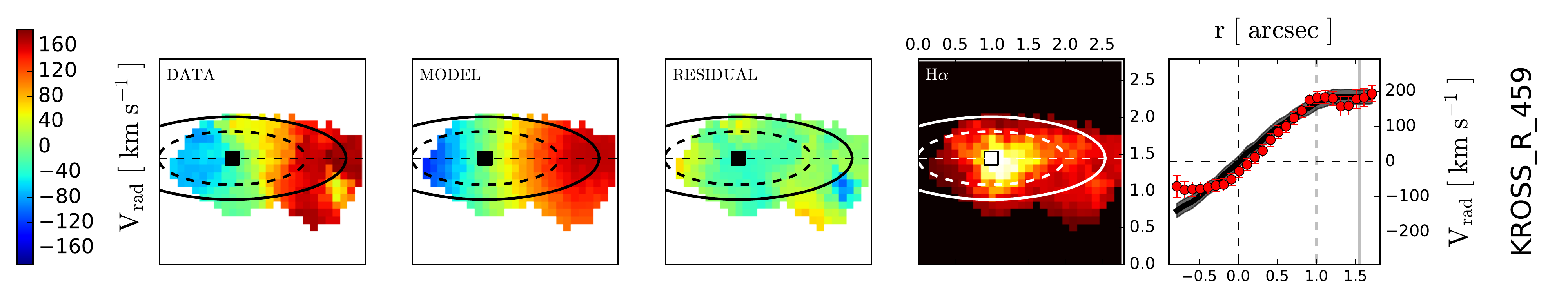}
\includegraphics[scale=0.45, trim=10 0 0 0, clip=true]{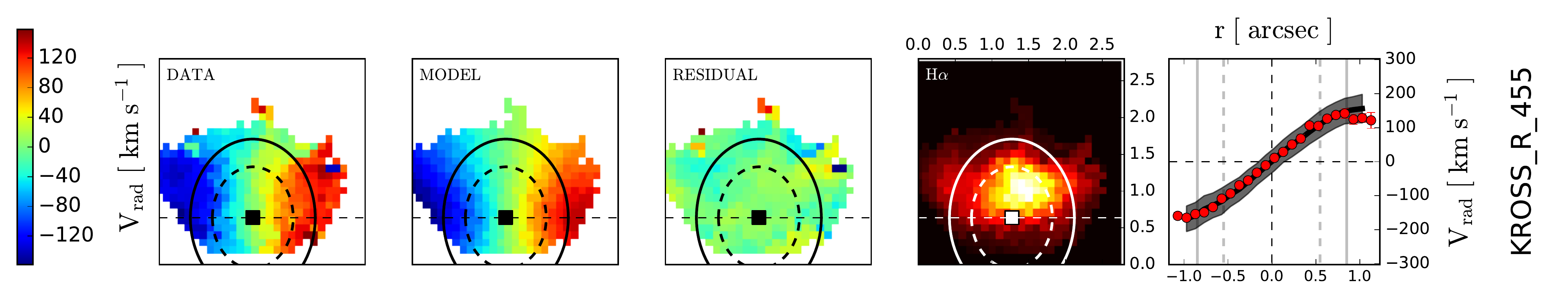}
\includegraphics[scale=0.45, trim=10 0 0 0, clip=true]{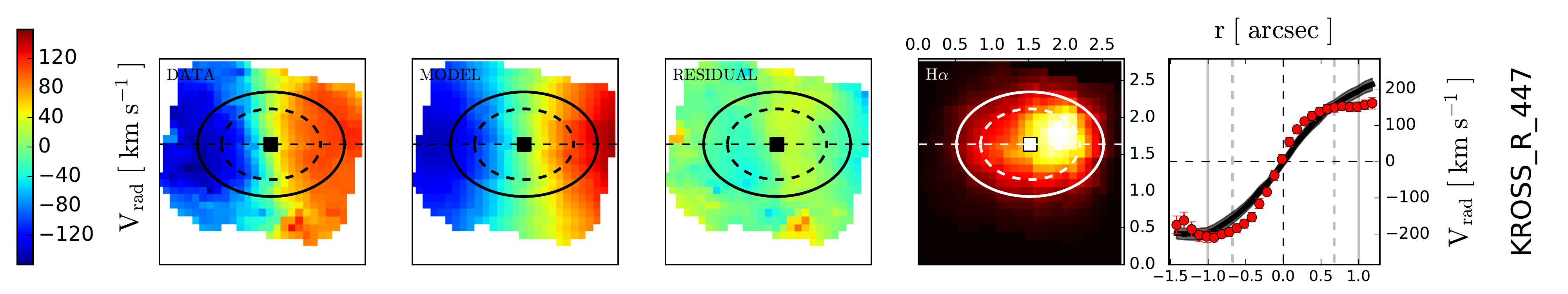}
\includegraphics[scale=0.45, trim=10 0 0 0, clip=true]{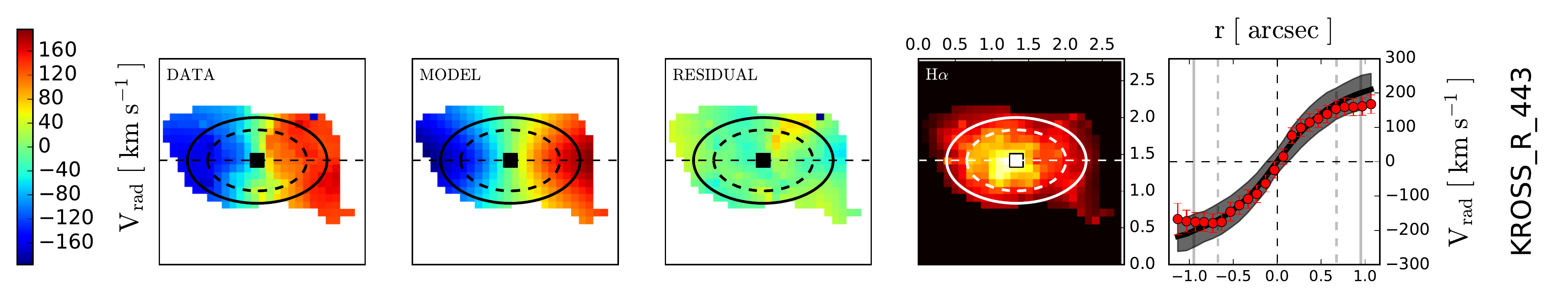}
\includegraphics[scale=0.45, trim=10 0 0 0, clip=true]{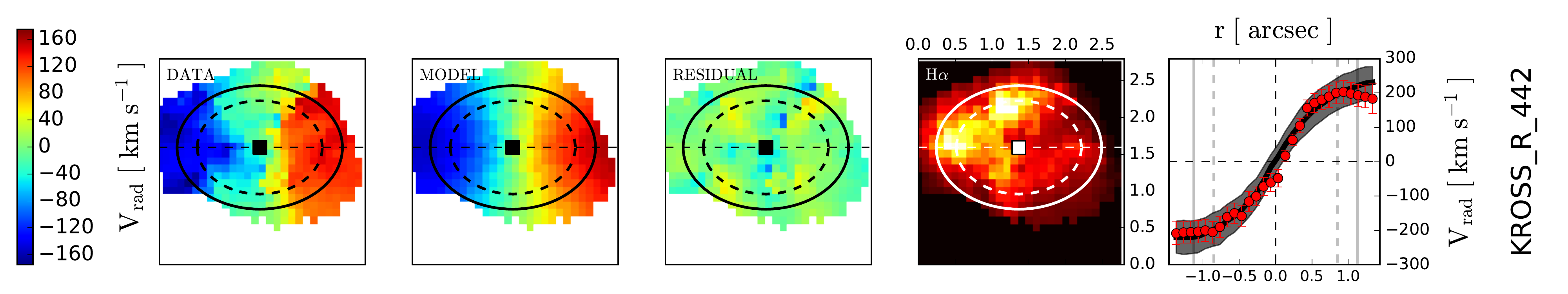}
\includegraphics[scale=0.45, trim=10 0 0 0, clip=true]{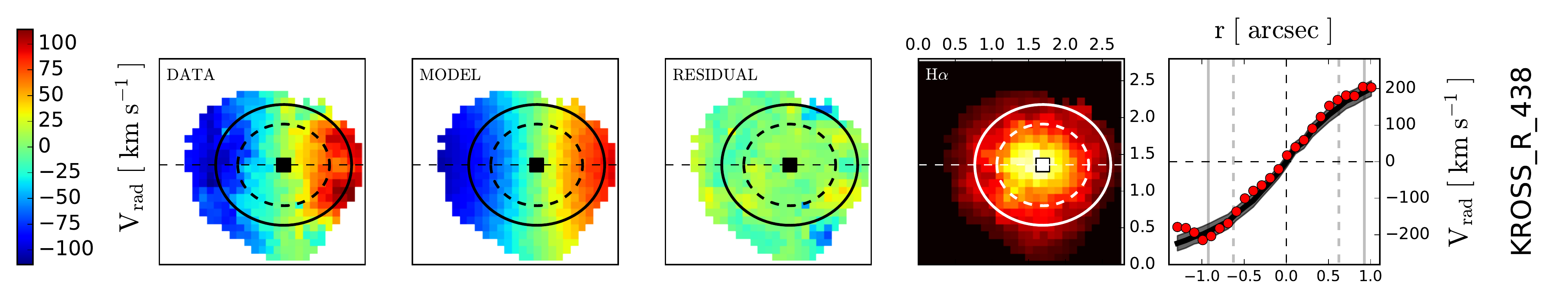}
\includegraphics[scale=0.45, trim=10 0 0 0, clip=true]{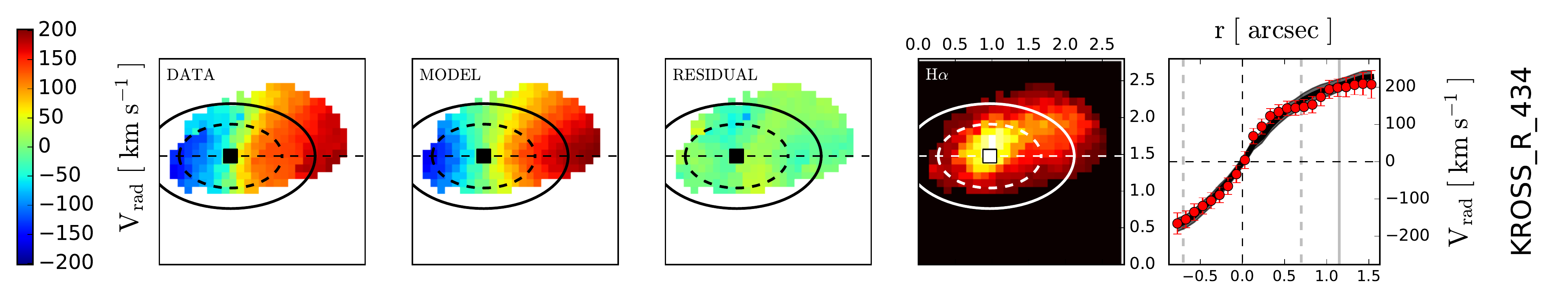}
\includegraphics[scale=0.45, trim=10 0 0 0, clip=true]{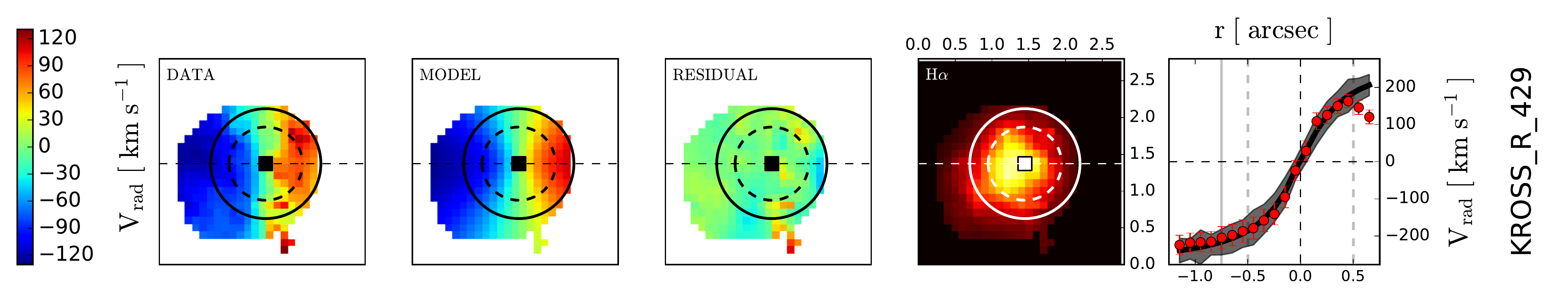}
\includegraphics[scale=0.45, trim=10 0 0 0, clip=true]{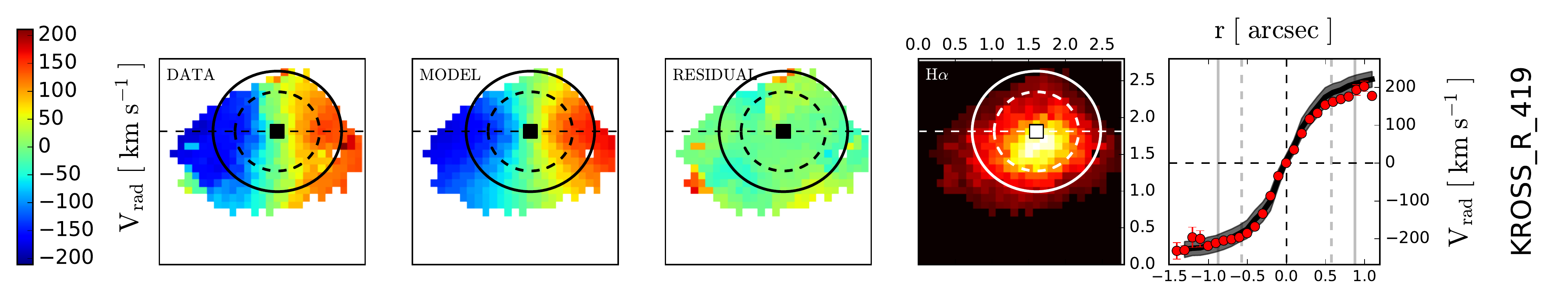}
\includegraphics[scale=0.45, trim=10 0 0 0, clip=true]{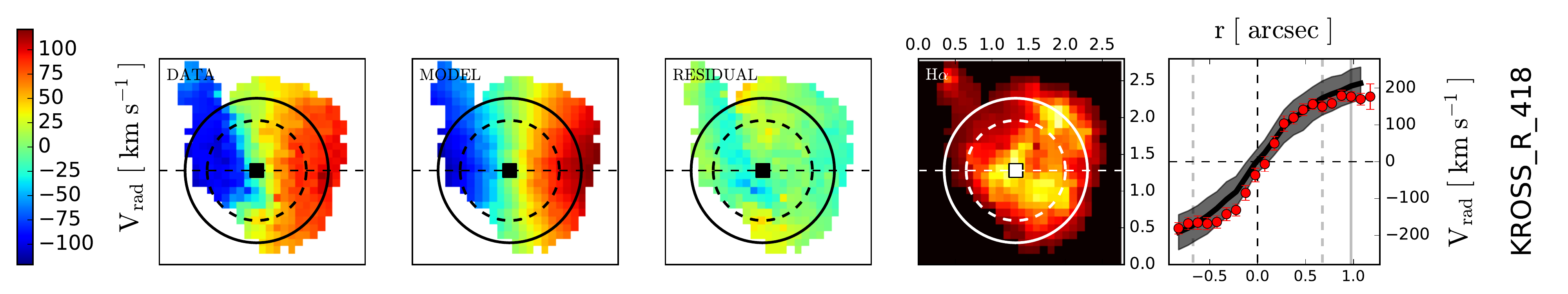}
\includegraphics[scale=0.45, trim=10 0 0 0, clip=true]{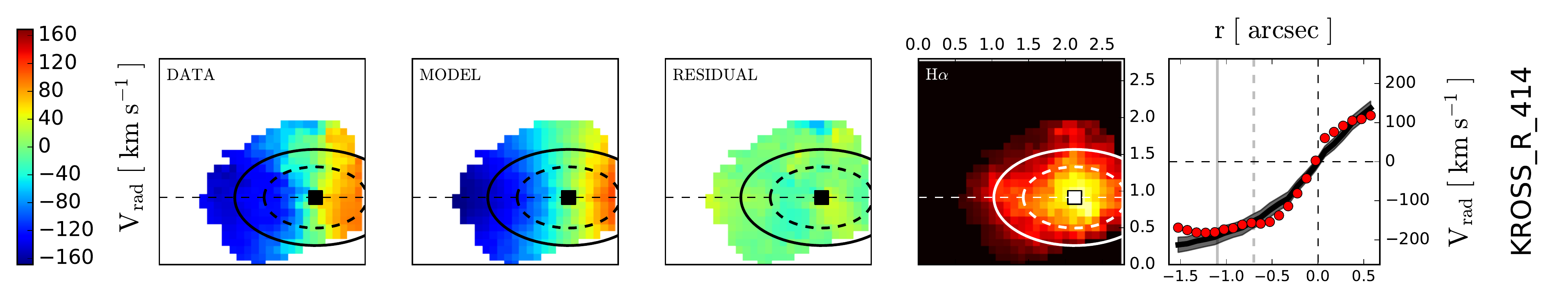}
\includegraphics[scale=0.45, trim=10 0 0 0, clip=true]{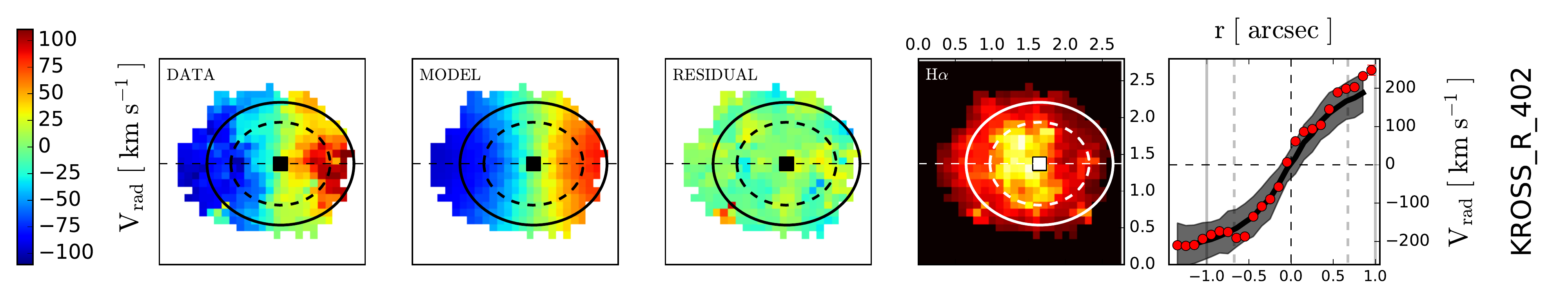}
\includegraphics[scale=0.45, trim=10 0 0 0, clip=true]{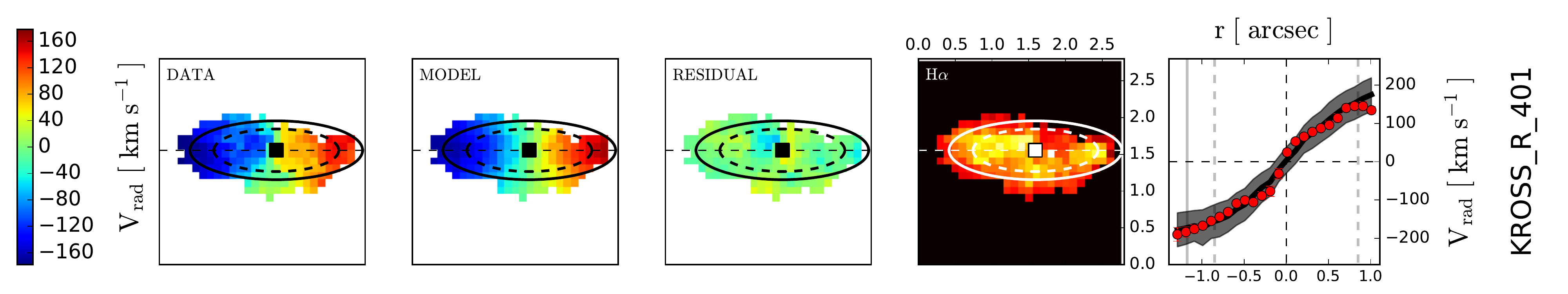}
\includegraphics[scale=0.45, trim=10 0 0 0, clip=true]{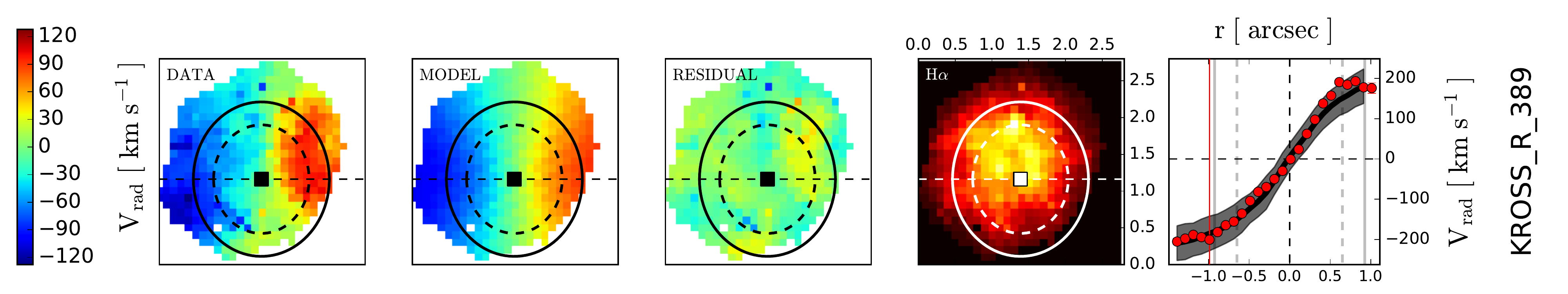}
\includegraphics[scale=0.45, trim=10 0 0 0, clip=true]{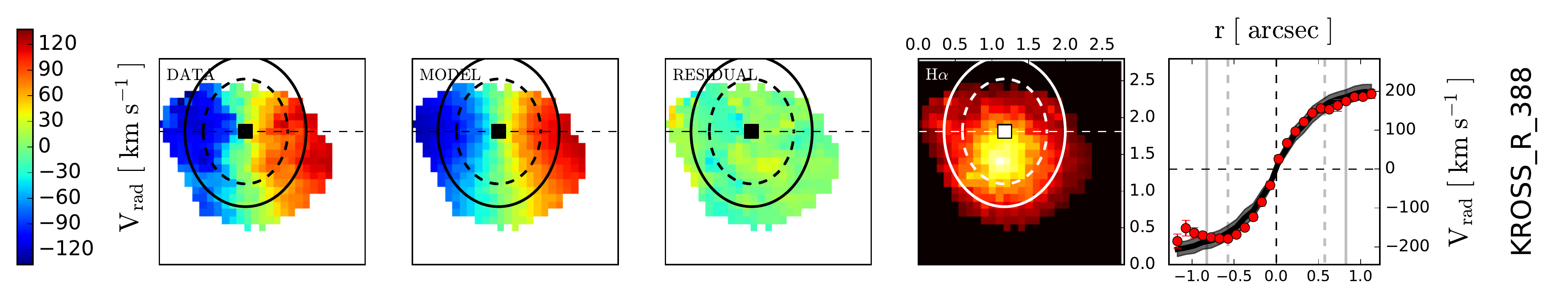}
\includegraphics[scale=0.45, trim=10 0 0 0, clip=true]{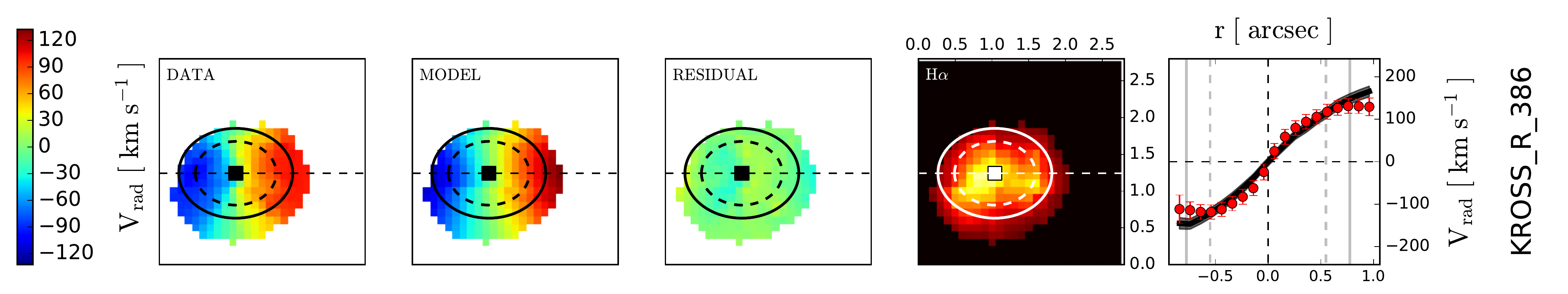}
\includegraphics[scale=0.45, trim=10 0 0 0, clip=true]{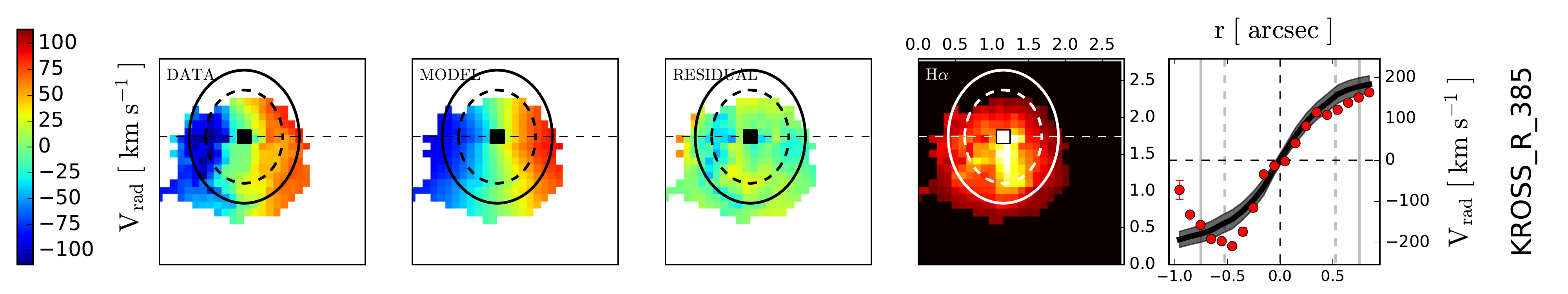}
\includegraphics[scale=0.45, trim=10 0 0 0, clip=true]{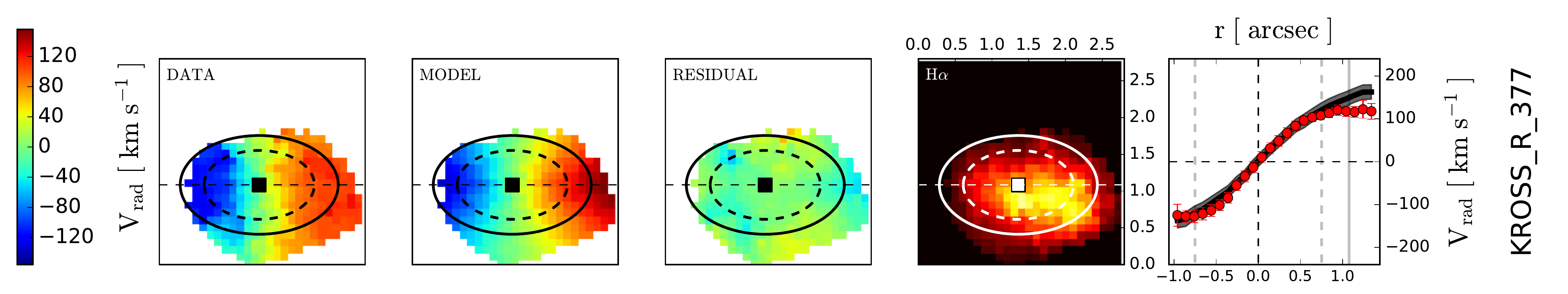}
\includegraphics[scale=0.45, trim=10 0 0 0, clip=true]{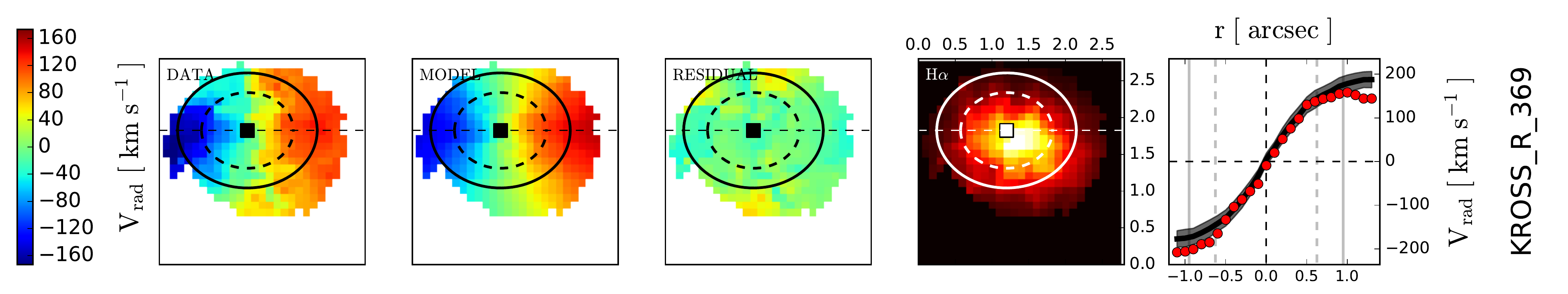}
\includegraphics[scale=0.45, trim=10 0 0 0, clip=true]{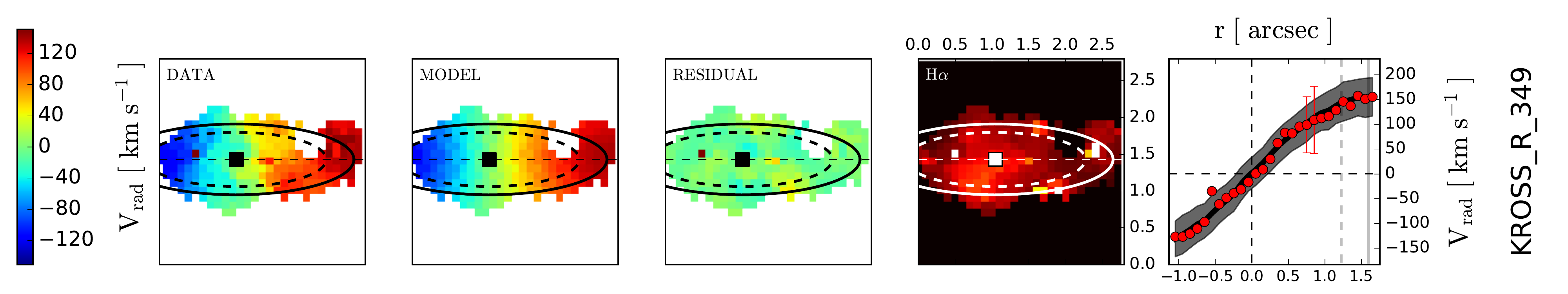}
\includegraphics[scale=0.45, trim=10 0 0 0, clip=true]{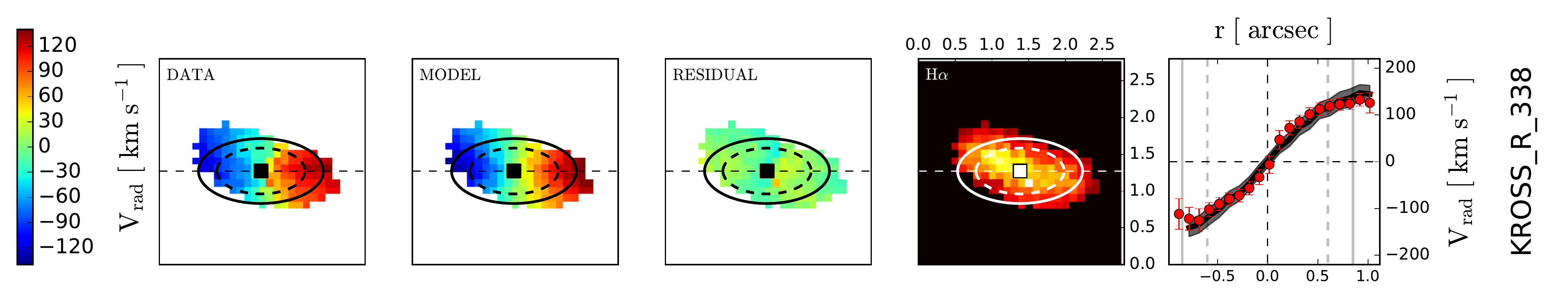}
\includegraphics[scale=0.45, trim=10 0 0 0, clip=true]{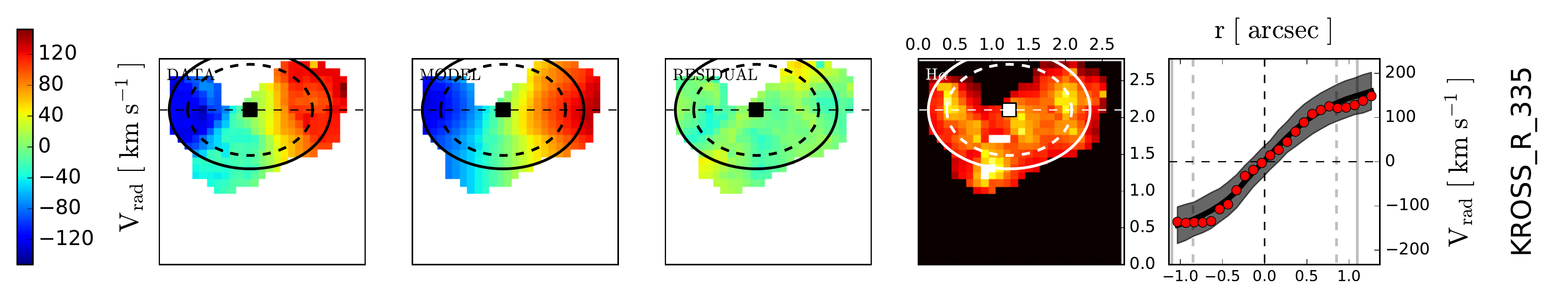}
\includegraphics[scale=0.45, trim=10 0 0 0, clip=true]{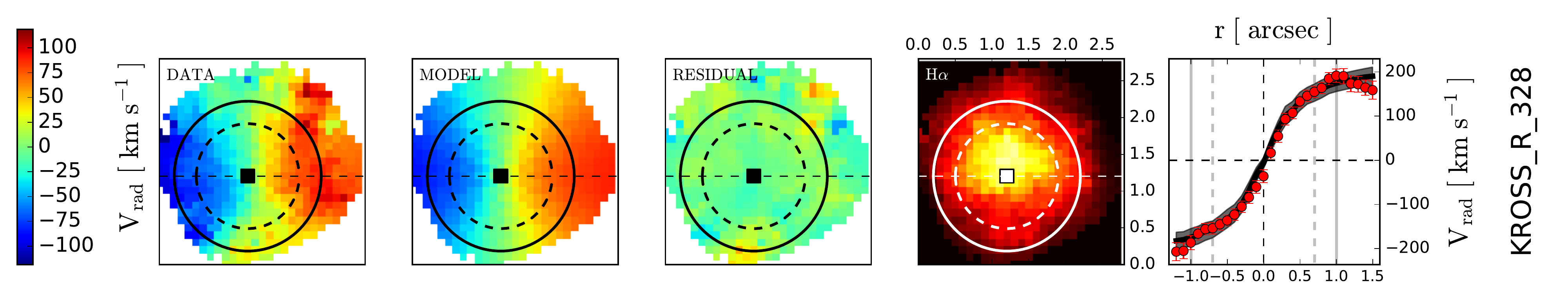}
\includegraphics[scale=0.45, trim=10 0 0 0, clip=true]{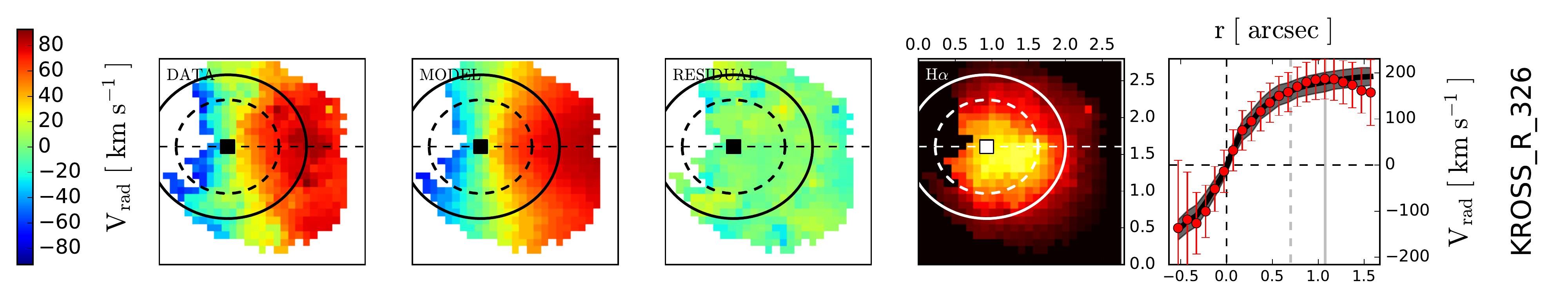}
\includegraphics[scale=0.45, trim=10 0 0 0, clip=true]{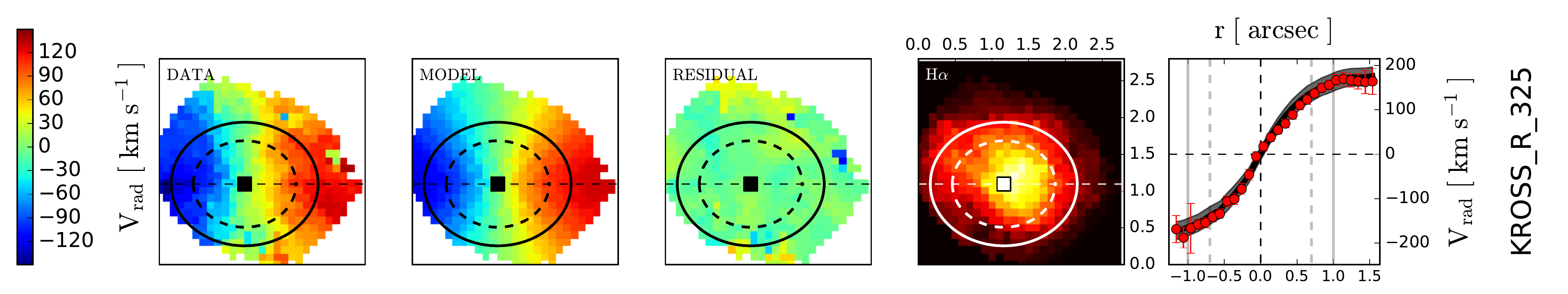}
\includegraphics[scale=0.45, trim=10 0 0 0, clip=true]{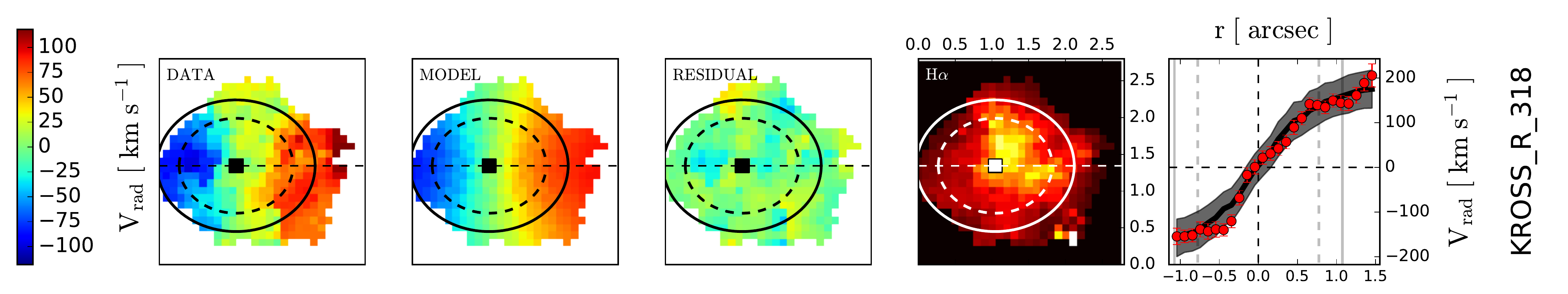}
\includegraphics[scale=0.45, trim=10 0 0 0, clip=true]{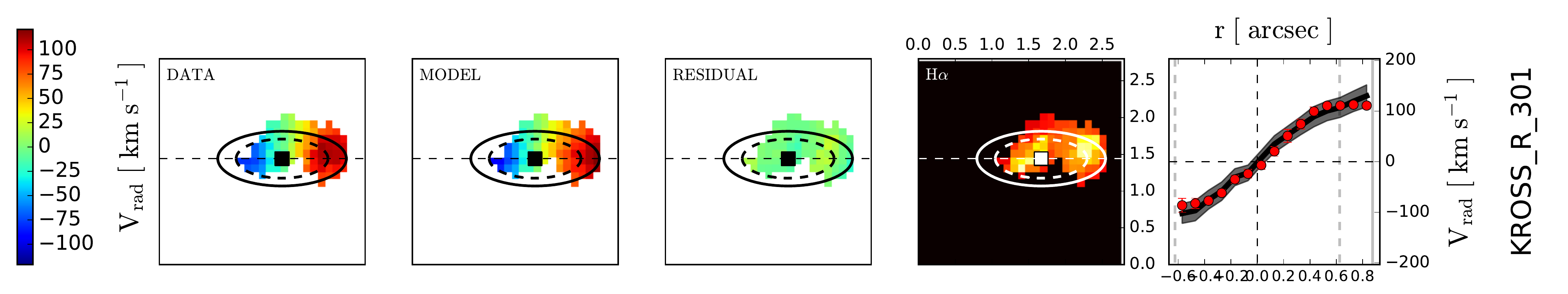}
\includegraphics[scale=0.45, trim=10 0 0 0, clip=true]{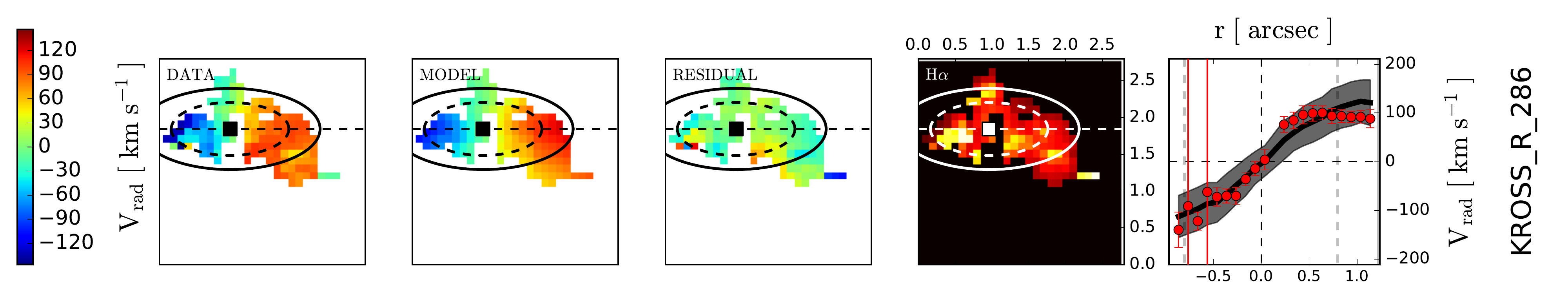}
\includegraphics[scale=0.45, trim=10 0 0 0, clip=true]{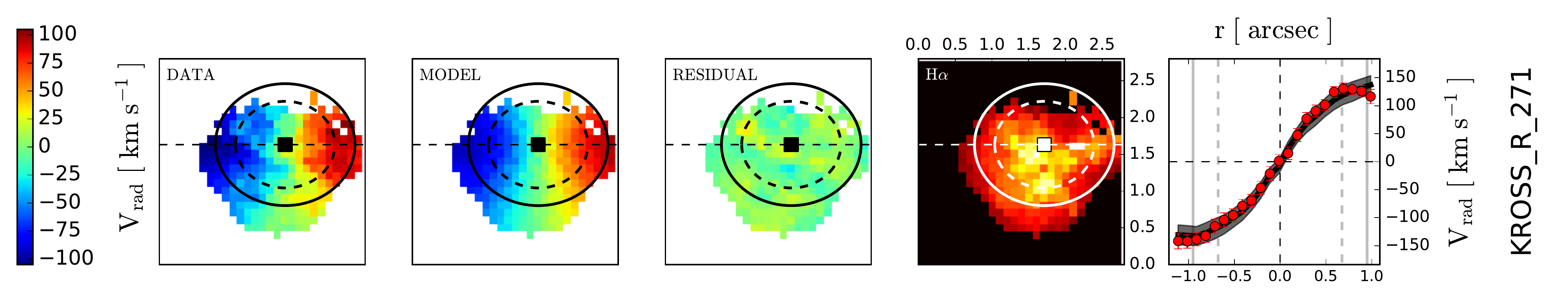}
\includegraphics[scale=0.45, trim=10 0 0 0, clip=true]{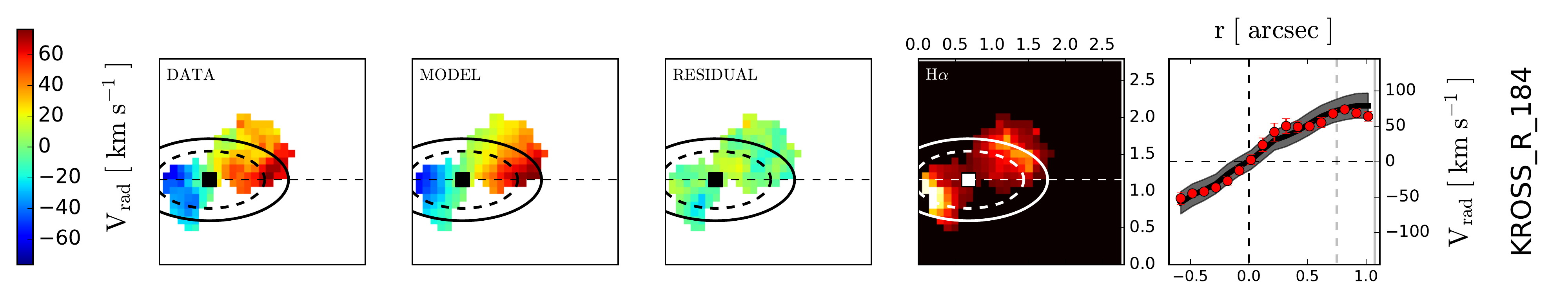}
\captionof{figure}{The best fitting model velocity fields (see \S\ref{subsec:modelvels}) for the KROSS {\it disky} sub-sample (see \S\ref{subsec:subsample}).}\label{fig:diskyfields}

\bsp	
\label{lastpage}
\end{document}